\documentclass[12pt]{article}
\usepackage{epsfig,amssymb,amsmath,psfrag,subfigure,cite}
\usepackage{hyperref}

\catcode`\@=11
\def \tr {\mathop{\rm tr}\nolimits}

\newcommand\lr[1]{{\left({#1}\right)}}

\newcommand \ket [1] {|{#1}\rangle}
\newcommand \bra [1] {\langle {#1}|}
\newcommand\re[1]{(\ref{#1})}
\def \qqquad {\qquad\quad}
\def \qqqquad {\qquad\qquad}

\newcommand{\cN}{{\cal N}}

\newcommand{\nt}{\notag\\}

\newcommand{\cZ}{{\cal Z}}
 
\newcommand{\p}[1]{(\ref{#1})}

\newcommand \vev [1] {\langle{#1}\rangle}

\newcommand{\pa}{\partial}
\newcommand{\ep}{\epsilon}
\renewcommand{\a}{\alpha}
\renewcommand{\b}{\beta}

\newcommand{\da}{{\dot\alpha}}

\newcommand{\bq}{\bar\theta}

 \newcommand{\cL}{{\cal L}}

\def\numberbysection{\@addtoreset{equation}{section}
                     \def\theequation{\thesection.\arabic{equation}}}

\numberbysection

\begin{document}

\thispagestyle{empty}

\null\vskip-43pt \hfill
\begin{minipage}[t]{50mm}
CERN-PH-TH/2014-270 \\
DCPT-14/79 \\
HU-EP-14/66\\
HU-MATH 2014-42\\
IPhT--T14/242 \\
LAPTH-239/14 
\end{minipage}

\vskip2.2truecm
\begin{center}
\vskip 0.2truecm

 {\Large\bf
 
Correlation functions of the chiral stress-tensor \\[2mm] multiplet in $\mathcal{N}=4$ SYM
}
\vskip 0.5truecm

\vskip 1truecm
{\bf  Dmitry Chicherin$^{a}$, Reza Doobary$^{b}$,  Burkhard Eden$^{c}$, Paul Heslop$^{b}$,  \\[2mm] Gregory P. Korchemsky$^{d}$, Lionel Mason$^{e}$,
Emery Sokatchev$^{a,f,g}$ \\
}

\vskip 0.4truecm
 $^{a}$ {\it LAPTH\,\footnote{Laboratoire d'Annecy-le-Vieux de Physique Th\'{e}orique, UMR 5108},   Universit\'{e} de Savoie, CNRS,
B.P. 110,  F-74941 Annecy-le-Vieux, France\\
  \vskip .2truecm
$^{b}$   Mathematics Department, Durham University,
Science Laboratories,
\\South Rd, Durham DH1 3LE,
United Kingdom \\
\vskip .2truecm
$^{c}$ {\it Institut f\"ur Mathematik und Physik, Humboldt-Universit\"at, Zum gro{\ss}en Windkanal 6, 12489 Berlin}
 \\
 \vskip .2truecm
$^{d}$ Institut de Physique Th\'eorique\,\footnote{Unit\'e Mixte de Recherche du CNRS, UMR 3681},
CEA Saclay, 
91191 Gif-sur-Yvette Cedex, France\\
\vskip .2truecm $^{e}$ Mathematics Department, Oxford University, Woodstock Road, OX2 6GG, United Kingdom  \\
\vskip .2truecm $^{f}$ Physics Department, Theory Unit, CERN, CH -1211, Geneva 23, Switzerland \\
\vskip .2truecm $^{g}$ Institut Universitaire de France,  103, bd Saint-Michel
F-75005 Paris, France 
                       } \\
\end{center}

\vskip -.2truecm

\centerline{\bf Abstract} 
\medskip
\noindent
{We give a new method for computing the correlation functions of the chiral part of the stress-tensor supermultiplet that 
relies on the reformulation of $\cN=4$ SYM in twistor space. It yields the correlation functions in the Born approximation 
as a sum of Feynman diagrams on twistor space that involve only propagators and no integration vertices. We 
use this unusual feature of the twistor Feynman rules to compute the correlation functions in terms of simple building blocks
which we identify as a new class of $\cN=4$ off-shell superconformal invariants. Making use of the duality between
correlation functions and planar scattering amplitudes, we demonstrate that these invariants represent an off-shell
generalisation of the on-shell   invariants defining tree-level scattering amplitudes in $\cN=4$ SYM.}

\newpage

\thispagestyle{empty}

{\small \tableofcontents}

\newpage
\setcounter{page}{1}\setcounter{footnote}{0}

\section{Introduction} 

In this paper, we continue the study of correlation functions of the operators in the stress-tensor supermultiplet $\mathcal T$ in $\mathcal N = 4$ SYM 
initiated in \cite{Eden:2011we,Eden:2012tu}. This supermultiplet plays a privileged role since it comprises all local conserved currents as well as the 
Lagrangian of the theory. Its correlation functions have a number of remarkable properties. 
The two- and  three-point functions are protected by superconformal symmetry and do not receive quantum corrections. The four-point  function $G_4=\vev{\mathcal T(1) \mathcal T(2) \mathcal T(3) \mathcal T(4)}$ 
is the first non-protected quantity.  At strong coupling it has been thoroughly
studied via the AdS/CFT correspondence \cite{Freedman:1998tz,Arutyunov:2000py}  whereas at weak coupling it has been  computed at one loop \cite{GonzalezRey:1998tk}, at two loops \cite{Eden:2000mv,Bianchi:2000hn} and recently at three loops \cite{Eden:2011we,Drummond:2013nda}. The operator product expansion of this correlation function has provided valuable data about the spectrum of anomalous dimensions of twist-two operators \cite{Dolan:2000ut}. 
The interest in these correlation functions, for an arbitrary number of points, has been renewed in the context of the recent studies of scattering amplitudes in $\mathcal N = 4$ SYM. The correlation functions have been found to 
be dual to the scattering amplitudes in a special light-like limit \cite{Alday:2010zy,Eden:2010zz,Mason:2010yk}. 

Computing the weak coupling corrections to these correlation functions
within the conventional Feynman diagram approach turned out to be a difficult task, even at low levels of the perturbative expansions. Already the evaluation of the two-loop correction to the four-point function needed judicious use of $\cN=1$ or $\cN=2$  supersymmetry \cite{Eden:2000mv,Bianchi:2000hn}. Going to higher orders became possible by using the Lagrangian insertion method combined with the recently discovered hidden permutation symmetry of $G_4$ that  mixes integration and external points \cite{Eden:2011we,Eden:2012tu}. More precisely, since the (on-shell chiral) Lagrangian of $\mathcal N=4$ SYM
appears as the top component in the chiral sector of the stress-tensor supermultiplet, the order $O(g^{2\ell})$ correction to $G_4$ can be related to the 
Born-level correlation function $G_{4+\ell}$ involving the insertion of $\ell$ additional chiral stress-tensor supermultiplets, integrated over their positions in the chiral 
superspace. The permutation symmetry follows from the Bose symmetry of the correlation function $G_{4+\ell}$.

This point illustrates the importance of the general multi-point correlation functions $G_n=\vev{\mathcal T(1)\dots \mathcal T(n)}$ 
of  the stress-tensor supermultiplet in the chiral sector. Another reason to study these is the above mentioned duality with scattering amplitudes. Knowing $G_n$ allows us to predict the general $n-$point tree-level superamplitude as well as the integrands of its perturbative corrections. 

The goal of the present paper is to develop a new approach to computing the correlation functions $G_n$ which makes efficient use of $\cN=4$ superconformal symmetry\footnote{Throughout the paper we always mean the {\it chiral half }of $\cN=4$ superconformal symmetry.}. Viewed as a function of the chiral odd variables $\theta$, $G_n$ admits 
the expansion 
\begin{align}\label{G-mod4}
G_n = G_{n;0} + G_{n;1} + \dots + G_{n; n-4} \,,
\end{align}
where $G_{n;p}$ is a homogenous polynomial in $\theta$ of degree $4p$. Notice that the expansion
terminates at $p=n-4$ (instead of the maximally allowed $p=n$) due to  $\mathcal N=4$ superconformal symmetry.
An important consequence of \re{G-mod4} is that for $n=4$ the correlation function coincides with its lowest
component, $G_4=G_{4;0}$, and so it does not depend on the Grassmann variables. 

Each term on the right-hand side of \re{G-mod4} should respect the $\cN=4$ superconformal symmetry. As a consequence, it can be expanded over a set of invariants $\mathcal I_{n;p}$ of this symmetry. As was shown in \cite{Eden:2011we}, for the bottom ($p=0$) and top ($p=n-4$) components the invariant is unique (up to an arbitrary function of conformal cross-ratios). For the remaining components in  \re{G-mod4} the number of invariants varies with $p$ and they have not been studied in the literature. One of the main goals of this paper is to provide a convenient basis for such invariants in twistor superspace.
 
Note that the expansion \re{G-mod4} is very similar to that of the $n-$particle scattering super-amplitude in $\mathcal N=4$ SYM. In fact, the two
quantities are related to each other in the limit in which the operators $\mathcal T(i)$ are located at the vertices of light-like
$n-$gon \cite{Alday:2010zy,Eden:2010zz,Eden:2011yp,Eden:2011ku,Mason:2010yk,Adamo:2011dq}. This duality yields  non-trivial relations between the invariants $\mathcal I_{n;p}$ and their on-shell
counter-parts defining the scattering amplitudes. It is in this sense that we can think of $\mathcal I_{n;p}$ as  the off-shell generalisation
of the on-shell (amplitude) invariants. In particular, in the simplest non-trivial case $p=1$, in the light-like
limit  the off-shell invariants $\mathcal I_{n;1}$ are related to the NMHV $R-$invariants \cite{Drummond:2008vq,Mason:2009qx}.  

Computing the higher components  $G_{n;p}$ in \re{G-mod4} and finding the corresponding off-shell superconformal invariants $\mathcal I_{n;p}$
proves to be a very non-trivial problem. In the conventional approach, the Born approximation to $G_{n;p}$ is given 
by a set of Feynman diagrams 
with many interaction vertices and the associated Feynman integrals. The number of diagrams and their complexity rapidly increase with the Grassmann
degree $p$. Moreover, the contribution of each individual diagram is neither gauge invariant nor (super)conformally covariant. The $\cN=4$
superconformal symmetry is only restored in the sum of all diagrams.  

In this paper we demonstrate  that these difficulties can be avoided by employing the reformulation of $\cN=4$ SYM in twistor space \cite{Boels:2006ir}.
We find a representation of the chiral part of the stress-tensor supermultiplet ${\cal T}$ as a four-fold fermionic integral of the main interaction term in the twistor Lagrangian.  In the judiciously chosen axial gauge, the self-dual sector of SYM is free and has no interaction vertices. Furthermore, all the interaction vertices are comprised in the non-polynomial expression for ${\cal T}$ in terms of the twistor superfield. As a result, the correlation function $G_{n;p}$ is given in the Born approximation by a new type of Feynman diagram  which only involves free propagators of twistor superfields but no interaction vertices.  The calculation of the twistor space Feynman diagrams is drastically simplified (no Feynman integrals!)  and yields very
concise expressions for $G_{n;p}$. We check by an explicit calculation that the results for $G_{n;1}$ obtained  by the new method agree with those of the conventional Feynman diagram approach.

Analysing the Feynman diagrams in twistor space, we introduce a new class of $\cN=4$ off-shell superconformal invariants and study their properties. 
The simplest invariant $R(1;234)$  is given by a nilpotent Grassmann polynomial of degree two in the odd variables $\theta$. It  depends on four points and an 
auxiliary (reference) supertwistor defining the axial gauge for the twistor action. This invariant serves as an elementary building block for constructing higher-point
invariants. Namely, the general $n-$point invariant $\mathcal I_{n;p}$ factorises into a product of $2p$ elementary $R-$invariants. We show that the correlation 
function \re{G-mod4} is given in the Born approximation  by a linear combination of such off-shell invariants with rational coefficient functions 
of the distances $x_{ij}^2\equiv (x_i-x_j)^2$. Although each invariant
depends on the  reference supertwistor, this dependence drops out  in their sum.

The paper is organised as follows. In Section 2 we define the correlation function of the stress-tensor multiplet in the chiral sector and 
summarise its properties.  In Section 3 we reformulate this correlation function in twistor space and develop a diagram technique
for computing its components $G_{n;p}$ of a given Grassmann degree $4p$. In Section 4 we present an explicit calculation of the
first non-trivial component $G_{n;1}$ and show that it satisfies all necessary consistency conditions (operator product expansion
and duality with the NMHV amplitude in the light-like limit). In Section 5, we apply the conventional Feynman diagram technique to 
compute the five-point correlation function $G_{5;1}$ in the Born approximation. In Section 6 we match the two approaches and demonstrate
that they lead to the same expressions for various components of the four- and five-point correlation functions. Section 7 contains
concluding remarks. Some technical details are summarised in four appendices.
 
\section{Correlation functions of the stress-tensor multiplet}

In this section, we define the correlation functions of the operators in the stress-tensor supermultiplet
in $\mathcal N=4$ SYM and discuss their general properties. A distinctive feature of this multiplet is 
that it comprises the stress-energy  tensor (hence the name) and 
the Lagrangian of the theory. They appear as coefficients in the expansion of the corresponding 
superfield $\mathcal T(x,\theta^A,\bar\theta_A)$ in powers of the odd coordinates $ \theta_{\alpha}^A$ and
$\bar\theta^{\dot\alpha}_{A }$ (with Lorentz spinor indices $\alpha=1,2$, $\dot\alpha=\dot 1,\dot 2$ and $SU(4)$ index  $A=1,\dots,4$). In addition, this
superfield is annihilated by half of the Poincar\'e supercharges and, as a consequence, it depends on 
half of the odd variables, both chiral and anti-chiral: 
\begin{align}\label{eq:2}
 \mathcal T = \mathcal T(x ,\theta^+,\bar\theta_{-},u) 
\,,\qqquad 
  \theta_{\alpha}^{+a} = \theta_{\alpha}^A u_A^{+a} \,,\qqquad 
 \bar\theta^{\dot\alpha}_{-a' } = \bar\theta^{\dot\alpha}_{A }  \bar u^A_{-a'}\,.
\end{align}
Here the odd coordinates $\theta^A$ and $\bar\theta_A$ appear projected with auxiliary bosonic
 variables $u_A^{+a}$ and $\bar u^A_{-a'}$ with $a=1,2$, $a'=1',2'$ (see Appendix \ref{app:conv} for details), or `harmonics' on the coset $SU(4)/(SU(2)\times SU(2)'\times U(1))$. The harmonics allow us to define the so-called Grassmann analytic (or just `analytic') superspace with odd coordinates $\theta^+$ and $\bar\theta_{-}$, without breaking the $R-$symmetry $SU(4)$. More details can be found in Refs.~\cite{HSS,HH,HH2} (see also footnote \ref{f5}).

For our purposes in this paper we shall restrict $ \mathcal T$ to its purely chiral sector by setting $\bar\theta^{\dot\alpha}_{-a' }=0$. 
Then the expansion of the superfield in powers of $\theta^+$ has the form\footnote{Here we use  the notation  $ (\theta^+)^2_{\alpha\beta} = \theta_{\alpha}^{+a}\theta_{\beta}^{+b}\epsilon_{ab}$, $
 (\theta^+)^{2\, ab}= \theta_{\alpha}^{+a}\theta_{\beta}^{+b}\epsilon^{\alpha\beta}$, $
 (\theta^+)^{3\, a}_\alpha = \theta_{\alpha}^{+b}\theta_{\beta}^{+c}\theta_{\gamma}^{+a}\epsilon_{bc}\epsilon^{\beta\gamma}$ and $(\theta^+)^4 = \theta_{\alpha}^{+a}\theta_{\beta}^{+b}\theta_{\gamma}^{+c}\theta^{+d}_\delta \epsilon_{bc}\epsilon_{ad} \epsilon^{\alpha\beta}\epsilon^{\gamma\delta}$.}
\begin{align}\notag\label{T-dec}
\mathcal T(x,\theta^+,0,u) & {}= O^{++++} (x) +  \theta_\alpha^{+a}  O^{+++,\alpha}_a(x)
+ (\theta^+)^2_{\alpha\beta}O^{++,\alpha\beta}(x)
\\[2mm] &{}
+ (\theta^+)^{2\, ab}O^{++}_{ab}(x)
+ (\theta^+)^{3\, a}_\alpha O^{+,\alpha}_a (x)+ (\theta^+)^4\mathcal L(x)\,,
\end{align}
where the lowest component (or superconformal primary) $O^{++++} =\tr(\phi^{++}\phi^{++})$  is a half-BPS operator
 built from the scalar fields $\phi^{++} = \phi^{AB} u_A^{+a}u_B^{+b}\epsilon_{ab}$
and the top component $\mathcal L(x)$ is the chiral form of the $\mathcal N=4$ SYM on-shell Lagrangian. The remaining components
can be obtained by successively applying the chiral $\mathcal N=4$ supersymmetry transformations to the lowest component \cite{Eden:2011yp}. Their explicit expressions are given in Eq.~\re{t4c} below. Notice that $\mathcal T$ carries four units of harmonic $U(1)$ charge, as indicated by the number of pluses in each term on the right-hand side.
  
In this paper we propose a new approach to evaluating the correlation functions of the stress-tensor multiplet
\begin{align}\label{Gn}
G_n = \vev{0| \mathcal T(1) \dots   \mathcal T(n) |0}\,,
\end{align}
where we used the short-hand notation $\mathcal T(i) = \mathcal T(x_i,\theta_i^+,0,u_i)$ so that $G_n$ depends
on $n$ copies of the chiral superspace coordinates $(x_i, \theta_i^+,u_i)$.  $\mathcal N=4$ superconformal symmetry imposes
strong constraints on $G_n$.  In particular, for $n=2$ and $n=3$, the super-correlation function  \re{Gn}
is a protected quantity, independent of the coupling constant. Moreover, it does not depend on the chiral odd
variables and coincides with the correlation function of the lowest component $\tr[\phi^{++}\phi^{++}]$ evaluated at Born level.

For $n\ge 4$ the correlation function \re{Gn}  depends on the coupling constant $g^2$. This dependence can be controlled 
through the Lagrangian insertion method which relies on the following relation
\begin{align}\notag\label{LI}
  {\partial\over\partial g^2} G_n &{} = \int d^4 x_{n+1}\, \vev{0| \mathcal T(1) \dots   \mathcal T(n)\mathcal L(x_{n+1}) |0}
\\\notag
&{}=\int d^4 x_{n+1}d^4\theta^+_{n+1}\,\vev{0| \mathcal T(1) \dots   \mathcal T(n)\mathcal T({n+1}) |0}
\\
&{}
\equiv\int d^4 x_{n+1}d^4\theta^+_{n+1}\,G_{n+1}\,.
\end{align}
Here in the second line we made use of the relation between the on-shell action of $\mathcal N=4$ SYM and the stress-tensor multiplet
\begin{align}
S_{\mathcal N=4} = \int d^4 x \,\mathcal L(x) = \int d^4 x\int d^4\theta^+ \,{\mathcal T}(x,\theta^+,0,u)
\end{align}
that follows from \re{T-dec}. Expanding the correlation functions in \re{LI} in the powers
of the coupling constant,  we find from \re{LI} that the order $O( g^{2\ell})$ correction to $G_n$ is determined
by the order $O( g^{2\ell-2})$ correction to $G_{n+1}$, integrated over the position of the $(n+1)-$th point. Successively
applying \re{LI} we can express the $O(g^{2\ell})$ integrand of $G_n$ in terms of the correlation function $G_{n+\ell}$
evaluated at the lowest order in the coupling, i.e., in the Born approximation.  

This property shows that in order to find any quantum correction to the above correlation function it is sufficient to evaluate \re{Gn} at Born level and for an arbitrary number of points $n$.
In this approximation $G_n$ is a rational function of the distances $x_{ij}^2\equiv (x_i-x_j)^2$.
This function can be reconstructed if we known the form of its singularities  
corresponding to null separations $x_{ij}^2=0$ between the operators in \re{Gn}.

The various components of the correlation 
function \re{G-mod4} have different dependence on the coupling constant $g^2$ and on the number of colours $N$. 
As follows from \re{Gn} and \re{T-dec}, the lowest component $G_{n;0}$ is given by the correlation function of 
scalar operators $\tr(\phi^{++} \phi^{++})$ and reduces, in the Born approximation, to a product of free scalar propagators.
Therefore, it does not depend on the coupling constant and scales as $G_{n;0}\sim {\rm dim} (SU(N))= N^2-1$. 
The higher components $G_{n;p}$ in \re{G-mod4} are given by more complicated correlation functions involving other members
of the supermultiplet \re{T-dec}. As we show later in the paper, their perturbative expansion necessarily involves 
interaction vertices whose number increases with $p$. Each vertex is accompanied by a power of the coupling constant $g$, so that 
$G_{n;p}$ scales in the Born approximation as 
\begin{align}\label{planar}
G_{n;p} = {N^2-1\over (2\pi)^{2n} } \left(g^{2} N \over 4\pi^2 \right)^p \widehat G_{n;p}\,,
\end{align}
with $\widehat G_{n;p}$ depending on the $n$ superspace points and on the parameter $1/N^2$ controlling
the non-planar corrections.  According to \cite{Eden:2012tu}, non-planar corrections only exist for $p\ge 4$ due to the
occurrence of the higher Casimir operators of the  gauge group $SU(N)$  in the individual Feynman diagrams.~\footnote{The simplest 
example is the quartic Casimir operator $d^{abcd} d^{abcd}/(N^2-1) =(N^4-6N^2+18)/(96 N^2)$ that first appears for $p=4$.}
The correlation function $G_{n;p}$ involves an overall factor which is a product of free scalar propagators, each
bringing a factor of $1/(2\pi)^2$. For the sake of simplicity of the formulae, in what follows we shall not display the normalisation factor in \re{planar}.
 
By construction,  the correlation functions $G_{n;p}$ have to respect (the chiral half of) $\cN=4$ superconformal
symmetry and  to satisfy the corresponding Ward identities. The general solution to these identities is given by a linear
combination of $\cN=4$ superconformal nilpotent invariants $\mathcal I_{n;p}$ whose number depends on the Grassmann degree 
$p$. As was shown in \cite{Eden:2011we}, for the top component of the correlation function with  $p=n-4$ the corresponding invariant 
$\mathcal I_{n;n-4}$ is unique leading to
\begin{align}\label{G-max}
  G_{n,n-4} =  {\mathcal I_{n;n-4} \over \prod_{1\le i<j\le n} x_{ij}^2}\,.
\end{align}
The explicit expression for $ \mathcal I_{n;n-4} $ can be found in \cite{Eden:2011we}. 

In this paper, we extend the relation \re{G-max} to the remaining components $G_{n;p}$ of the correlation function  \re{G-mod4} 
with $p<n-4$. Namely, we shall construct the set of $\cN=4$ superconformal invariants $\mathcal I_{n;p}$ and determine
their contributions to $G_{n;p}$.

\section{Correlation functions on twistor space}

In this section, we present a new approach to computing the correlation functions \re{Gn} that relies on the 
reformulation of $\mathcal N=4$ SYM as a gauge theory on twistor space based on a twistor action.  The twistor space Feynman diagrams that arise from this twistor action provide an off-shell generalization of the MHV diagrams of \cite{Cachazo:2004kj} that give rise to scattering amplitudes.  The framework extends to null polygonal Wilson loops \cite{Bullimore:2010pj,Mason:2010yk} and other correlators \cite{Adamo:2011dq,Adamo:2011pv} giving dual conformal invariant versions of MHV diagrams for the amplitude or standard ones for the Wilson loop.  

Here we show how to obtain  Feynman rules on twistor space for the correlation functions  \re{Gn}  that avoid many of the 
difficulties of conventional space-time Feynman diagrams. 
The main advantage of the twistor rules as opposed to the conventional ones
is that the contribution of each diagram manifests the $\mathcal
N=4$ superconformal symmetry up to the choice of a reference twistor that has been used to define the axial gauge. Its contribution remains invariant under a superconformal transformation acting on all external data, if we in addition transform the reference supertwistor linearly. In the sum over all diagrams, dependence on the reference supertwistor drops out as we shall prove below.
  There are also relatively few diagrams compared to the conventional ones, particularly at low MHV degree.

\subsection{Twistor space approach}
Non-projective twistor space 
is the fundamental representation space of the complexified spinor covering of the super conformal group $SL(4|4;\mathbb{C})$. We first explain how the bosonic  conformal group in this form acts on space-time and how it relates to bosonic twistor space and then  build up to the full supersymmetric correspondence.

As mentioned above, the correlation functions \re{Gn} at Born level are rational functions of the distances $x_{ij}^2$. Therefore, they admit analytic continuation to complex space-time coordinates. This is an advantage because the action of the complexified conformal group $SL(4;\mathbb{C})$ on the correlation functions can be greatly simplified by employing the embedding formalism, in which complexified compactified Minkowski space is realised as a light-cone in complex projective space $\mathbb{CP}^5$ with homogenous coordinates $X^{IJ} \sim c X^{IJ}$
(with $I,J=1,\dots,4$)
\begin{align}
(X\cdot X) \equiv X_{IJ} X^{IJ} =0\,,  \label{quadric}
\end{align}
where $X_{IJ} =\frac12 \epsilon_{IJKL} X^{KL}$ and $X^{IJ} = - X^{JI}$. The complex coordinates $x_{\alpha\dot\alpha}$ 
define a particular parameterisation of $X^{IJ}$
\begin{align}\label{X-x}
X^{IJ} =  \left[\begin{array}{cc}  \epsilon_{\alpha\beta} & -i x_\alpha^{\dot\beta}  \\ i x_\beta^{\dot\alpha} & -  x^2 \epsilon^{\dot\alpha\dot\beta} \end{array}\right] ,
\end{align}
with $x_\alpha^{\dot\beta}= x_{\alpha\dot\alpha}\epsilon^{\dot\alpha\dot\beta}$ and $x^2=\frac12 x_\alpha^{\dot\beta} x^\alpha_{\dot\beta}$.
Conformal transformations of $x_{\alpha\dot\alpha}$ correspond to global $SL(4;\mathbb{C})$ transformations of $X^{IJ}$.

Bosonic  twistor space is the complex projective space $\mathbb{CP}^3$ whose homogenous
coordinates $Z^I\sim c Z^I$ (with $I=1,\dots,4$) transform in the fundamental representation of the cover $SL(4;\mathbb{C})$ of the conformal group. A space-time point  $X^{IJ}$ corresponds to a line  in twistor space given by the incidence relation
\begin{align}\label{inc}
X_{IJ} Z^J = 0\,.
\end{align}
For a given point $X^{IJ}$ this relation defines a line in twistor space since \eqref{quadric} is the condition that $X_{IJ}$ has rank two. Choosing two arbitrary points on this line, $Z_1^J$ and $Z_2^J$, we
can reconstruct $X^{IJ}$ as
\begin{align}\label{X-Z}
X^{IJ} = Z_1^I Z_2^J - Z_1^J Z_2^I  = \epsilon^{ab} Z_a^I Z_b^J \, .
\end{align}
Combining \re{X-x} and \re{X-Z} we obtain that each point in complexified Minkowski space-time $x_{\alpha\dot\alpha}$ is mapped into
a line $X_{IJ} Z^J(\sigma) = 0$ in twistor space~\footnote{More precisely this is a line in projective twistor space $\mathbb{CP}^3$ or equivalently a  two-plane in (non-projective) twistor space $\mathbb{C}^4$. So  Minkowski space is the  Grassmannian of two-planes in $\mathbb{C}^4$, $Gr(2,4)$.}
\begin{align}
Z^I(\sigma) = Z_1^I \sigma^1 + Z_2^I \sigma^2 \equiv Z_a^I \sigma^a\,,
\end{align}
with $\sigma^a=(\sigma^1,\sigma^2)$ being local coordinates on the line. 
For $n$ points $x_i$, defining the space-time coordinates of the 
operators in the correlation function \re{Gn}, the corresponding configuration in twistor space consists of $n$ (non-intersecting) lines 
whose moduli are determined by the corresponding projective coordinates $X_i^{IJ}$,  as shown in Fig.~\ref{fig-sample} below. Then, the (square of the) distance between two operators is given by
\begin{align}\notag
x_{ij}^2 \sim \frac12 (X_i\cdot X_j) &{} = \frac14 \epsilon_{IJKL} X_i^{IJ} X_j^{KL} 
\\
& {}= \frac14 \epsilon_{IJKL}  \epsilon^{ab} Z_{i,a}^I Z_{i,b}^J
\epsilon^{cd} Z_{j,c}^K Z_{j,d}^L \equiv \langle Z_{i,1} Z_{i,2} Z_{j,1} Z_{j,2}\rangle\,,
\end{align}
where $Z_{i,a}$ and $Z_{j,a}$ (with $a=1,2)$ are two pairs of points belonging to two lines with moduli $X_i$ and $X_j$, respectively.
If two lines intersect, we can choose $Z_{i,2}^I=Z_{j,1}^I$ leading to $x_{ij}^2=0$. Thus, the light-like limit of the correlation function, $x_{ij}^2\to 0$, 
corresponds to the limit of intersecting lines.

To deal with correlation functions in $\mathcal N=4$ SYM in the chiral sector, we have to extend the twistor space to include four odd 
coordinates
\begin{align}\label{ZZ}
\mathcal Z  = (Z^I,\chi^A)\,,\qqqquad \text{(with $I,\, A=1,\dots,4$)}\,,
\end{align}
subject to the equivalence relation $\mathcal Z  \sim c \mathcal Z $. The odd twistor 
coordinates $\chi^A$ satisfy an incidence relation analogous to \re{inc}. Using the parameterisation \re{X-x} we
can rewrite the relation between a point in chiral Minkowski (super)space-time $(x^{\dot\alpha\alpha},\theta^{A\alpha})$
and a line  in twistor superspace as
\begin{align}\label{line}
Z^I = (\lambda_\alpha, i x^{\dot\alpha\beta}\lambda_\beta)\,,\qqqquad \chi^A = \theta^{A,\beta} \lambda_\beta\,,
\end{align} 
with $\lambda_\alpha$ being homogeneous coordinates on the line in
twistor space.\footnote{Again, more precisely this identifies chiral Minkowsksi superspace with the space of lines in {\em{projective}} supertwistor space $\mathbb{CP}^{3|4}$, or equivalently the space of two-planes in  non-projective supertwistor space $\mathbb{C}^{4|4}$, that is the Grassmannian $Gr(2,4|4)$. Similarly analytic superspace, on which the stress-energy tensor naturally sits, is the super-Grassmannian of $(2|2)$ planes in $\mathbb{C}^{4|4}$, $Gr(2|2,4|4)$~\cite{HH}. The modding out of a super-plane accounts for the halving of the odd degrees of freedom (for example in~\eqref{eq:2}). \label{f5}}
{The $\cN=4$ superconformal transformations correspond to global $GL(4|4)$ rotations of the supertwistor $\mathcal Z$.} 

\subsection{$\mathcal N=4$ SYM on twistor space}  
  
The fields of $\mathcal N=4$ SYM theory are described on projective twistor space $\mathbb{PT}$ by a superfield $\mathcal{A}$ that takes values in $(0,1)$-forms with values in the Lie algebra of the gauge group.  Expanding in the fermionic coordinates $\chi^A$ we obtain
\begin{align}\notag
\mathcal A(Z,\bar Z,\chi) &{}= a(Z,\bar Z) + \chi^A \tilde \gamma_A(Z,\bar Z) + \frac12 \chi^A\chi^B \phi_{AB}(Z,\bar Z)
\\
&{}+ \frac1{3!} \epsilon_{ABCD}\chi^A\chi^B\chi^C \gamma^{D}(Z,\bar Z)
 +\frac1{4!} \epsilon_{ABCD}\chi^A\chi^B\chi^C\chi^D  g(Z,\bar Z)  \, .
\end{align}  
The coefficients in front of $\chi^n$ are antiholomorphic  $(0,1)-$differential forms on the supertwistor space $\mathbb{CP}^{3|4}$, homogeneous of degree $n$ that are
related to the various component fields of $\mathcal N=4$ SYM by the Penrose transform: $g$ and $a$ give
rise to self-dual and anti self-dual part of the field strength tensor, $\tilde \gamma_A$ and $\gamma^{D}$ are mapped
into gaugino fields and $\phi_{AB}$ produce the scalar fields. 

The twistor action of $\mathcal N=4$ SYM takes the form
\begin{align} \label{S-tw}
 &{} S[\mathcal A] = \int_{\mathbb{CP}^{3|4}} \mathcal D ^{3|4} \mathcal Z \wedge \tr\left(\frac12 \mathcal A \,\bar{\partial}\mathcal A  
 - \frac13 \mathcal A^3 \right) +   \int d^{4} x \,d^8 \theta \, L_{\rm int}(x,\theta)\,,
\end{align}  
where $\mathcal D ^{3|4} \mathcal Z = \frac1{4!} \epsilon_{IJKL} Z^I d Z^J d Z^K d Z^L d^4 \chi$ is the integration measure on the complex
projective space and
\begin{align}\label{logdet}
L_{\rm int}(x,\theta)=g^2  \left[ \ln \det (\bar{\partial} - \mathcal A ) -\ln \det  \bar{\partial} \right].
\end{align}
The separation of the action $S[\mathcal A]$ into the sum of two terms corresponds to expansion of $\mathcal N=4$ theory around 
the self-dual sector. Indeed, the holomorphic Chern-Simons action is equivalent, in the appropriate gauge, to the self-dual part of the $\mathcal N=4$ action.
The second term on the right-hand side of \re{S-tw} describes the interaction induced by the non self-dual part of the action.
It involves the logarithm of the chiral determinant of the Dirac operator evaluated on the line in twistor space defined in \re{line}, and then integrated over all lines.
 
To perform calculations using \re{S-tw} it is convenient to choose an axial gauge in which the component of $\mathcal{A}$ in the direction of a fixed reference twistor $\mathcal{Z}_*$ vanishes.  In this gauge, the cubic term in the holomorphic Chern-Simons action
vanishes and the remaining quadratic term defines the propagator  
\begin{align}\label{A-prop}
\vev{\mathcal A^a (\mathcal Z_1)\mathcal A^b (\mathcal Z_2)} = \bar{\delta}^{2|4}(\mathcal Z_1,\mathcal Z_2,\mathcal Z_*)\delta^{ab}\,,
\end{align}
where we have displayed the $SU(N)$ indices of the fields $\mathcal A=\mathcal A^a T^a$ (with $T^a$ being the $SU(N)$ generators in the fundamental representation) and explicitly denoted the supertwistor $\mathcal Z_*$  that defines the axial gauge.
Here 
\begin{align}\label{bar-delta}
 \bar{\delta}^{2|4}(\mathcal Z_1,\mathcal Z_2,\mathcal Z_*) = \int {ds\over s} {dt\over t}  \bar\delta^{4|4} (s\mathcal Z_1+t\mathcal Z_2+\mathcal Z_*)
\end{align}
is a projective delta function. It is a homogenous $(0,2)-$form on twistor space that enforces the condition for its arguments to be collinear in the
projective space. In the axial gauge, all interaction vertices are produced by $L_{\rm int}$. Its expansion in powers of superfields looks like
\begin{align}\notag\label{Sint}
 L_{\rm int} (x,\theta) &{}= - g^2  \sum_{n\ge 2} {1 \over n}   \tr\left[\bar{\partial}^{-1}\mathcal A\dots \bar{\partial}^{-1}\mathcal A \right]
 \\
&{} =- g^2   \sum_{k\ge 2} {1 \over k}   \int  { \tr\left[\mathcal A(\mathcal Z(\sigma_1))\wedge D \sigma_1 \dots \mathcal A(\mathcal Z(\sigma_k))\wedge D \sigma_k \right] \over \vev{\sigma_1\sigma_2}\dots
 \vev{\sigma_k\sigma_1}}\,,
\end{align}
where $D\sigma_i = \vev{\sigma_i, d\sigma_i}\equiv \epsilon_{ab}\sigma_i^a d\sigma_i^b$ is the projective measure and 
\begin{align}\label{bra}
\vev{\sigma_i \sigma_j}  =  \epsilon_{ab} \sigma_i^a  \sigma_j^b\,.
\end{align}
In the second relation in \p{Sint} the superfields are integrated along the line in twistor space 
$\mathcal Z(\sigma_i)=\mathcal Z_1 \sigma_i^1 + \mathcal Z_2 \sigma_i^2$  parameterised by coordinates $\sigma_i^a\equiv (\sigma_i^1,\sigma_i^2)$ 
with two reference points $\mathcal Z_1$ and $\mathcal Z_2$ of the form \re{ZZ} and \re{line} with the same $x^{\alpha\dot\alpha}$ and $\theta^{A\alpha}$ 
but different $\lambda_\alpha$.

Making use of \re{Sint} and \re{A-prop} we can apply the conventional Feynman diagram technique to compute the correlation functions of operators built 
from supertwistor fields at weak coupling. To establish the correspondence with \re{Gn} we have to work out the representation of the stress-tensor superfield $\mathcal T(x,\theta^+,u)$ in twistor space.
Our main contention is that
\footnote{Other previous works discussing composite operators within the twistor framework are the proof of the correlator/amplitude duality for the Konishi multiplet \cite{Adamo:2011dq,Adamo:2011cd} and the recent papers \cite{Koster:2014fva,Brandhuber:2014pta} where the ${\cal N}=4$ one-loop dilatation operator in the $SO(6)$ sector is rederived. In either case the realisation of the operators is necessarily different from our approach because they are not connected to the Lagrangian by supersymmetry, which we use extensively.}
\begin{align}\label{T-conven}
\mathcal T(x,\theta^+,u)=\int d^4 \theta^- L_{\rm int}(x,\theta)\,,
\end{align}
where $\theta^{-a' \alpha}=\theta^{A\alpha}u^{-a'}_A$ and $\theta^{+a \alpha}=\theta^{A\alpha}u^{+a}_A$ are the projected fermionic coordinates required in the definition of $\mathcal{T}$.  We first remark that, although $L_{\rm int}$ is not gauge invariant because of the chiral gauge anomaly in $\ln \det (\bar{\partial} - \mathcal A )$ in \re{logdet}, the fermionic integration in \re{T-conven} annihilates the anomalous gauge variation.\footnote{This is a refinement of the discussion following Eq.(3.6) of \cite{Boels:2006ir}.  There, the variation of $L_{\rm int}$ under a gauge transformation  is seen to be quintic in the $\theta$'s. A more detailed examination shows that the Grassamann integral in \p{T-conven} does not find a matching $\theta-$structure in this gauge variation. }

To justify \re{T-conven}, denote the corresponding operator as $\mathcal T_{\mathcal A}(x,\theta^+,u)$ 
and examine another equivalent representation for the correlation function \re{Gn} 
\begin{align}\label{Gn-tw}
G_n=\vev{0|\mathcal T_{\mathcal A}(1) \dots \mathcal T_{\mathcal A}(n) |0}_{\mathcal A}\,,
\end{align}
where we inserted the subscript ${\mathcal A}$ to indicate that the expectation value is evaluated with the action given by \re{S-tw}.
To determine the explicit expression for $\mathcal T_{\mathcal A}(x,\theta^+,u)$ we shall require that, in the twistor space approach, the 
derivative of the correlation function \re{Gn-tw} with respect to the coupling constant $\partial G_n/\partial {g^2}$ has to be related 
to $G_{n+1}$ as in the last relation in \re{LI}.

Since the dependence of the twistor action \re{Sint} on the coupling constant only resides in $L_{\rm int}$ we obtain
\begin{align}\label{Gn-A}
  {\partial\over\partial g^2} G_n &{} =\int d^4 x_{n+1} d^8\theta_{n+1} \vev{0| \mathcal T_{\mathcal A}(1) \dots \mathcal T_{\mathcal A}(n)
 L_{\rm int}(x_{n+1},\theta_{n+1})
  |0}_{\mathcal A}\,.
\end{align}
This relation is remarkably similar to \re{LI}. However, an important difference is that, in distinction to $ L_{\rm int}(x_{n+1},\theta_{n+1})$ the stress-tensor superfield $\mathcal T(x_{n+1},\theta^+_{n+1}, u_{n+1})$ entering the second line in \re{LI} only depends on half of the $\theta_{n+1}^{A\alpha}$ variables while it has an additional dependence on the harmonic variables $u_{n+1}$. To match the sets of variables these two operators depend on, we employ the harmonics to decompose $\theta_{n+1}^{A\alpha}$ into the two projections 
\begin{align}\label{halves}
 \theta_{n+1}^{+a,\alpha} = \theta_{n+1}^{A\alpha}u_{n+1,A}^{+a}\,,\qqqquad  \theta_{n+1}^{-a',\alpha} = \theta_{n+1}^{A\alpha}u_{n+1,A}^{-a'}\,,
\end{align}
with $A=(+a,-a')$, and then integrate out $\theta_{n+1}^{-a',\alpha}$ using the identity $\int d^8\theta_{n+1} = \int d^4 \theta_{n+1}^+ \int d^4 \theta_{n+1}^-$. Appealing to the analogy with \re{LI} we identify the resulting
operator as representing the stress-tensor superfield in twistor space
\begin{align}\label{TA}
\mathcal T_{\mathcal A}(n+1) = \int d^4 \theta_{n+1}^{-}\, L_{\rm int}(x_{n+1},\theta_{n+1})\,,
\end{align}
whereby \re{Gn-A} takes the same form as \re{LI}. Notice that the dependence of $\mathcal T_{\mathcal A}(n+1)$ on the harmonic variable
$u_{n+1}$ enters through the integration measure $\int d^4 \theta_{n+1}^{-}$. 

We combine the relations \re{TA} and \re{Gn-tw} to obtain
the following representation for the correlation function in twistor space
\begin{align}\label{G-int}
G_n = \int d^4 \theta_{1}^{-}\dots d^4 \theta_{n}^{-} \,\vev{0|L_{\rm int}(1) \dots L_{\rm int}(n) |0}_{\mathcal A}\,,
\end{align}
where $L_{\rm int}(i)\equiv L_{\rm int}(x_{i},\theta_{i})$ and $\theta_i^{-a',\alpha} = \theta_i^{A\alpha} (u_i)_A^{-a'}$. As before, we 
will be interested in computing this correlation function to lowest order in the coupling constant. In this approximation, we can neglect
the dependence of the twistor action \re{S-tw} on the coupling constant and retain only the first (Chern-Simons) term on the right-hand 
side of \re{S-tw}. In addition, we recall that in the axial gauge the Chern-Simons term reduces to the kinetic term, quadratic in 
twistor superfield $\mathcal A$. As a consequence, calculating \re{G-int} we can treat  $\mathcal A$ as a free field. In this way, replacing
$L_{\rm int}(i)$ by its expression \re{Sint} we have to perform all possible Wick contractions of the superfields $\mathcal A$ and
express the correlation function \re{G-int} as a product of propagators defined in \re{A-prop}. This leads to the set of Feynman rules
formulated in the next subsection.

\subsection{Feynman rules from twistor space}

\begin{figure}[h!t]
\psfrag{Z1}[cc][cc]{$\scriptstyle Z_{i,1}$}  \psfrag{Z2}[cc][cc]{$\scriptstyle Z_{i,2}$} 
\psfrag{Z3}[cc][cc]{$\scriptstyle Z_{j,1}$}  \psfrag{Z4}[cc][cc]{$\scriptstyle Z_{j,2}$} 
\psfrag{i}[cc][cc]{$i$} \psfrag{j}[cc][cc]{$j$}  \psfrag{3}[cc][cc]{$3$}  \psfrag{4}[cc][cc]{$4$}  \psfrag{n}[cc][cc]{$n$} 
\psfrag{j1}[cc][cc]{$j_1$} \psfrag{j2}[cc][cc]{$j_2$} \psfrag{j3}[cc][cc]{$j_3$} \psfrag{j4}[cc][cc]{$j_k$} 
\psfrag{vertex}{$\displaystyle {\tr[T^{a_{j_1}} T^{a_{j_2}}\cdots  T^{a_{j_k}}] \over \vev{\sigma_{i j_1}\sigma_{i j_2}}\vev{\sigma_{i j_2}\sigma_{i j_3}}\cdots \vev{\sigma_{i j_k}\sigma_{i j_1}}  }$ }\psfrag{dots}[cc][cc]{$\dots$}
\psfrag{propagator}{$\displaystyle \delta^{a_{i}a_{j}} \delta^{4|4}(\cZ_* + \sigma_{ij}^{\alpha} \cZ_{i,\alpha}+ \sigma_{ji}^{\alpha} \cZ_{j,\alpha})$}
 \centerline{\includegraphics[width = 0.65 \textwidth]{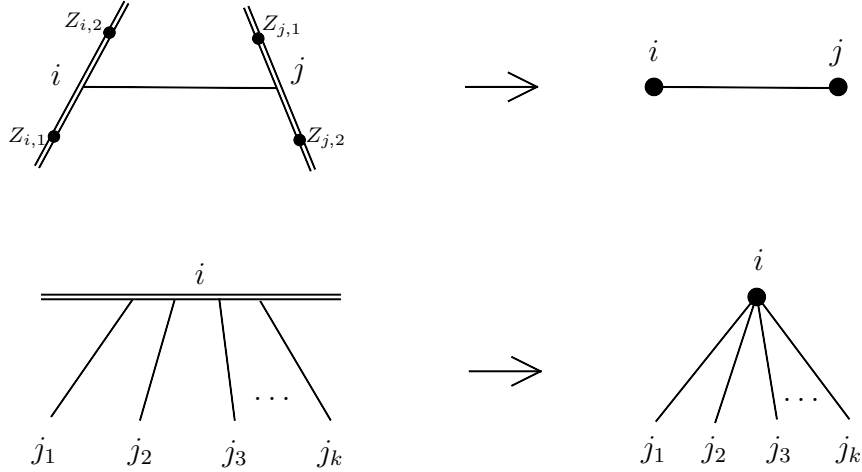}}
\caption{\small Propagators and vertices in twistor space.}
\label{fig-rules}
\end{figure}

According to the definition \re{Sint}, each operator $L_{\rm int}(x_i,\theta_i)$ lives on a line in twistor space 
\begin{align}
\mathcal Z_i(\sigma) = \mathcal Z_{i,1} \sigma^{1}+ \mathcal Z_{i,2} \sigma^{2}\equiv \mathcal Z_{i,\alpha} \sigma^{\alpha}\,,
\end{align}
with two reference points $\mathcal Z_{i,1}$ and $\mathcal Z_{i,2}$ satisfying the incidence relations involving $x_i$ and $\theta_i$.
Then, each term in the sum in the second relation in \re{Sint} can be viewed as a line in twistor space; the $k$ legs attached to it represent the twistor superfields $\mathcal A(\mathcal Z_i(\sigma))$. For our purposes it will also be
convenient to treat the same diagram as defining a new effective interaction vertex as shown in Fig.~\ref{fig-rules}. 
\medskip 
  
\begin{figure}[h!t]
\psfrag{1}[cc][cc]{$\scriptstyle 1$} \psfrag{2}[cc][cc]{$\scriptstyle 2$}  \psfrag{3}[cc][cc]{$\scriptstyle 3$}  \psfrag{4}[cc][cc]{$\scriptstyle 4$}  \psfrag{n}[cc][cc]{$\scriptstyle n$} 
\centerline{\includegraphics[width = 0.65 \textwidth]{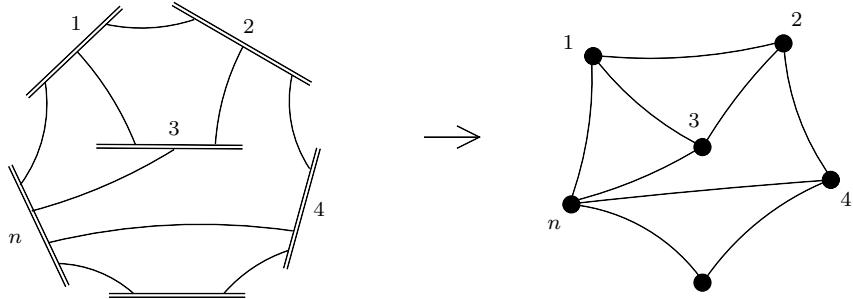}}
\caption{\small Feynman diagram on twistor space contributing to an $n-$point correlation function. A double line with label $i$ represents
a line in twistor space with moduli $(x_i,\theta_i)$. Solid lines stand for propagators of twistor superfields and dots denote
effective interaction vertices. }
\label{fig-sample}
\end{figure}

\noindent
Then, the correlation function \re{G-int} 
is given by a set of diagrams in which an arbitrary number of propagators are stretched between $n$ lines, or equivalently connect
$n$ effective vertices (see Fig.~\ref{fig-sample}). 

Let us consider the propagator connecting two lines with indices $i$ and $j$. 
Denoting the local parameters of the points on these two lines by $\sigma_{ij}^\alpha$ and $\sigma_{ji}^\alpha$, respectively, we can write its contribution as
\begin{align}\notag
& {}\int \vev{\sigma_{ij} d\sigma_{ij}} \int \vev{\sigma_{ji} d\sigma_{ji}} \, \delta^{2|4}(\mathcal Z_i(\sigma_{ij}),\mathcal Z_j(\sigma_{ji}),\mathcal Z_*)  
(\dots)
\\ \notag
&{} \qquad =  \int \vev{\sigma_{ij} d\sigma_{ij}} \int \vev{\sigma_{ji} d\sigma_{ji}} \int {ds\over s} {dt\over t}  \, \delta^{4|4}(s\mathcal Z_i(\sigma_{ij})+t\mathcal Z_j(\sigma_{ji})+\mathcal Z_*)  
(\dots)
\\
&{}\qquad=   \int  d^2\sigma_{ij} \int d^2 \sigma_{ji} \, \delta^{4|4}(\cZ_* + \sigma_{ij}^{\alpha} \cZ_{i,\alpha}+ \sigma_{ji}^{\alpha} \cZ_{j,\alpha})  
(\dots)\,,
\end{align}
where the expression inside $(\dots)$ corresponds to the rest of the diagram and we made use of \re{bar-delta} in the second relation.
Here in the third relation we replaced the integration variables $\sigma^\alpha_{ij} \to s \sigma^\alpha_{ij} $ and $ \sigma^\alpha_{ji} \to t \sigma^\alpha_{ji}$
taking into account that the expression inside $(\dots)$ is a homogenous function of $\sigma_{ij}$ and $\sigma_{ji}$ of degree $(-2)$.

\begin{figure}[h!t]
\psfrag{Z1}[cc][cc]{$\scriptstyle Z_{i,1}$}  \psfrag{Z2}[cc][cc]{$\scriptstyle Z_{i,2}$} 
\psfrag{Z3}[cc][cc]{$\scriptstyle Z_{j,1}$}  \psfrag{Z4}[cc][cc]{$\scriptstyle Z_{j,2}$} 
\psfrag{i}[cc][cc]{$i$} \psfrag{j}[cc][cc]{$j$}  \psfrag{3}[cc][cc]{$3$}  \psfrag{4}[cc][cc]{$4$}  \psfrag{n}[cc][cc]{$n$} 
\psfrag{j1}[cc][cc]{$j_1$} \psfrag{j2}[cc][cc]{$j_2$} \psfrag{j3}[cc][cc]{$j_3$} \psfrag{j4}[cc][cc]{$j_k$} 
\psfrag{vertex}{$-\displaystyle {\tr[T^{a_{j_1}} T^{a_{j_2}}\cdots  T^{a_{j_k}}] \over \vev{\sigma_{i j_1}\sigma_{i j_2}}\vev{\sigma_{i j_2}\sigma_{i j_3}}\cdots \vev{\sigma_{i j_k}\sigma_{i j_1}}  }$ }\psfrag{dots}[cc][cc]{$\dots$}
\psfrag{propagator}{$\displaystyle \delta^{a_{i}a_{j}} \delta^{4|4}(\cZ_* + \sigma_{ij}^{\alpha} \cZ_{i,\alpha}+ \sigma_{ji}^{\alpha} \cZ_{j,\alpha})$}
\centerline{\includegraphics[width = 0.35 \textwidth]{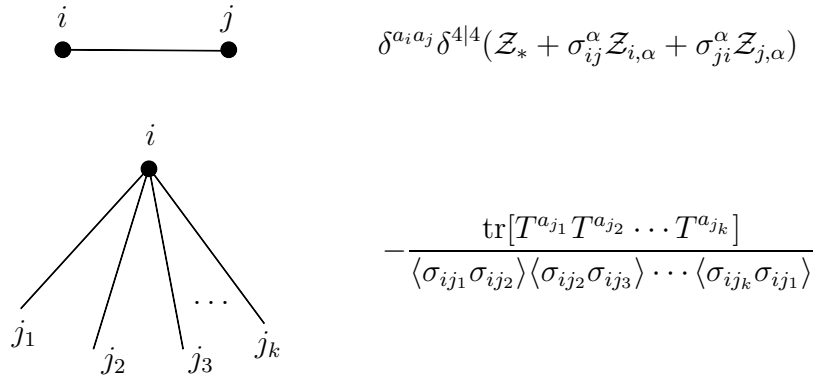}\hspace*{50mm}}
\caption{\small Feynman rules for propagators and vertices in twistor space.}
\label{fig-rules2}
\end{figure}

Then, the Feynman rules taking us from a graph as shown in Fig.~\ref{fig-sample} to a contribution to the correlation function \re{G-int} are as follows:
\begin{itemize}

\item To each line connecting vertices $i$ and $j$ we
associate two pairs of spinor variables $\sigma_{ij}^\alpha$ and $\sigma_{ji}^\alpha$ (with $\alpha=1,2$). They define the coordinates
of the end points $\sigma_{ij}^{\alpha} \cZ_{i,\alpha}$ and $\sigma_{ji}^{\alpha} \cZ_{j,\alpha}$ belonging
to the $i$th and $j$th lines, respectively,  in projective twistor space~\footnote{Such an assignment of the $\sigma_{ij}$ variables would be ambiguous if two vertices were connected
by more than one line.  As we show below (see Eq.~\re{more}), this never happens for $n-$point correlation functions if $n>2$.};

\item A propagator connecting vertices $i$ and $j$ produces a graded delta function
$\delta^{a_ia_j}\delta^{4|4}(\cZ_* + \sigma_{ij}^{\alpha} \cZ_{i,\alpha}+ \sigma_{ji}^{\alpha} \cZ_{j,\alpha})$ with $a_i$ and $a_j$ being  $SU(N)$ colour
indices;

\item Each vertex comes with a Parke-Taylor-like denominator 
 accompanied by  the $SU(N)$ colour factor, $-\tr[T^{a_{j_1}} T^{a_{j_2}}\cdots  T^{a_{j_k}}]/\prod_{\ell=1}^{k}
 \vev{\sigma_{ij_\ell}\sigma_{ij_{\ell+1}}}$  (with $j_{k+1}\equiv j_1$
and 
 $\vev{\sigma_{ij_\ell}\sigma_{ij_{\ell+1}}} $ given by \re{bra}).
In virtue of $\tr T^{a_j}=0\phantom{^\beta}$, we must have at least two lines coming from each vertex;

\item Finally, at each vertex $i=1,\dots,n$ we have to perform an integration $\int d^2 \sigma_{ij_1}\dots d^2 \sigma_{ij_{k}}$ over the $\sigma-$parameters of all lines attached
to that vertex and, in addition, integrate out half of the Grassmann variables by $\int d^4 \theta_{i}^{-}$. 
 \end{itemize}
These rules are summarised in Fig.~\ref{fig-rules2}.
 
To compute an $n-$point correlation function using these Feynman rules we have to examine all diagrams with exactly $n$ vertices and an arbitrary number of propagators.
Since each vertex has at least two lines attached  to it, the minimal number of propagators is $n$. Let us denote the total number of propagators as $n+p$ (with $p\ge 0$)
and examine the Grassmann degree of the corresponding diagram.  Each propagator increases the Grassmann degree by four units whereas each vertex reduces
it by four units due to the integration $\int d^4 \theta_{i}^{-}$. Thus, the Grassmann degree of a diagram containing $n$ vertices and $n+p$ propagators is $4 \, p$. 
This counting is in perfect agreement with the general form of the correlation function \re{G-mod4}. It also allows us to identify each term in the expansion \re{G-mod4}
with the contribution of a particular class of diagrams:
\begin{align}\label{rule}
G_{n;p} = \text{Sum of diagrams with $n$ vertices and $n+p$ propagators} 
\end{align}
    
\subsection{Lowest component}

To illustrate the formalism, we apply the Feynman rules formulated in the previous subsection to compute the simplest $G_{n;0}$ component of the correlation 
function \re{G-mod4}. According to \re{rule}, $G_{n;0}$ is given by the sum of diagrams with $n$ vertices and $n$ propagators. A distinctive feature
of such diagrams is that all vertices are bivalent. In what follows we shall only consider connected twistor 
diagrams.~\footnote{The disconnected twistor diagrams describe contributions to the correlation function which reduce to products of correlators with lower number of points.}
A particular example of such a diagram is the graph in which vertices $i$ and $i+1$ are connected by a single line. All remaining diagrams can be obtained by permuting the labels of the vertices. According to the Feynman rules in Fig.~\ref{fig-rules}, the contribution of the $i$th  vertex involves  $-1/(\vev{\sigma_{i,i-1}\sigma_{i,i+1}}\vev{\sigma_{i,i+1}\sigma_{i,i-1}})=1/\vev{\sigma_{i,i-1}\sigma_{i,i+1}}^{2}$. We combine it with the propagators to obtain
\footnote{{Here we do not display the factor $(N^2-1)$ coming from the contraction of the $SU(N)$ colour indices since it is included in \re{planar}.} \label{foot}}
\begin{align}\label{G-MHV}
G_{n;0} = \prod_{i=1}^{n}\int d^4 \theta_{i}^{-} \int {d^2 \sigma_{i,i-1} d^2 \sigma_{i,i+1} \over \vev{\sigma_{i,i-1}\sigma_{i,i+1}}^2} \delta^{4|4}(\cZ_* + \sigma_{i,i-1}^{\beta} \cZ_{i,\beta}+ \sigma_{i-1,i}^{\beta} \cZ_{i-1,\beta}) + \text{($S_n-$perm)},
\end{align}
where ($S_n-$perm) denotes the additional terms needed to restore the Bose symmetry of the correlation function.

We recall that $\cZ_{i,1}$ and $\cZ_{i,2}$ denote two points on a line in supertwistor space. They have the general form \re{ZZ} and \re{line} with the local
coordinates $\lambda_{1,\beta}$ and  $\lambda_{2,\beta}$, respectively. The correlation function \re{G-MHV} should not depend on the choice of these coordinates.
Indeed, the change of the local coordinates corresponds to the $GL(2)$ rotation $\lambda_{\gamma,\beta}\to g_\gamma{}^\delta \lambda_{\gamma,\beta}$, or
equivalently $\cZ_{i,\beta}\to g_\gamma{}^\delta\cZ_{i,\delta}$. This variation can be compensated in \re{G-MHV} by the change of the integration variable
$\sigma_{ik}^\beta \to (g^{-1})^\beta{}_\delta \sigma_{ik}^\delta$. We can make use of this symmetry to choose $\cZ_{i,\beta}$ in the following form
\begin{align}\label{Z-gauge}
 \cZ_{i,\beta} = (Z_{i,\beta}^{\,I},\ \theta^{\,A}_{i,\beta})\,,\qqqquad Z_{i,\beta}^{\,I}=(\epsilon_{\alpha\beta},\ i   x_{i,\beta} ^{\,\dot\alpha})\,,
\end{align}
with $I=(\alpha,\dot\alpha)$.
It is also convenient to parameterise the axial gauge supertwistor as
\begin{align}\label{Z*}
  \cZ_{* } = (Z_{* }^{\,I},\ \theta^{\,A}_{*}) \,.
\end{align}   
We substitute \re{Z-gauge} into \re{G-MHV} and perform the integration over $\theta_{i}^{-}$ to obtain (see Eq.~\re{dd} below)  
\begin{align}
G_{n;0} = \prod_{i=1}^{n}\,y_{i,i+1}^2  \int {d^2 \sigma_{i,i-1} d^2 \sigma_{i,i+1}  } \delta^{4 }( Z_* + \sigma_{i,i-1}^{\beta}  Z_{i,\beta}+ \sigma_{i-1,i}^{\beta}  Z_{i-1,\beta}) + \text{($S_n-$perm)}\,.
\end{align} 
Here $ y_{i,i+1} = y_i - y_{i+1}$ with $y_i$ being the local coordinates on the harmonic coset introduced in \p{u-y}. We notice that the total number of delta functions in this integral matches the number of integration variables. Therefore, the integral is localised at the values of the $\sigma-$parameters
satisfying $Z_* + \sigma_{i,i-1}^{\beta}  Z_{i,\beta}+ \sigma_{i-1,i}^{\beta}  Z_{i-1,\beta}=0$. Equivalently
\begin{align}\notag\label{si}
& \sigma_{i,i-1}^{\alpha} = \epsilon^{\alpha\beta} { \vev{Z_{i,\beta} Z_* Z_{i-1,1}Z_{i-1,2} } \over \vev{Z_{i-1,1}Z_{i-1,2}Z_{i,1}Z_{i,2}} }\,,
\\
& \sigma_{i-1,i}^{\alpha} = \epsilon^{\alpha\beta} { \vev{Z_{i-1,\beta} Z_* Z_{i,1}Z_{i,2} } \over \vev{Z_{i-1,1}Z_{i-1,2}Z_{i,1}Z_{i,2}} }\,,
\end{align}
where we used the notation $\vev{Z_1 Z_2 Z_3 Z_4} =\epsilon_{IJKL} Z_1^I Z_2^J Z_3^K Z_4^L$.
In this way, we finally obtain  
\begin{align}\label{Gn0}
 G_{n;0} = \prod_{i=1}^{n}\,{y_{i,i+1}^2\over x_{i,i+1}^2}+ \text{($S_n-$perm)}\,.
\end{align}
Notice that the dependence on the reference supertwistor $\cZ_*$ disappeared in $G_{n;0}$ as it should for a gauge invariant quantity.

The result \re{Gn0} perfectly meets our expectations. In the conventional approach, $G_{n;0}$ coincides with the correlation function of $n$ operators
$\tr[\phi^{++}\phi^{++}]$, the lowest component of the stress-tensor multiplet \re{T-dec}. Then, to lowest order in the coupling constant, $G_{n;0}$ is
given by a product of $n$ free scalar propagators $\vev{\phi^{++}(i)\phi^{++}(j)} = y_{ij}^2/x_{ij}^2$, properly symmetrised to respect Bose symmetry.

\subsection{Twistor Feynman rules for higher components}
\label{sec:twist-feynm-rules}

To compute higher components of the correlation function $G_{n;p}$ we have to examine all diagrams containing $n$ vertices and $n+p$ propagators. 
We can apply the Feynman rules formulated in the previous sections to write down their contribution as a productsof $n+p$ graded
delta functions of the form $\delta^{4|4}(\cZ_* + \sigma_{ij}^{\alpha} \cZ_{i,\alpha}+ \sigma_{ji}^{\alpha} \cZ_{j,\alpha})$. However, this approach
is not very efficient in that it involves integrating a function of Grassmann degree $4(n+p)$
over $4n$ odd variables $\int d^4 \theta_{i}^{-}$ to arrive at the function $G_{n;p}$ of Grassmann degree $4p$. So, in this subsection we instead perform the explicit  integration over the variables $\theta_{i}^{-}$ at the
level of the twistor Feynman rules and thus derive a simpler set of rules. 

To begin with, we split each propagator up into a product of bosonic and fermionic delta functions,
\begin{align}\label{eq:1}
\delta^{4 }( Z_* + \sigma_{ij}^{\alpha}  Z_{i,\alpha}+ \sigma_{ji}^{\alpha}  Z_{j,\alpha})  \delta^4(\theta_*+\sigma_{ij}^\alpha\theta_{i,\alpha} + \sigma_{ji}^\alpha\theta_{j,\alpha})\,.
\end{align}
To integrate over $\theta_{i}^{-}$, we employ the harmonics $u_i$  to decompose the variables $\theta_{i}$ into two halves \re{halves}  (see Appendix~\ref{app:conv}), 
\begin{align}
\theta^A_{i} = \theta_{i}^{+a} \bar u_{i,+a}^A+\theta_{i}^{-a'}\bar u_{i,-a'}^A\,.
\end{align}
Then, multiplying the argument of the fermionic delta function by the $4\times 2$ matrices $u_{i,A}^{+a}$ and $u_{j,A}^{+a}$ we find after some
algebra
\begin{align}\label{dd}
\delta^4(\theta_*+\vev{\sigma_{ij}\theta_{i}} + \vev{\sigma_{ji} \theta_{j}}) = y_{ij}^2\,\delta^2 \left(\vev{\sigma_{ij}\theta_i^-} + A_{ij}\right) \delta^2 \left(\vev{\sigma_{ji}\theta_j^-} + A_{ji}\right)\,,
\end{align}
with $y_{ij}^2=\frac14  \epsilon^{ABCD} u_{i,A}^{+a}\epsilon_{ab}u_{i,B}^{+b} u_{j,C}^{+c}\epsilon_{cd}u_{j,D}^{+d}$.
Here the functions 
\begin{align}\label{A-mat}
A_{ij}^{a'}  = \left[\vev{\sigma_{ji}\theta_j^{+b}} +\vev{\sigma_{ij}\theta_i^{+c}}(U_{ij})_{+c}^{+b}+\theta_*^A u_{j,A}^{+b}\right] (U_{ij}^{-1})_{+b}^{-a'}
\end{align}
depend only on $\theta^+_i$ and $\theta^+_j$, and the matrices $U_{ij}$ are defined as
\begin{align}
 (U_{ij})_{+c}^{+b} = \bar u_{i,+c} ^A u_{j,A}^{+b}\,,\qqqquad (U_{ij})_{-a'}^{+b} = \bar u_{i,-a'} ^A u_{j,A}^{+b}
 \,.
\end{align}
The function $A_{ji}^{a'}$ can be obtained from $A_{ij}^{a'}$ by exchanging the indices $i\leftrightarrow j$.
It is often convenient to use a parameterisation of the harmonic variables $u_i$ in terms of the local coordinates  $y_i$ on the harmonic coset  defined in \re{u-y}. In this case, $(U_{ij})_{+c}^{+b}=\delta_c^b$ and $ (U_{ij})_{-a'}^{+b}=(y_{ij})_{a'}^{b}$, so that the
expression \re{A-mat} significantly simplifies,
\begin{align}\label{A-y}
A_{ij}^{a'}  = \left[\vev{\sigma_{ji}\theta_j^{+b}} +\vev{\sigma_{ij}\theta_i^{+b}} +\theta_*^A u_{j,A}^{+b}\right] (y_{ij}^{-1})_{b}^{a'}\,.
\end{align}

Notice that the dependence on $\theta^-_i$ and $\theta^-_j$  on the right-hand side of \re{dd} resides in the first and second
delta functions, respectively. This suggests associating the first delta function with the vertex $i$ and the
second one with the vertex $j$. Then, if the vertex $i$ has $k$ propagators attached to it, we take into account the additional $\sigma-$dependent factor
coming from the Feynman rules in Fig.~\ref{fig-rules2} to arrive at the integral
\begin{align}\label{22}
  R(i;j_1j_2\dots j_k)=-\int d^4{\theta^-_i}  {  \delta^2(\vev{\sigma_{ij_1} \theta^-_i}
  + A_{ij_1})\delta^2(\vev{\sigma_{ij_2}\theta^-_i} +A_{ij_2})
   \dots  \delta^2(\vev{\sigma_{ij_k} \theta^-_i} +A_{ij_k}) \over \vev{\sigma_{i j_1} \sigma_{i j_2}}  \,\vev{\sigma_{i j_2} \sigma_{i j_3}}\,\dots \,\vev{\sigma_{i j_k} \sigma_{i j_1}} }\ .
\end{align}
Here the index $i$ labels the vertex and the indices $j_1,\dots,j_k$ enumerate the outgoing lines. By construction, this integral has Grassmann degree $(2k-4)$.
As we shall see in the next section, the quantity $R(i;j_1j_2\dots j_k)$ plays a crucial
role in our analysis. 

Relation \re{22} depends on the parameters $\sigma_{ij}^\alpha$ and $\sigma_{ji}^\alpha$. Their values can be determined using 
the bosonic part  of the propagator \re{eq:1}. Namely,  solving the equation $Z_*^I + \vev{\sigma_{ij} Z_i^I}+ \vev{\sigma_{ji} Z_j^I} =0$ we obtain
\begin{align}\label{sigma}
\sigma_{ij}^\alpha = \epsilon^{\alpha\beta}{\vev{Z_{i,\beta} Z_*Z_{j,1} Z_{j,2}} \over \vev{Z_{i,1} Z_{i,2} Z_{j,1} Z_{j,2}}}  \,,\qqqquad
\sigma_{ji}^\alpha = \epsilon^{\alpha\beta}{\vev{Z_{j,\beta} Z_*Z_{i,1} Z_{i,2}} \over \vev{Z_{i,1} Z_{i,2} Z_{j,1} Z_{j,2}}} \,,
\end{align}
c.f. \re{si}.
Finally, for each propagator \re{eq:1}  the bosonic delta function allows us to do the $\sigma-$integration yielding 
\begin{align}\label{eq:22}
y_{ij}^2 \int d^2\sigma_{ij} d^2\sigma_{ji}  \, \delta^{4|0} (Z^*+\sigma_{ij} Z_{i}^{}+\sigma_{ji} Z_{j}^{})
 = {y_{ij}^2\over \vev{Z_{i,1} Z_{i,2} Z_{j,1} Z_{j,2}}}=  {y_{ij}^2 \over x_{ij}^2}\ ,
\end{align}
where the additional factor of $y_{ij}^2$ comes from \re{dd}.

In summary, we arrive at the following twistor Feynman rules shown in Fig.~\ref{fig-rules3}:

\begin{itemize}
\item  A line connecting vertices $i$ and $j$ is associated with the   propagator $d_{ij}=y_{ij}^2 / x_{ij}^2$;

\item  Bivalent vertices are  associated with $R(i;j_1 j_2)\tr[T^{a_{j_1}} T^{a_{j_2}}]=R(i;j_1 j_2)\delta^{a_{j_1}a_{j_2}}$;

\item  Higher valency vertices are associated with   
$ R(i;j_1\dots j_{k})\tr[T^{a_{j_1}}\dots T^{a_{j_k}}]$  evaluated for the $\sigma-$parameters given by \re{sigma}.
  
\begin{figure}[h!t]
\psfrag{Z1}[cc][cc]{$\scriptstyle Z_{i,1}$}  \psfrag{Z2}[cc][cc]{$\scriptstyle Z_{i,2}$} 
\psfrag{Z3}[cc][cc]{$\scriptstyle Z_{j,1}$}  \psfrag{Z4}[cc][cc]{$\scriptstyle Z_{j,2}$} 
\psfrag{i}[cc][cc]{$i$} \psfrag{j}[cc][cc]{$j$}  \psfrag{3}[cc][cc]{$3$}  \psfrag{4}[cc][cc]{$4$}  \psfrag{n}[cc][cc]{$n$} 
\psfrag{j1}[cc][cc]{$j_1$} \psfrag{j2}[cc][cc]{$j_2$} \psfrag{j3}[cc][cc]{$j_3$} \psfrag{j4}[cc][cc]{$j_k$} 
\psfrag{vertex}{$\displaystyle  \tr[T^{a_{j_1}} T^{a_{j_2}}\cdots  T^{a_{j_k}}]  R(i;j_1j_2\dots j_k)$ }\psfrag{dots}[cc][cc]{$\dots$}
\psfrag{propagator}{$\displaystyle  \delta^{a_{i}a_{j}} d_{ij} =  \delta^{a_{i}a_{j}} {y_{ij}^2\over x_{ij}^2}$}
\centerline{\includegraphics[width = 0.35 \textwidth]{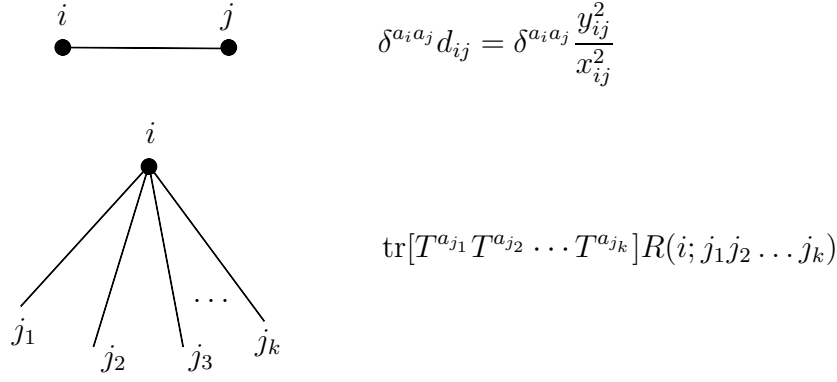}\hspace*{50mm}}
\caption{\small Feynman rules for propagators and vertices in analytic superspace.}
\label{fig-rules3}
\end{figure}

\end{itemize}

\subsection{Properties of the $R-$vertices}

Let us summarise the properties of $R(i;j_1j_2\dots j_k)$. In twistor diagrams, this function is accompanied by the colour factor 
$\tr[T^{a_{j_1}}\dots T^{a_{j_k}}]$ with the same ordering of external lines.
As follows from the representation \re{22}, $R(i;j_1j_2\dots j_k)$ is invariant under a cyclic shift of the $j-$indices and changes sign under a `mirror' exchange of the indices, $j_\ell\to j_{k-\ell+1}$,
\begin{align}\label{sym}
 R(i;j_1j_2\dots j_{k-1} j_k) = R(i;j_2j_3\dots j_{k}j_1) = (-1)^k R(i;j_{k}j_{k-1}\dots j_2 j_1) \,.
\end{align}
For $k=3$ external lines, this relation implies that $R(i;j_1j_2 j_3)$ is completely antisymmetric under the exchange of external legs,
\begin{align}\label{anti}
 R(i;j_1j_2 j_3) = -R(i;j_1j_3 j_2) = - R(i;j_3j_2 j_1) = R(i;j_2 j_3 j_1)\,.
\end{align}
In the special case $j_2=j_3$, corresponding to a graph in which the two external legs are attached to the same vertex,  this relation  implies 
\begin{align}\label{R3=0}
 R(i;j_1j_2 j_2) =0\,.
\end{align}

Let us examine the explicit expression for $ R(i;j_1j_2\dots j_k)$ for the lowest values of $k$.
For a bivalent vertex, $k=2$, the integration in \re{22} yields
\begin{align}\label{R-2pt}
 R(i;j_1j_2)=1\,.
\end{align}
For a valency three vertex, $k=3$, we can make use of the Schouten identity 
\begin{align}
  \label{eq:6}
  \sigma_{ij_1}^\alpha \langle \sigma_{ij_2} \sigma_{ik}\rangle +
  \sigma_{ij_2}^\alpha\langle \sigma_{ik} \sigma_{ij_1}\rangle+
  \sigma_{ik}^\alpha\langle \sigma_{ij_1} \sigma_{ij_2}\rangle = 0   \ 
\end{align}
to rewrite the argument of one of the three delta functions on the support of the other two 
in such a way that it becomes $\theta^-_i$ independent. In this way, we obtain
\begin{align}\label{eq:10}
R(i;j_1j_2j_3)= -{\delta^2\Big(\langle\sigma_{ij_1}\sigma_{ij_2}\rangle
A_{ij_3}+\langle\sigma_{ij_2}\sigma_{ij_3}\rangle
A_{ij_1}+\langle\sigma_{ij_3}\sigma_{ij_1}\rangle A_{ij_2}\Big) \over  \vev{\sigma_{i j_1} \sigma_{i j_2}}  \,\vev{\sigma_{i j_2} \sigma_{i j_3}}\,\vev{\sigma_{i j_3} \sigma_{i j_1}}}\ .
\end{align}
For vertices of higher valency, we can recursively apply the same trick, reducing a $k-$valent vertex to a product of $3-$ and $(k-1)-$valent vertices. 
Specifically, we rewrite the last delta function on the right-hand side of \re{22} as a combination of the first and the $(k-1)$st to get 
\begin{align}\label{eq:28}
  R(i;j_1j_2 \dots j_k)=R(i;j_1j_2\dots j_{k-1})\, R(i;j_1j_{k-1}j_k)\ .
\end{align}
Continuing recursively we can express the $k-$valent vertex as a product of $(k-2)$ copies of $3-$valent vertices  
\begin{align}\label{eq:11}
  R(i;j_1j_2 \dots j_k)=R(i;j_1j_2j_3)\, R(i;j_1j_3j_4)\,\dots\,R(i;j_1j_{k-1}j_k)\ .
\end{align}
Note that the index $j_1$ plays a special role here as it appears in every factor on the right-hand side. We can obtain another equivalent representation for  
$R(i;j_1j_2 \dots j_k)$ by making use of the symmetry properties \re{sym}. Combining \re{eq:11} with \re{R3=0} we find that the $R-$vertex vanishes if two indices 
of external lines coincide
\begin{align}\label{more}
R(i;j_1j_1j_3 \dots j_k)= R(i;j_1j_2\dots j_1 \dots j_k)=0\,.
\end{align}
In terms of twistor diagrams this relation implies that diagrams with (at least) two propagators stretched between any two twistor lines do not contribute
to the correlation function. 

We observe that the denominator in \re{22} has the same form as in the Parke-Taylor MHV amplitude upon identifying the variables $\sigma_{ij}$ with the holomorphic
variables $\lambda_j$ that define the on-shell momenta of the particles. As a consequence, we can use the properties of the MHV amplitude to obtain non-trivial relations for 
$R(i;j_1j_2\dots j_k)$. In particular, the $U(1)$ decoupling relation for MHV amplitudes \cite{Dixon:1996wi} translates into
\begin{align}\label{U(1)}
 R(i;j_1j_2\dots j_{k-1} j_k) + R(i;j_1j_3\dots j_{k}j_2) +\dots + R(i;j_1j_k\dots j_{k-2}j_{k-1}) =0\,,
\end{align}
where the sum runs over cyclic permutations of the indices $j_2,\dots,j_{k-1},j_k$. This relation can be verified using the Schouten identity \re{eq:6}.

The $R-$vertices satisfy another set of non-trivial relations. In the simplest case of three-point vertices it takes the form
\begin{align}\label{3pt-new}
R(i;j_1j_2j_3) =  R(i;j_4j_2 j_3) + R(i; j_1 j_4j_3) + R(i;j_1 j_2 j_4) \,, 
\end{align}
with $j_1,\dots,j_4$ being arbitrary. The proof of this relation can be found in Appendix~\ref{App:R}. We can then use \re{3pt-new} and 
\re{eq:11} together to obtain an analogous relation for four points
\begin{align}\notag\label{4pt-cyc}
 R(i;j_1j_2j_3j_4) {}& =R(i;j_5j_2j_3j_4) +R(i;j_1j_5j_3j_4) + R(i;j_1j_2j_5j_4) +R(i;j_1j_2j_3j_5)
\\[2mm]
{}& + R(i;j_5j_1j_2) R(i;j_5j_3j_4) + R(i;j_5j_2j_3) R(i;j_5j_4j_1)\,.
\end{align}
It is straightforward to generalise it to an arbitrary number of points
\begin{align}\label{npt-cyc}
 R(i;j_1j_2\dots j_k)= R(i;j_{k+1}j_2\dots j_k)+\frac12\sum_{p=2}^{k-2} R(i;j_{k+1}j_1\dots j_p)R(i;j_{k+1}j_{p+1}\dots j_k) +\text{cyclic($j_1 j_2\dots j_k$)}\,,
\end{align}
where the expression on the right-hand side is symmetrised with respect to cyclic permutations
of the indices $j_1,j_2,\dots,j_k$. 

\section{Next-to-lowest component}

As we have shown in the previous section, the lowest component of the correlation function \re{G-mod4} reduces to a product
of free scalar propagators \re{Gn0}. In this section, we shall compute the first component $G_{n;1}$ of \re{G-mod4} with non-trivial 
dependence on the Grassmann variables. We recall that $G_{n;1}$ is a homogenous function of  $\theta_i^+$ (with $i=1,\dots,n$) of  
degree four. 

In the conventional approach, to obtain $G_{n;1}$ we have to replace the superfields $\mathcal T(i)$ in \re{Gn} by their expansion \re{T-dec} in powers
of $\theta^+_i$ and to single out the contribution involving products of four Grassmann variables. In this way, $G_{n;1}$ is given by a
sum of $n-$point correlation functions involving various components of the stress-tensor supermultiplet. Each of these component correlation functions has conformal symmetry, but $\cN=4$ supersymmetry  is not manifest. The main advantage of the twistor
space approach is to offer an efficient way of finding $G_{n;1}$ without the need of computing individual component correlation functions; $\cN=4$ supersymmetry is manifest.\footnote{We recall that the price to pay for this is the presence of the reference twistor $\cZ_*$, in addition to the external data. The important point however is that $\cZ_*$ drops out from the final expressions, due to gauge invariance.}
 
\begin{figure}[h!t]
\psfrag{i}[cc][cc]{$\scriptstyle i$}
\psfrag{j}[cc][cc]{$\scriptstyle j$}
\psfrag{a}[cc][cc]{(a)}  \psfrag{b}[cc][cc]{(b)}  \psfrag{c}[cc][cc]{(c)}  \psfrag{d}[cc][cc]{(d)} \psfrag{e}[cc][cc]{(e)} 
\psfrag{j1}[cc][cc]{$\scriptstyle j_1$} \psfrag{j2}[cc][cc]{$\scriptstyle j_2$} \psfrag{j3}[cc][cc]{$\scriptstyle j_3$} \psfrag{j4}[cc][cc]{$\scriptstyle j_4$} 
\psfrag{k1}[cc][cc]{$\scriptstyle k_1$} \psfrag{k2}[cc][cc]{$\scriptstyle k_2$}
\psfrag{l1}[cc][cc]{$\scriptstyle l_1$} \psfrag{l2}[cc][cc]{$\scriptstyle l_2$}
\psfrag{m1}[cc][cc]{$\scriptstyle m_1$} \psfrag{m2}[cc][cc]{$\scriptstyle m_2$}
\psfrag{dots}[cc][cc]{$\vdots$}
 \centerline{\includegraphics[width =   \textwidth]{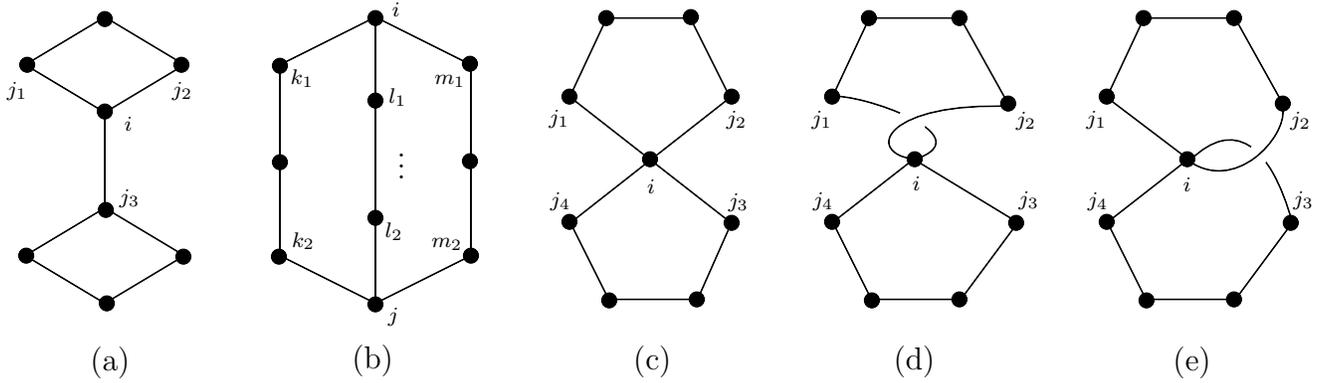}}
\caption{\small Topologies of twistor diagrams that contribute to $G_{n;1}$.}
\label{fig-topos}
\end{figure}

According to \re{rule},  the correlation function $G_{n;1}$ is given by the sum of all twistor diagrams containing $n$ vertices and $(n+1)$ edges.
Since each vertex is at least $2-$valent, such diagrams may have either two $3-$valent vertices, or a single $4-$valent 
vertex with the remaining vertices being $2-$valent.  Thus, we distinguish different topologies of 
twistor diagrams shown in Fig.~\ref{fig-topos}. The last three diagrams correspond to different embeddings of the colour-ordered quartic vertex.

\medskip
 
Let us first consider the contribution of the diagram shown in Fig.~\ref{fig-topos}(a). It involves two chains of propagators
attached to two cubic vertices with indices $i$ and $j_3$. Applying the Feynman rules, we find that the contribution 
of this diagram to the correlation function vanishes  
\begin{align}\label{Ga}
G_{n;1}^\text{(a)} \sim \delta^{a_{j_1}a_{j_2}}\tr[T^{a_{j_1}} T^{a_{j_2}}T^{a_{j_3}}] R(i;j_1j_2j_3)  = 0 \,,
\end{align}
where $\delta^{a_{j_1}a_{j_2}}$ comes from the product of propagators connecting $2-$valent vertices $j_1$ and $j_2$.
Here we took into account that $ T^{a} T^{a}=C_F=(N^2-1)/N$ is the quadratic Casimir of the  gauge group $SU(N)$ 
and, as a consequence, the colour trace in the above relation vanishes, $\tr T^{a_{j_3}} =0$.
 
The diagram shown in  Fig.~\ref{fig-topos}(b) contains three chains of propagators attached to two vertices with indices $i$
and $j$. Explicitly, its contribution is
\begin{align}\notag
{\rm Fig.~\ref{fig-topos}(b)}={}&  R(i; k_1l_1m_1) \tr[T^{a_{k_1}} T^{a_{l_1}}T^{a_{m_1}}]
\times 
R(j; k_2 l_2 m_2)  \tr[T^{a_{k_2}} T^{a_{l_2}}T^{a_{m_2}}]
\\
\times 
& \delta^{a_{k_1}a_{k_2}}
\left(  {y_{ik_1}^2 \over x_{ik_1}^2} \dots {y_{k_2j}^2 \over x_{k_2j}^2} \right) 
\times
\delta^{a_{l_1}a_{l_2}}\left(  {y_{il_1}^2 \over x_{il_1}^2} \dots {y_{l_2j}^2 \over x_{l_2j}^2} \right) 
\times
\delta^{a_{m_1}a_{m_2}}\left(  {y_{im_1}^2 \over x_{im_1}^2} \dots {y_{m_2j}^2 \over x_{m_2j}^2} \right) \,,
\end{align}
where the dots stand for the product of the remaining propagators constituting the three chains.  
As opposed to the previous case, the colour factor of this diagram is different from zero. In the correlation function, the above expression should be symmetrised with respect to the indices of all vertices in order to respect the Bose symmetry. In particular, since the cubic vertex is antisymmetric under the exchange of external legs,
 $R(i; k_1l_1m_1)= - R(i; l_1k_1m_1)$, its colour factor  $\tr[T^{a_{k_1}} T^{a_{l_1}}T^{a_{m_1}}]$ should also have the same property for the contribution of the diagram to be Bose symmetric. This allows us to replace $\tr(T^{a_{k_1}} T^{a_{l_1}}T^{a_{m_1}}) \to 
  \tr([T^{a_{k_1}} ,T^{a_{l_1}}]T^{a_{m_1}})$ yielding
\begin{align}
  \delta^{a_{k_1}a_{k_2}}\delta^{a_{l_1}a_{l_2}}\delta^{a_{m_1}a_{m_2}}\tr\left([T^{a_{k_1}}, T^{a_{l_1}}]T^{a_{m_1}}\right)\tr\left([T^{a_{k_2}}, T^{a_{l_2}}]T^{a_{m_2}}\right) = -2(N^2 -1) N\,,
\end{align}
where we used $ [T^a, T^b] = i \sqrt{2} f^{abc} T^c$ with $f^{abc} f^{abc'} =N  \delta^{cc'} $ for the gauge group $SU(N)$  and  $\tr(T^a T^b) = \delta^{ab}$. In this way, we find the contribution of the diagram in Fig.~\ref{fig-topos}(b) (see footnote \ref{foot})
\begin{align}\label{Gb}
G_{n;1}^{\rm (b)}=-   R(i; k_1l_1m_1)  
R(j; k_2 l_2 m_2)  
d_{ik_1\dots k_2 j}d_{il_1\dots l_2 j}d_{im_1\dots m_2 j} + \text{($S_n-$perm)}\,,
\end{align}
where the notation was introduced for the product of 
scalar propagators
\begin{align}\label{D-fun}
d_{ik_1\dots k_2 j} =  {y_{ik_1}^2 \over x_{ik_1}^2} \dots {y_{k_2j}^2 \over x_{k_2j}^2}\,.
\end{align} 
The diagrams shown in Fig.~\ref{fig-topos}(c)--(e) contain two chains of propagators that are attached to the quartic vertex in three different ways. Their contribution to the correlation function is  
\begin{align} \label{cde}
\text{Fig.~\ref{fig-topos}(c+d+e)}= d_{ij_1\dots j_2 i}  {}& d_{ij_4\dots j_3 i}  \big[ C_{\rm c}\, R(i;j_1j_2j_3j_4) 
  +C_{\rm d}\, R(i;j_1j_3j_4j_2) 
+C_{\rm e}\, R(i;j_1j_4j_2j_3) \big] \, .
\end{align}
The colour factors are
\begin{align}\notag\label{CC}
C_{\rm c} &= \delta^{a_{j_1}a_{j_2}}\delta^{a_{j_3}a_{j_4}} \tr\lr{T^{a_{j_1}} T^{a_{j_2}}T^{a_{j_3}}T^{a_{j_4}}} =N C_F^2 \, ,
\\ \notag
C_{\rm d} &= \delta^{a_{j_1}a_{j_2}}\delta^{a_{j_3}a_{j_4}}  \tr\lr{T^{a_{j_2}} T^{a_{j_1}}T^{a_{j_3}}T^{a_{j_4}}} =N C_F^2 \, ,
\\
C_{\rm e} &= \delta^{a_{j_1}a_{j_2}}\delta^{a_{j_3}a_{j_4}}  \tr\lr{T^{a_{j_1}} T^{a_{j_4}}T^{a_{j_2}}T^{a_{j_3}}} 
=N C_F(C_F-N) \, ,
\end{align}
where $C_F=(N^2-1)/N$ is the quadratic Casimir of $SU(N)$ in the fundamental representation.
Notice that $C_{\rm e}$ is suppressed at large $N$ by a factor of $1/N^2$, compared to $C_{\rm c}$ and $C_{\rm d}$. This reflects the fact that the former diagram is non-planar whereas the latter two are planar. 

Substituting \re{CC} into \re{cde} we expect to encounter both planar and non-planar   
contributions. It turns out that the non-planar diagram \ref{fig-topos}(e) cancels against the $1/N^2$ suppressed contributions of the diagrams in Fig.~\ref{fig-topos}(c)+(d) in such a way that their total sum remains planar in the large $N$ limit, in perfect agreement with \re{planar}. To show this, we apply the relation \re{U(1)} for $k=4$ to replace 
$R(i;j_1j_4j_2j_3)=-R(i;j_1j_2j_3j_4) -R(i;j_1j_3j_4j_2) $ in \re{cde} leading to
\begin{align}\notag\label{planar1}
{}& C_{\rm c}\, R(i;j_1j_2j_3j_4) 
  +C_{\rm d}\, R(i;j_1j_3j_4j_2) 
+C_{\rm e}\, R(i;j_1j_4j_2j_3) 
\\[2mm]\notag
{}& \qquad  =(C_{\rm c}-C_{\rm e})R(i;j_1j_2j_3j_4) + (C_{\rm d}-C_{\rm e})R(i;j_1j_3j_4j_2) 
\\[2mm]
{}& \qquad =  (N^2-1) N \big[ R(i;j_1j_2j_3j_4) + R(i;j_2j_1j_3j_4) \big]\,,
\end{align}
where in the last relation we made use of the identity $R(i;j_1j_3j_4j_2)=R(i;j_2j_1j_3j_4)$, Eq.~\re{sym}. Comparing with \re{CC} we
observe that all terms proportional to $C_F^2$ cancel out in the sum over all diagrams and the only terms that survive are those involving 
the colour factor $C_F N$. This property is reminiscent of the so-called non-abelian exponentiation of Wilson loops~\cite{Gatheral:1983cz,Frenkel:1984pz}. 

We combine \re{cde} and \re{planar1} to obtain the contribution of the diagrams Fig.~\ref{fig-topos}(c),(d),(e) to the correlation function
(see footnote \ref{foot})
\begin{align}\label{Gc}
G_{n;1}^\text{(c)+(d)+(e)} =  R(i;j_1j_2j_3j_4)\,d_{ij_1\dots j_2 i}  {}& d_{ij_4\dots j_3 i}+ \text{($S_n-$perm)}\,.
\end{align}
This expression involves a quartic vertex which can be expressed in terms of cubic vertices using \re{eq:11}
\begin{align}\label{4vertex}
  R(i;j_1j_2 j_3 j_4)=R(i;j_1j_2j_3)\, R(i;j_1j_3j_4) = R(i;j_2j_4j_1)\, R(i;j_2j_3j_4)\ .
\end{align}
Finally, we combine relations \re{Ga}, \re{Gb} and \re{Gc} to obtain the following representation for 
the next-to-lowest component of the correlation function:
\begin{align}\notag\label{Gn1}
G_{n;1} =  {}& -   R(i; k_1l_1m_1)  
R(j; k_2 l_2 m_2)  
d_{ik_1\dots k_2 j}d_{il_1\dots l_2 j}d_{im_1\dots m_2 j} 
\\[2mm]
{}& + R(i;j_1j_2j_3)\, R(i;j_1j_3j_4)d_{ij_1\dots j_2 i}  d_{ij_4\dots j_3 i} + \text{($S_n-$perm)} \,.
\end{align}
Here the indices $i,j,k,l,m$ label $n$ different points and the sum runs over their permutations.

The following comments are in order concerning the properties of \re{Gn1}.

A remarkable feature of \re{Gn1} is that the whole dependence on the Grassmann variables is encoded in the simple cubic $R-$vertex given by \re{eq:10}. According to its definition, Eqs.~\re{eq:10} and \re{A-y}, the function $R(i;j_1j_2j_3)$ is a homogenous polynomial in $\theta_i^+$ of degree $2$, so that $G_{n;1}$ has Grassmann degree $4$ as it should be.

Recall that the dependence of the correlation function \re{Gn1} on the super-coordinates of the operators
$(x_i,\theta_i^+)$ enters into $R(i;j_1j_2j_3)$ through the commuting  spinors $\sigma_{ij}$ and the function $A_{ij}$ 
given by \re{sigma} and \re{A-y}, respectively. They depend  in turn on the supertwistor coordinates
defined in \re{Z-gauge} as well as on the reference supertwistor $\mathcal Z_*$. Notice that each term
on the right-hand side of \re{Gn1} depends on $\mathcal Z_*$ but this dependence should cancel in the total
sum in order for $G_{n;1}$ to be gauge invariant. We   demonstrate the independence of the correlation function 
\re{Gn1} of the reference supertwistor $\mathcal Z_*$ in the next section.

For $n=4$ the relation \re{Gn1} takes the form
\begin{align}\notag\label{G4-exp}
G_{4;1} {}& =  -  \prod_{1\le i<j\le 4} d_{ij}  
 \left[  R(1;324) R(2;314)/d_{34}+  R(1;234) R(3;214)/d_{24}+  R(1;243) R(4;213)/d_{23} \right.
\\ 
{}&  \ \ +  R(2;134) R(3;124)/d_{14} +  R(2;143) R(4;123)/d_{13} +  R(3;142) R(4;132)/d_{12} \big],
\end{align}
with $d_{ij}=y_{ij}^2/x_{ij}^2$. However, $G_{4;1}$ should vanish due to $\cN=4$ superconformal
symmetry (see Eq.~\re{G-mod4}). Therefore, the linear combination inside the square brackets in this relation should vanish. We  demonstrate this in Sect.~\ref{sec:example-component-} by an explicit calculation. 

\subsection{The light-like limit}
\label{sec:lightlike-limit}

As another test of \re{Gn1} we consider the limit of the correlation function $G_n$ in which the $n$ operators become sequentially light-like separated. In chiral superspace, this corresponds to $x_{i,i+1}^2\to 0$ and $\theta_{i,i+1}^{A,\alpha} (x_{i,i+1})_{\alpha\dot\alpha}\to 0$ for $i=1,\ldots,n$ and  the periodic boundary condition $i+n\equiv i$ is assumed.
In this limit we expect the correlation function to be related to
the square of the $n-$particle superamplitude~\cite{Eden:2011yp,Adamo:2011dq,Eden:2011ku}
\begin{align}
G_n \stackrel{x_{i,i+1}^2 \to 0}{\sim} G_{n;0} \left(1 + R_n^{\rm NMHV} + \ldots +  R_n^{\rm \overline{MHV}}\right)^2\,,
\end{align}
where $R_n^{\rm NMHV}$ is given by the ratio of the NMHV and MHV  $n-$particle amplitudes and similarly for the other components. For $G_n$ computed in the Born approximation, the amplitudes can be replaced by their tree level expressions. In this way, we find for the next-to-lowest component
\begin{align}\label{G-R}
\lim_{x_{i,i+1}^2 \to 0} G_{n;1}/G_{n;0}= 2  R_n^{\rm NMHV} \,.
\end{align}
The NMHV ratio function $R_n^{\rm NMHV}$ is known to have an enhanced dual (super)conformal symmetry 
\cite{Drummond:2008vq} and is given by a sum of five-point on-shell invariants (see Eq.~\re{sum-NMHV} below). 
The duality relation \re{G-R} then suggests that the ratio of the correlation functions $G_{n;1}/G_{n;0}$
should also have an enhanced symmetry, at least in the light-like limit.

Let us first examine the asymptotic behaviour of the lowest component $G_{n;0}$ in the light-like limit.
It is easy to see from \re{Gn0} that, in the sum over all $S_n$ permutations, only one term provides the 
leading singularity, 
\begin{align} \label{Gn0-lim}
G_{n;0} \stackrel{x_{i,i+1}^2 \to 0}{\sim}  \prod_{i=1}^{n}\,{y_{i,i+1}^2 \over x_{i,i+1}^2}
\equiv  d_{12\dots n}\,,
\end{align}
where  the $d-$function was introduced in \re{D-fun}.

For the next-to-lowest component $G_{n;1}$
the light-like limit can be imposed diagram by diagram.
Since each edge $(ij)$ connecting the vertices with the corresponding
labels comes with a factor $y^2_{ij}/x^2_{ij}$, we observe that only those graphs containing the edges 
$(12), (23) ,\dots, (n1)$ provide the leading contribution  in the light-like limit 
$x_{i,i+1}^2\to 0$;  all other graphs will be subleading. So in this limit only graphs containing a simply connected $n-$gon will survive. This $n-$gon clearly yields the same product of free scalar propagators
$y^2_{i,i+1}/x^2_{i,i+1}$ as the leading term in $G_{n;0}$, and therefore it provides a non-vanishing contribution to the ratio $G_{n;1}/G_{n;0}$ in the light-like limit. 

Examining the diagrams shown in Fig.~\ref{fig-topos}(b) -- (e) we notice that, since the total number of vertices in the diagrams equals $n$,  graphs (c), (d) and (e) cannot contain a simply connected $n-$gon and are thus subleading in the light-like limit.  For graph (b) to contain an $n-$gon, one of the chains connecting the cubic vertices $i$ and $j$ should not contain any vertices. In other words, the graphs that contribute to $G_{n;1}$ in the light-like limit have the form of an $n-$gon with one additional propagator stretched between vertices $i$ and $j$. Using \re{Gn1} their contribution is brought to the form
\begin{align} \label{Gn1-lim}
G_{n;1}  \stackrel{x_{i,i+1}^2 \to 0}{\sim}  d_{12\dots n} \sum_{i\neq j} R_{ij*} \,,
 \end{align}
where $R_{ij*}$ is given by the product of two cubic vertices
\begin{align}\label{RR}
 R_{ij*} = {y_{ij}^2\over x_{ij}^2} R(i; i-1\, j \,i+1)  
R(j; j-1 \, i \,j+1)\,.
\end{align} 
Here we explicitly indicated the dependence of $R_{ij*}$ on the reference supertwistor $\mathcal Z_*$.

Taking into account \re{Gn0-lim} and \re{Gn1-lim}, we find for the ratio of correlation functions in the light-like limit
\begin{align}\label{G-rat}
 \lim_{x_{i,i+1}^2 \to 0} G_{n;1}/G_{n;0}= 2  \sum_{i<j} R_{ij*} \, .
\end{align}
To simplify the expression for $R_{ij*}$ it is convenient to return to the integral representation of $G_{n;1}$ based
on the Feynman rules in twistor space in Fig.~\ref{fig-rules2}. Then, 
\begin{align} \label{Gn1-new}
R_{ij*} =
\int  \frac{d^2 \sigma_{ij}d^2
  \sigma_{ji} \,\vev{\sigma_{ii-1}\sigma_{ii+1}} \vev{\sigma_{jj-1}\sigma_{jj+1}}}{\vev{\sigma_{ii-1}\sigma_{ij}}\vev{\sigma_{ij} \sigma_{ii+1}} \vev{\sigma_{jj-1}\sigma_{ji}}\vev{\sigma_{ji} \sigma_{jj+1}}}
   \delta^{4|4}(\cZ_* +
  \sigma_{ij}^{\alpha}\cZ_{i,\alpha} +
  \sigma_{ji}^{\alpha}\cZ_{j,\alpha}) \, .
\end{align}
To reproduce \re{RR} it suffices to split the delta function in this relation into bosonic and fermionic parts, Eq.~\re{eq:1}, 
and to apply relations \re{dd} and \re{eq:22}. 

The parameters $\sigma_{ii-1}$ and $\sigma_{i-1 i}$ in \re{Gn1-new} are given
by the general expressions \re{si} which become singular in the light-like limit since $\vev{Z_{i-1,1}Z_{i-1,2}Z_{i,1}Z_{i,2}}=x_{i-1,i}^2 \to 0$.
Nevertheless, we can use the invariance of \re{Gn1-new} under rescalings of  $\sigma$  to put
\begin{align}\label{spec}
& \sigma_{i,i-1}^{\beta} = \epsilon^{\beta\alpha}   \vev{Z_* Z_{i-1,1}Z_{i-1,2} Z_{i,\alpha}}  \,,
\qqqquad \sigma_{i,i+1}^{\beta} = \epsilon^{\beta\alpha}   \vev{Z_* Z_{i+1,1}Z_{i+1,2} Z_{i,\alpha}}  \,,
\end{align}
and similarly for $\sigma_{j,j-1}$ and $\sigma_{j,j+1}$.
We recall that  in twistor space the light-like limit, $x_{ii+1}^2\rightarrow 0$ and $\theta_{i,i+1}^{A,\alpha}(x_{i,i+1})_{\alpha\dot\alpha}\to 0$, is equivalent to the intersection of the
corresponding twistor lines $\cZ_{i\alpha}$ and $\cZ_{i{+}1\,\alpha}$. The local $GL(2)$ invariance (corresponding to the reparameterisation freedom on each twistor line) allows us to choose this intersection to occur in the following convenient manner
\begin{align}
  \label{eq:33}
  \cZ_{i,2}=\cZ_{i+1,1} \equiv \cZ_i  \,, \qqqquad (\text{$i=1\dots n$})\,,
\end{align}
where $\cZ_i = (Z_i, \chi_i^A)$ with $Z_i=(\lambda_i^\alpha, x_i^{\dot\alpha\beta}\lambda_{i\beta})$ and
 $\chi_i^A=\theta_i^{A,\beta}\lambda_{i\beta}$. Substituting the bosonic part of this relation into \re{spec} we find  
\begin{align}
  \label{eq:34}
  \sigma_{i\, i{+}1}^{\a=1}= \sigma_{ i\,i-1}^{\a=2} =0\,, \qqqquad   \sigma_{i\,
  i{+}1}^{\a=2}  =- \sigma_{i{+}1\,i}^{\a=1}   \ .
\end{align}
Denoting $\sigma_{ij}^\alpha=(s_1,s_2)$ and $\sigma_{ji}^\alpha=(t_1,t_2)$ we finally obtain from \re{Gn1-new}
\begin{align}\notag
 R_{ij*} {}& =\int {ds_1 ds_2 dt_1 dt_2\over s_1 s_2 t_1 t_2} \delta^{4|4}(\cZ_* +
 s_1 \cZ_{i-1} +s_2 \cZ_i +t_1 \cZ_{j-1} +t_2 \cZ_j)
 \\
 {}& = {\delta^4(\chi_* \vev{i-1 i j-1 j}+ \chi_{i-1} \vev{ i j-1 j *} + \ldots  + \chi_j \vev{* i-1 i j-1} ) \over \vev{i-1 i j-1 j}\vev{ i j-1 j *} 
 \vev{j -1 j * i-1} \vev{ j * i-1 i}\vev{* i-1 i j-1}}\,,
\end{align}
with $\vev{i-1 i j-1 j}\equiv \vev{Z_{i-1}Z_ iZ_{j-1} Z_j}$,
which is precisely the invariant defining the NMHV tree-level amplitude \cite{Drummond:2008vq,Mason:2009qx}
\begin{align}\label{sum-NMHV}
R_n^{\rm NMHV}  =  \sum_{i<j} R_{ij*} \,.
\end{align}
Comparing this relation with \re{G-rat} we observe perfect agreement with \re{G-R}. In addition, \re{RR}
yields the factorisation of the NMHV (on-shell) invariant $R_{ij*}$ into a product of two (off-shell) cubic vertices in the light-like limit.

\subsection{Independence of the reference twistor}
\label{sec:canc-spur-poles}

In the previous section we have shown  that the correlation function $G_{n;1}$ can be built from the cubic vertices $R(i;j_1j_2 j_3)$. These 
vertices depend on the four supertwistors corresponding to the external points $i,j_1,j_2,j_3$ as well as on the reference supertwistor $\mathcal Z_*$. 
They are constructed using the Feynman rules in Fig.~\ref{fig-rules2} that have manifest $\mathcal N=4$ superconformal covariance as long as 
we transform the reference twistor too. In this sense the symmetry  of $R(i;j_1j_2 j_3)$ is actually broken by the presence of the fixed constant reference supertwistor. However,
the symmetry is restored in $G_{n;1}$ since it must not depend on the reference twistor (that is, on the  gauge choice). In this section we confirm that this is indeed the case. 
 
 As follows from \re{eq:10}, the dependence of $R(i;j_1j_2 j_3)$ on the reference twistor enters through the parameters $\sigma_{ij}$ given by \re{sigma}.
Viewed as a function of $\mathcal Z_*$, the vertex $R(i;j_1j_2 j_3)$ has spurious poles located at 
$ \vev{\sigma_{i j_1} \sigma_{i j_2}}  \,\vev{\sigma_{i j_2} \sigma_{i j_3}}\,\vev{\sigma_{i j_3} \sigma_{i j_1}}=0$.
We shall argue  that the absence of spurious poles is equivalent to the $\mathcal Z_*-$independence of $G_{n;1}$.
Let us show how the spurious poles cancel in the sum of all twistor diagrams shown in Fig.~\ref{fig-topos}.

More specifically, consider a particular spurious pole located at $\vev{\sigma_{12} \sigma_{13}}=0$. 
\footnote{Of course we can choose any three points for the spurious pole condition.}
We can use \re{sigma} to verify the following
identity
\begin{align}\label{66}
  \vev{\sigma_{12} \sigma_{13}} x_{12}^2x_{13}^2
  =\vev{\sigma_{23} \sigma_{21}}  x_{23}^2x_{21}^2 =\vev{\sigma_{31} \sigma_{32}} 
  x_{13}^2x_{23}^2 \equiv (123) \,,
\end{align}
where $(123)$ is totally antisymmetric under the exchange of any pair of points.
It implies that the same spurious pole corresponds to $\vev{\sigma_{12} \sigma_{13}}=\vev{\sigma_{23} \sigma_{21}}=\vev{\sigma_{31} \sigma_{32}} =0$,
or equivalently
\begin{align}\label{la}
(\sigma_{13})^\alpha=z_1 (\sigma_{12})^\alpha\,,\qquad 
(\sigma_{21})^\alpha=z_2 (\sigma_{23})^\alpha\,,\qquad 
(\sigma_{32})^\alpha=z_3 (\sigma_{31})^\alpha\,.
\end{align}
The complex parameters $z_i$ in this relation are not independent however. We take into account the identity (see Eq.~\re{Iden1} in Appendix~\ref{D} for its
derivation)
\begin{align}\label{iden1}
 {(\sigma_{13}^\alpha \sigma_{21}^\beta)}+{
  (\sigma_{12}^\alpha \sigma_{23}^\beta)}-{
  (\sigma_{13}^\alpha \sigma_{23}^\beta)} = 0 \,,\qquad \text{for $(123)=0$}
\end{align}
and substitute \re{la}  to get
\begin{align}
 z_1 + 1/z_2 - z_1/z_2 = 0\,.
\end{align}
To obtain an analogous relation between $z_1$ and $z_3$ we permute the indices $2$ and $3$ on both sides of \re{iden1} and 
take into account that $(132)=-(123)$. In this way, we obtain
\begin{align}
  \label{eq:43}
  z_2 = \frac{z_1-1}{z_1}\,, \qqqquad z_3 =
  \frac1{1-z_1}\ .
\end{align}
Examining the expression for the cubic vertex \re{eq:10} for different values of the indices, we find that the spurious pole at $(123)=0$ 
appears in three different vertices,  
\begin{align}
  \label{eq:38}
  R(1;23i),\qqqquad   R(2;31j),\qqqquad   R(3;12k),
\end{align}
where $i$, $j$ and $k$ are arbitrary points (different from $1,2,3$). We use \re{eq:10} and \re{la} to compute the residues at the spurious 
pole
\begin{align}\notag\label{eq:44}
   \lim_{(123)\rightarrow 0} (123) {y_{12}^2\over x_{12}^2}
   {y_{13}^2\over x_{13}^2}R(1;23i) {}& = \frac1z_1 y_{12}^2 y_{13}^2 \delta^2(z_1 A_{12}
   - A_{13}) \, ,
\\\notag
   \lim_{(123)\rightarrow 0} (123) {y_{12}^2\over x_{12}^2}
   {y_{23}^2\over x_{23}^2}R(2;31j) {}&=  \frac1z_2 y_{12}^2 y_{23}^2 \delta^2(z_2 A_{23}
   - A_{21})\, ,
 \\
   \lim_{(123)\rightarrow 0} (123) {y_{13}^2\over x_{13}^2}
   {y_{23}^2\over x_{23}^2}R(3;12k) {}&=  \frac1z_3 y_{13}^2 y_{23}^2 \delta^2(z_3 A_{31}
   - A_{32})\, ,
\end{align}
where $A_{ij}$ are given by \re{A-y} and \re{A-mat}. 

Let us show that the sum of the three residues \re{eq:44} vanishes. To simplify
the calculation, we make use of the superconformal symmetry of the $R-$vertex to fix the gauge 
\begin{align}
  \label{eq:39}
  \theta^+_1=\theta^+_2=\theta^+_3=0\,, \qqqquad   y_1=0,\ y_2=1,\ y_3\to\infty\ .
\end{align}
The generic values of these coordinates can be restored via a finite $\mathcal N=4$ superconformal transformation.
In this gauge, the $A_{ij}$ in \re{eq:44} simplify to $A_{ij}^{a'}  =  \theta_*^A u_{j,A}^{+b} (y_{ij}^{-1})_{b}^{a'}$.
Splitting $\theta_*^A = (\theta_*^a, \theta_*^{a'})$ and expressing  
$ u^+_j$ in terms of the variables $y_j$ as described in \re{u-y}, we find
\begin{align} \notag
 A_{12}&= \theta'_*-\theta_*\,, && A_{23} = -\theta_*\,, && A_{31} = -\theta'_* y_3^{-1}\,,
 \\[2mm]
 A_{13}&=-\theta_*\,, && A_{21} = -\theta'_*\,, && A_{32} =  (\theta_*-\theta'_*)y_3^{-1}\,.
\end{align}
Substituting these relations into \re{eq:44} and taking into account \re{eq:43}, we find that the delta functions on the right-hand side of \re{eq:44} are proportional to
\begin{align}\label{r}
   r_{123} =  y_{3}^2
   \delta^2\big(\Theta_* \big)\,,\qqqquad \Theta_*=(1-z_1) \theta_*+z_1 \theta'_*  .
 \end{align}
Next, we evaluate the sum of the residues of the three $R-$vertices at the spurious pole $(123)=0$ 
and find that it vanishes,
\begin{align}\notag\label{zero}
  \lim_{(123)\rightarrow 0}  (123)\bigg[ {y_{12}^2\over x_{12}^2}
   {y_{13}^2\over x_{13}^2}R(1;23i)+{y_{12}^2\over x_{12}^2}
   {y_{23}^2\over x_{23}^2}R(2;31j)+{y_{13}^2\over x_{13}^2}
   {y_{23}^2\over x_{23}^2}R(3;12k)\bigg]{}&
   \\
  =    r_{123}\bigg[\frac1{z_1} - \frac1{z_1(1-z_1)} + \frac1{(1-z_1)}\bigg]  {}&   = 0\,.
\end{align}
Here the three terms in the second relation correspond to the three terms in the first line.  Notice that the residues of the vertices \re{eq:38} at the spurious pole do not depend on the choice of the points $i$, $j$, $k$ and are proportional
to each other.

We can now apply \re{zero} to show the cancellation of spurious poles in the sum of the diagrams contributing
to the correlation function $G_{n}$. As we explained in Sect.~\ref{sec:twist-feynm-rules}, these diagrams involve vertices of 
different valency.  According to \re{eq:11}, they can all be expressed in terms of the cubic $R-$vertices. Examining
all possible vertices we find that the spurious pole at $(123)=0$ is only present in the vertices of the following types: 
$R(1;23a.. b)$, $R(2;31c.. d)$ and $R(3;12e.. f)$ with indices $a,b,c,d,e,f$ labeling the other external points. Indeed,
we can use \re{eq:28} to obtain the following representation 
\begin{align}\label{eq:50}
  R(1;23a.. b)&=R(1;3a.. b)R(1;23b)=R(1;2a.. b)R(1;23a)\,, \notag\\
  R(2;31c.. d)&=R(2;1c.. d)R(2;31d)=R(2;3c.. d)R(2;31c)\,, \notag\\
  R(3;12e.. f)&=R(3;2e.. f)R(3;12f)=R(3;1e.. f)R(3;12e)\,,
\end{align}
where the cubic vertices are of the form \re{eq:38} and thus contain a spurious pole at $(123)=0$. 

\begin{figure}[h!]
\psfrag{1}[cc][cc]{$\scriptstyle 1$}\psfrag{2}[cc][cc]{$\scriptstyle 2$}\psfrag{3}[cc][cc]{$\scriptstyle 3$}
\psfrag{a}[cc][cc]{$\scriptstyle a$}\psfrag{b}[cc][cc]{$\scriptstyle b$}\psfrag{c}[cc][cc]{$\scriptstyle c$}
\psfrag{d}[cc][cc]{$\scriptstyle d$}\psfrag{e}[cc][cc]{$\scriptstyle e$}\psfrag{f}[cc][cc]{$\scriptstyle f$}
\psfrag{A}[cc][cc]{(a)}\psfrag{B}[cc][cc]{(b)}\psfrag{C}[cc][cc]{(c)}
\psfrag{D}[cc][cc]{(d)}
\centering
 \includegraphics{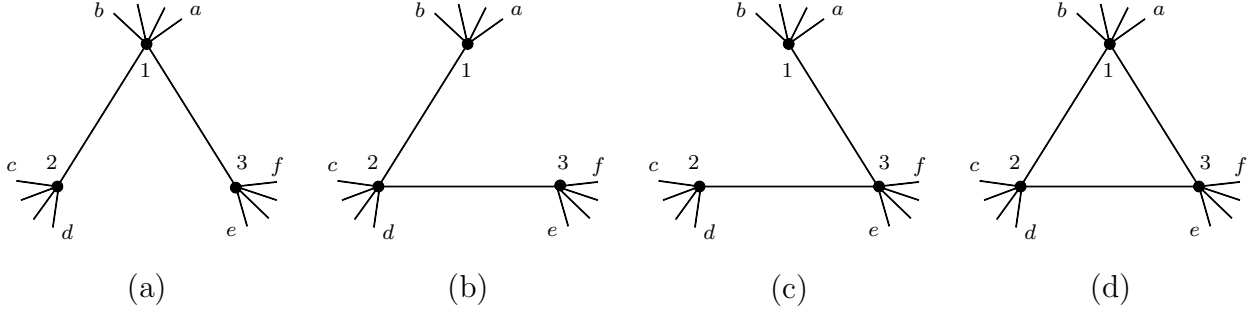}
  \caption{All subgraphs with a potential spurious pole at 
    $(123)=0$. The spurious pole is present in 
    graphs (a), (b) and (c) but cancels in their sum. The 
    graph (d) in fact has no spurious pole at 
    $(123)=0$. In the above
    diagrams the number of legs coming out of each of the vertices
    1,2,3 is arbitrary and we can even have just one leg coming out. 
    For example, we can have $a=b$ or $c=d$ etc. In the 
    graph (d) we can even have no additional legs from the  vertices. } 
\label{fig2}
\end{figure}

Let us consider the graphs shown in Fig.~\ref{fig2}. They can be viewed as part of a bigger diagram in which points $a,b,c,d,e,f,\dots$
label other vertices. The first three graphs in Fig.~\ref{fig2} have the same number of propagators, hence their contribution
to the correlation function has the same Grassmann degree. A special feature of these graphs is that they involve vertices
of the form \re{eq:50} and thus have spurious poles. Moreover, these are the only diagrams that are singular for $(123)=0$.
There is however another graph (see Fig.~\ref{fig2}(d))  that contains the same singular vertices \re{eq:50}. We will show below
that its contribution remains finite for $(123)=0$.

The total contribution of the graphs shown in Fig.~\ref{fig2} (a)-(c) is
\footnote{Here we assume the planar limit.}
 \begin{align}\label{eq:51}
&  d_{12}d_{13} R(1;23a.. b) R(2;1c.. d)  R(3;1e.. f)+ d_{12}d_{23} R(1;2a.. b) R(2;31c.. d)  R(3;2e.. f)\notag \\
&  + d_{13}d_{23}  R(1;3a.. b) R(2;3c.. d) R(3;12e.. f)\,,
\end{align}
where $d_{ij}=y_{ij}^2/x_{ij}^2$ is a scalar propagator.  We apply \re{eq:50} to rewrite the first term in the last relation as 
\begin{align}\notag
R(1;23a.. b) R(2;1c.. d)  R(3;1e.. f) {}&= R(1;2a.. b)R(1;23a)R(2;1c.. d)  R(3;1e.. f) 
\\[2mm]
{}& = R(1;2a.. b)R(1;23a)R(2;3c.. d)  R(3;1e.. f) + \text{(reg.)}\,,
\end{align}
where  `reg' denotes terms regular for $(123)=0$. Here in the second relation we took into account that the residues of $R(1;23a)$
and $R(2;31d)$ at $(123)=0$ are proportional to each other and  are independent of the points $a$ and $d$ (see Eq.~\re{zero}), leading to
\begin{align}\notag
 \lim_{(123)\rightarrow 0}  (123)  R(1;23a)R(2;1c.. d) {}&= \xi \lim_{(123)\rightarrow 0}  (123)  R(2;31d)R(2;1c.. d) 
 \\ \notag
 {}&= \xi \lim_{(123)\rightarrow 0}  (123)  R(2;31c)R(2;3c.. d) 
  \\
 {}&=  \lim_{(123)\rightarrow 0}  (123)  R(1;23a)R(2;3c.. d),
\end{align}
where $\xi= (z_1-1) d_{23}/d_{13}$ and we applied \re{eq:50} in the second line. 
The remaining terms in \re{eq:51} can be simplified likewise. In this way, we evaluate the residue of \re{eq:51} at $(123)=0$ and find that it is proportional to the same linear combination of cubic vertices as in \re{zero},
\begin{align}\notag \label{zero1}
   \lim_{(123)\rightarrow 0} (123)  {}&\times \text{Eq}.\re{eq:51} =  R(1;2a..b)R(2;3c..d)R(3;1e..f) 
   \\
 {}&  \times \lim_{(123)\rightarrow 0} (123) \big[d_{12} d_{13} R(1;23a) +d_{12} d_{23} R(2;31c) +d_{13} d_{23} R(3;12e)   \big] = 0\,.
\end{align}
We conclude that the spurious pole is indeed absent in the sum of all diagrams  in 
Fig.~\ref{fig2}(a)-(c).

Finally, there exists the possibility of having a subgraph of the type shown in Fig.~\ref{fig2}(d). Its contribution contains the 
product of three vertices 
\begin{align}
d_{12} d_{23} d_{13} R(1;23a..b) R(2;31c..d) R(3;12e..f)\,,
\end{align}
each of which having a spurious pole at $(123)=0$. Denoting $(123)=\epsilon$ we find  for $\epsilon\to 0$ 
\begin{align}\label{3R}
R(1;23a..b)\sim R(1;23 a) \sim {1\over \epsilon} \delta^2\big(\Theta_* +\epsilon f_1
+ O(\epsilon^2) \big)\,.
\end{align}
Here in the first relation we applied \re{eq:50} and in the second relation made use of \re{eq:44} and \re{zero}.
As compared with \re{r}, we included in \re{3R} the subleading $O(\epsilon)$ correction parameterised
by some odd function $f_1$ whose explicit form will not be important for our purposes. For $\epsilon=0$, the
delta function on the right-hand side of \re{3R} coincides with $r_{123}$ defined in \re{r}.  The two remaining $R-$vertices
in \re{zero1} also satisfy \re{3R} with $f_1$ replaced by some functions. Then, for the product of three $R-$vertices
we find for $\epsilon\to 0$
\begin{align}\notag
\text{Eq.\re{3R}} {}&\sim {1\over \epsilon^3} \delta^2\big(\Theta_* +\epsilon f_1) \big)\delta^2\big(\Theta_* +\epsilon f_2
 \big)\delta^2\big(\Theta_* +\epsilon f_3\big)
 \\
{} &={1\over \epsilon^3}\delta^2\big(\Theta_* +\epsilon f_1) \big)\delta^2\big(\epsilon (f_1-f_2)
 \big)\delta^2\big(\epsilon (f_1-f_3)\big) \sim  O(\epsilon)\,,
\end{align}
so that the contribution of the graph in Fig.~\ref{fig2}(d) vanishes for $(123)\to 0$.

Note that the above discussion is not sensitive to the number of legs attached to vertices $1,2$ and $3$ 
(see Fig.~\ref{fig2}). In particular, it also applies when there is only one additional line coming out of each vertex, 
e.g. we could have $a=b$ and/or $c=d$ and/or $e=f$. In this case,  $R(1;2a..b)$, $R(2;3c..d)$ and $R(3;1e..f)$ in 
\re{zero1} describe bivalency vertices which equal 1 according to \re{R-2pt}.

\medskip
\begin{figure}[h!]
\psfrag{1}[cc][cc]{$\scriptstyle 1$}\psfrag{2}[cc][cc]{$\scriptstyle 2$}\psfrag{3}[cc][cc]{$\scriptstyle 3$}
\centering
 \includegraphics{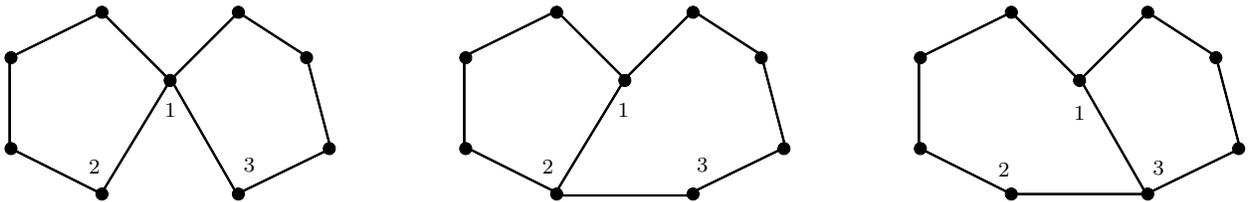}
  \caption{Example of diagrams contributing to $G_{n;1}$ and having a spurious pole at $(123)=0$.
  This pole cancels in the sum of three diagrams.} 
\label{fig:example}
\end{figure}

The mechanism of cancellation of spurious poles described in this subsection is rather general as it applies to 
any component of the correlation function $G_n$.  In application to the next-to-lowest component $G_{n;1}$
defined by the diagrams shown in Fig.~\ref{fig-topos} given by \re{Gn1}, we can restrict ourselves to the graphs in 
Fig.~\ref{fig2} containing vertices of valency $2$, $3$ and $4$ only. As an example, we show in Fig.~\ref{fig:example} 
the set of diagrams which contribute to $G_{n;1}$ and whose sum is free from spurious pole at $(123)=0$.
It is straightforward to extend the analysis of spurious poles to the higher components of $G_{n}$.

In this subsection we have demonstrated  that the correlation function $G_n$ is free from spurious poles depending on  the reference supertwistor $\mathcal Z_*$.
This property combined with the fact that $G_n$ is a rational homogeneous  function of  $\mathcal Z_*$  of degree $0$ implies that it is $\mathcal Z_*$ independent.
  
\subsection{Short-distance limit} 
 
In the previous subsection  we have shown that all spurious poles cancel in the correlation function $G_n$.
As a consequence, the only singularities that $G_n$ can have are those coming from short distances $x_i\to x_j$.
We shall refer to them as physical poles. 

The short distance asymptotics of $G_n$ is controlled by the operator product expansion of the stress-tensor 
multiplets $\mathcal T(1)\mathcal T(2)$. Each operator depends on the set of coordinates $(x_i,\theta_i^+,u_i(y_i))$
and the short distance Euclidean limit $1\to 2$ amounts to $x_1\to x_2$, $\theta_1^+\to \theta_2^+$ and $y_1\to y_2$.
In this limit we have  
\begin{align}\label{OPE}
\mathcal T(1) \mathcal T(2) = {N^2-1\over 2} \lr{y_{12}^2\over x_{12}^2}^2 \mathcal I +  2\,{y_{12}^2\over x_{12}^2} {\mathcal T}(1) + \dots\,,
\end{align}
where the dots denote terms suppressed by powers of $x_{12}^2$ and $y_{12}^2$. The first term on the right-hand side 
of \re{OPE} involves the identity operator and it describes the disconnected contribution to the correlation function 
$G_n$ for $1\to 2$.  Applying \re{OPE}, we find the leading asymptotic behaviour of the connected part of the correlation 
function $G_n$ for $1\to 2$ to be
\begin{align}\label{lim}
G_{n}\stackrel{1\to 2}{\sim} 2  {y_{12}^2\over x_{12}^2} \,G_{n-1} \,.
\end{align}
Examining the twistor diagrams contributing to $G_n$, we find that the physical pole $y_{12}^2/x_{12}^2$ only comes from the 
diagrams in which vertices $1$ and $2$ are connected by a propagator. Then, in order to verify \re{lim} 
it is sufficient to show that  in the short-distance
limit the product of two $R-$vertices at points $1$ and $2$ reduces to a single $R-$vertex.

\begin{figure}[th!]
\psfrag{g}[cc][cc]{$2 d_{12}$}\psfrag{cyclic}[cc][cc]{cyclic}\psfrag{+}[cc][cc]{$+$}
\psfrag{to}[cc][cc]{$\stackrel{1\to 2}{\sim}$}
\psfrag{1}[cc][cc]{$\scriptstyle 1$}\psfrag{2}[cc][cc]{$\scriptstyle 2$}
\psfrag{a}[cc][cc]{$\scriptstyle j_1$} \psfrag{b}[cc][cc]{$\scriptstyle j_2$}\psfrag{c}[cc][cc]{$\scriptstyle j_3$} \psfrag{d}[cc][cc]{$\scriptstyle j_4$}  
\centerline {\includegraphics[width = 0.75 \textwidth]{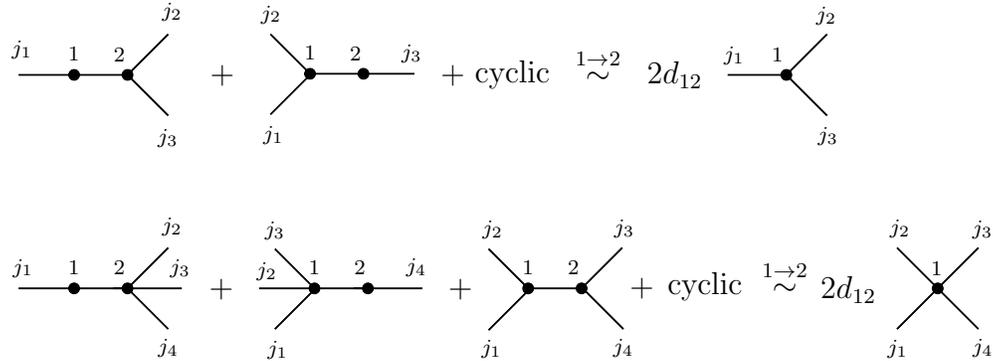}}
\caption{\small The OPE relations for $3-$ and $4-$point vertices. The expressions on the left-hand side are symmetrised with respect to cyclic shifts of the labels of the external legs.}
\label{fig-OPE}
\end{figure}

For the lowest component $G_{n;0}$,  the relation
\re{lim} follows immediately from \re{Gn0}. For the next-to-lowest component $G_{n;1}$, we have to examine different
contributions where the vertices $1$ and $2$ have valency $2$, $3$ and $4$. If both vertices have valency $2$, the contribution
of the corresponding  graph to $G_{n;1}$ automatically verifies \re{lim}. When one of the vertices has valency $2$ and
the other has valency $3$, the corresponding contribution to $G_{n;1}$ reads (see Fig.~\ref{fig-OPE})
\begin{align}\label{OPE3}
d_{12}\big[ R(1;2j_1)R(2;1j_2 j_3) + R(1;2j_1 j_2)R(2;1j_3) + \text{cyclic($j_1 j_2 j_3$)} \big] \,,  
\end{align}
where $d_{12}=y_{12}^2/x_{12}^2$. This expression is invariant under cyclic shifts of the indices of the external legs $j_1,j_2$ and $j_3$.
It can be simplified using \re{3pt-new} and \re{R-2pt},
\begin{align}
\text{Eq.~\re{OPE3}} = d_{12} \left[ R(1;j_1j_2j_3) + R(2;j_1j_2j_3)\right] \stackrel{1\to 2}{\sim} 2 d_{12} R(1;j_1j_2j_3) \,,
\end{align}
where in the last relation we took into account that the difference $R(1;j_1j_2j_3) - R(2;j_1j_2j_3)$ vanishes in the limit $1\to 2$. 
Thus,  in the short-distance limit the product of two vertices of valency $2$ and $3$ reduces to a single
valency $3$ vertex leading to \re{lim}.

Finally, we have to examine the product of two vertices of total valency $6$ (see the second line in Fig.~\ref{fig-OPE}). Their contribution to the correlation
function is given by the expression
\begin{align}\label{OPE4}
d_{12}\big[ R(1;2j_1)R(2;1j_2 j_3 j_4) + R(1;2 j_1 j_2 j_3)R(2;1j_4)  + R(1;2j_1 j_2) R(2;1j_3 j_4)+\text{cyclic($j_1 j_2 j_3 j_4$)}\big],
\end{align}
which is symmetric under cyclic shifts of the external legs $j_1,\dots,j_4$. Using \re{22} it is straightforward to verify  that
each term in  the square brackets remains finite for $1\to 2$. Moreover, the resulting expression can be simplified
with the help of \re{4pt-cyc} (applied for $i=j_5=1$)
\begin{align}
\text{Eq.~\re{OPE4}} \stackrel{1\to 2}{\sim} 2 d_{12} R(1; j_1 j_2 j_3 j_4)\,,
\end{align}
in perfect agreement with \re{lim}.

The above relations can be extended to the product of vertices of an arbitrary total valency $k$. In this case,
\re{OPE3} and \re{OPE4} should be generalised to include the sum of products of vertices of valency $(p+1)$ and
$(k-p+1)$ with $p=1,\dots,k-1$. Then, in the short distance limit $1\to 2$, we can apply the identity \re{npt-cyc} for
$i=j_{k+1}=1$ to show that the sum collapses into  $2 d_{12} R(1;j_1\dots j_k)$,
leading to \re{lim}.
 
To conclude, in this section we have demonstrated that the expressions for the correlation function $G_n$
obtained within the twistor space approach satisfy two consistency conditions: they are independent of
the reference supertwistor and have the correct asymptotic behaviour in the light-like and short distance limits. 
In the following two sections, we shall compare these results with the analogous expressions for $G_n$ computed
using the conventional Feynman rules in Minkowski space and shall demonstrate
their perfect agreement.

\section{Correlation functions from Feynman diagrams}

In this section we outline the calculation of the correlation function $G_n$ in the conventional
Feynman diagram approach. More precisely, we shall concentrate on computing the next-to-lowest 
component $G_{n;1}$  in the Born approximation. As was explained above, $G_{n;1}$ has 
Grassmann degree $4$ and its perturbative expansion starts at order $O(g^2)$.

\subsection{Next-to-lowest component}

To evaluate $G_{n;1}$, we use the superfield expansion \re{T-dec} of the stress-tensor multiplet $\mathcal T$ in \re{Gn} and retain the contributions of Grassmann degree $4$. This yields a representation 
for $G_{n;1}$ as a collection of correlation functions involving various components of $\mathcal T$. Each
correlation function has conformal symmetry but not the $\mathcal N=4$ supersymmetry. The latter
is realised in the form of Ward identities that  these correlation functions satisfy.

The stress-tensor multiplet has the form \re{T-dec} with components given by the following gauge invariant composite operators \cite{Eden:2011yp}
\begin{align} \notag \label{t4c}
{}&O^{++++} = \tr(\phi^{++}\phi^{++}), 
\\[1.5mm] \notag 
{}&O^{+++,\alpha}_a =2\sqrt{2} i  \tr\big(
\psi^{+\alpha}_a \phi^{++}\big),
\\[1.5mm] \notag 
{}&O^{++,\alpha\beta}= \tr \left(
 \psi^{+c(\alpha}\psi_c^{+\beta)}-
i {\sqrt{2}}  F^{\alpha\beta}\phi^{++}\right),
\\ \notag 
{}&O^{++}_{ab}=-   \tr   \left(\psi^{+\gamma}_{(a}\psi^{+}_{b)\gamma}
- {g}{\sqrt{2}} [\phi_{(a}^{+C},\bar\phi_{+b,C)}]\phi^{++}\right),
\\ \notag 
{}&O^{+,\alpha}_a=-\frac{4}3   \tr\left(F_\beta^\alpha \psi_a^{+\beta}
+  ig [\phi_a^{+B}, \phi_{BC}]\psi^{C\alpha}\right),
\\
{}& \mathcal L  = \frac13\tr \left\{ -\frac12  F_{\alpha\beta}F^{\alpha\beta}  + {\sqrt{2}}  g \psi^{\alpha A} [\phi_{AB},\psi_\alpha^B] - \frac18 g^2 [\phi^{AB},\phi^{CD}][\phi_{AB},\phi_{CD}] \right\}\,,
\end{align}
where  the shorthand notations were introduced for the scalar and gaugino fields projected with $SU(4)$ harmonic variables
\begin{align}\notag
{}&  
\phi^{+\, B}_a =\ep_{ab}  u^{+b}_A\phi^{A B}\,,&&
\bar\phi_{+b,A} = \bar{u}^{B}_{+b}\phi_{AB} \,, && \phi^{++} = - \frac{1}{2} u^{+a}_A \ep_{ab} u^{+b}_B \phi^{AB} \,,
\\
{}&  \psi^{+\,\a}_a = \ep_{ab} u^{+b}_A \psi^{\a A}\,,&& \psi^{+\,a\a} = u^{+a}_A \psi^{\a A}  \,.
\end{align}
Here $\phi^{AB} = \frac12 \epsilon^{ABCD}\phi_{CD}$,
and we adopt the conventions for the raising-lowering of indices summarised in Appendix~\ref{app:conv}. We also use weighted symmetrisation $A_{(\alpha\beta)} = \frac12(A_{\alpha \beta} + A_{\beta \alpha})$. 

The correlation function $G_{n;1}$ depends on the analytic superspace Grassmann variables $\rho_i\equiv \theta_i^+$ with $i=1,\dots,n$. 
It can be expanded over eight different nilpotent polynomials in $\rho_i$ of degree $4$, covariant under Lorentz and $R-$symmetry
transformations,
\begin{align} \notag \label{14}
G_{n;1}{}&=\sum_i \rho^{4}_i f(i) +
\sum_{i \neq j} \rho_{i\a}^{a} (\rho^{3}_j)_\b^{b} f^{\a\b}_{ab}(i,j) + 
\sum_{i \neq j} (\rho^{2}_i)^{(\a \b)} (\rho^{2}_j)^{(\gamma \delta)} f_{(\a\b)(\gamma \delta)}(i,j)  
\\ \notag
{}&+ \sum_{i \neq j} (\rho^{2}_i)^{(\a \b)} (\rho^{2}_j)^{(cd)} f_{(\a\b)(cd)}(i,j) + 
\sum_{i \neq j} (\rho^{2}_i)^{(ab)} (\rho^{2}_j)^{(cd)} f_{(ab)(cd)}(i,j) 
\\ \notag
{}&+\sum_{i \neq j \neq k} \rho_i^{\a a} \rho_j^{\b b} (\rho^{2}_k)^{(\gamma\delta)} f_{\a\b(\gamma\delta),ab}(i,j,k)
+ \sum_{i \neq j \neq k} \rho_i^{\a a} \rho_j^{\b b} (\rho^{2}_k)^{(cd)} f_{\a\b,ab(cd)}(i,j,k) 
\\
{}&+\sum_{i \neq j \neq k \neq l} \rho_i^{\a a} \rho_j^{\b b} \rho_k^{\gamma c} \rho_l^{\delta d} f_{\a\b\gamma\delta,abcd}(i,j,k,l)\,,
\end{align}
where we introduced the notation for 
\begin{align}\label{3.33}
(\rho^3){}_{\a}^{\, a} = \rho^{b}_\a \rho^{\b}_b \rho^{a}_\b\,, \qquad \rho^4 = \rho^{b}_\a \rho^{\b}_b \rho^{c}_\b \rho^{\a}_c\,, 
\qquad   
(\rho^2){}_{(\a \b)} = \rho_\alpha^{a}\epsilon_{ab}\rho_\beta^{b}\,, \qquad 
(\rho^2){}^{(ab)} = \rho_\alpha^{a}\epsilon^{\alpha\beta}\rho_\beta^{b}\,.
\end{align}
The functions $f$, $f_{\a\b,ab}$, $f_{(\a\b)(\gamma \delta)}$, $ f_{(\a\b)(cd)}$, $f_{(ab)(cd)}$, 
$f_{\a\b(\gamma\delta),ab}$, $f_{\a\b,ab(cd)}$, $f_{\a\b\gamma\delta,abcd}$ are polynomials in the 
variables $y_i$ and are rational functions in the variables $x_i$. They correspond to the 
correlation functions of  the operators \re{t4c}, e.g.
\begin{align}\notag\label{f-ex}
 f(1) {}&= \vev{0| \cL(1) O^{++++}(2) \dots O^{++++}(n)|0} \,,
\\[2mm]
f^{\a\b}_{ab}(1,2) {}& = \vev{0| O^{+++,\alpha}_a(1)O^{+,\beta}_b(2) O^{++++}(3) \dots O^{++++}(n) |0}\,.
\end{align} 
In what follows we shall calculate the eight coefficient functions in \re{14} at order $O(g^2)$ by means of the standard  $\cN=4$ SYM Feynman rules. 

\subsection{$T-$block approach} 
 
We use the explicit component field form of the Lagrangian 
of $\cN=4$ SYM \footnote{The operator $\cL$ in \re{t4c} coincides (up to a normalisation factor) with the chiral form of the $\cN=4$ SYM {\it on-shell} Lagrangian.}
\begin{align}\notag
\mathcal{L}_{\mathcal{N}=4}= \tr \left\{ -\frac14  \left(F_{\alpha\beta}F^{\alpha\beta} + \bar{F}_{\dot\alpha\dot\beta}\bar{F}^{\dot\alpha\dot\beta}\right) +
\frac14 D_{\alpha\dot\alpha} \phi^{AB} D^{\dot\alpha\alpha} \phi_{AB} + \frac18 g^2 [\phi^{AB},\phi^{CD}][\phi_{AB},\phi_{CD}]   \right. \\ 
 + 2 i\bar{\psi}_{\dot\alpha A} D^{\dot\alpha\alpha} \psi^{A}_{\alpha} - {\sqrt{2}}  g \psi^{\alpha A} [\phi_{AB},\psi_\alpha^B] + 
{\sqrt{2}}  g \bar{\psi}_{\dot\alpha A} [\phi^{AB},\bar{\psi}^{\dot\alpha}_B]  \bigg\}\,, \label{fLa}
\end{align}
where all fields are in the adjoint representation of the  gauge group $SU(N)$ , e.g. $\phi_{AB} = \phi_{AB}^a T^a$,
$F_{\alpha\beta}=F_{\alpha\beta}^a T^a$, $\psi^{\alpha A}=\psi^{\alpha A a} T^a$, with the generators $T^a$
being $N\times N$ traceless matrices normalised as $\tr( T^a T^b)= \delta^{ab}$.

We do the calculation in coordinate space. The scalar and gaugino propagators have the form
\begin{align}\notag
{} & \langle\phi^{++}(x_1,u_1)\;\phi^{++}(x_2,u_2) \rangle = \frac{1}{(2\pi)^2} \frac{y_{12}^2}{x_{12}^2}\,,
\\
{}& \langle\psi^{A}_{\a}(x_1)\;\bar{\psi}^{B}_{\da}(x_2) \rangle = - \frac{1}{(2\pi)^2} \pa_{\a \da}\frac{1}{x_{12}^2} \delta^{AB} \,,
\end{align}
with the $SU(N)$ indices suppressed. It is convenient to introduce the normalisation factor 
\begin{align}
c_{n} = \frac{g^2 N(N^2-1)}{(2\pi)^{2n+2}}\,.
\end{align}
As we will see in a moment, it appears in the expression for the individual diagrams. The same normalisation factor enters \re{planar} for $p=1$.

To illustrate our approach, we first compute the coefficient function $f^{\a\b}_{ab}(1,2)$ for $n=4$ points. According to \re{f-ex},
it is given by the four-point correlation function involving two scalar operators $O^{+++}$ and the operators $O^{+++,\alpha}_a$ and $O^{+,\beta}_b$
defined in \re{t4c}. To lowest order in the coupling, $f^{\a\b}_{ab}(1,2)$ receives contribution from the following Feynman diagrams 
(and their permutations $3 \rightleftarrows 4$) 
$$\psfrag{1}[cc][cc]{$\scriptstyle 1$}\psfrag{2}[cc][cc]{$\scriptstyle 2$}\psfrag{3}[cc][cc]{$\scriptstyle 3$}\psfrag{4}[cc][cc]{$\scriptstyle 4$}
\psfrag{g}[cc][cc]{ }
\begin{array}{c}\includegraphics[width = 3 cm]{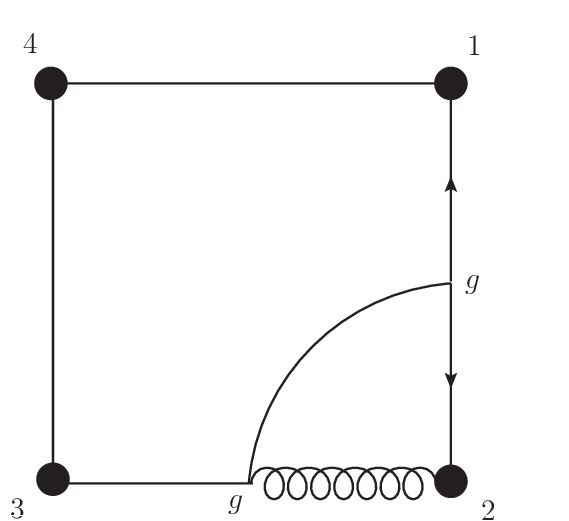} \\ (\Gamma_{4;1}) \end{array} 
\begin{array}{c}\includegraphics[width = 3 cm]{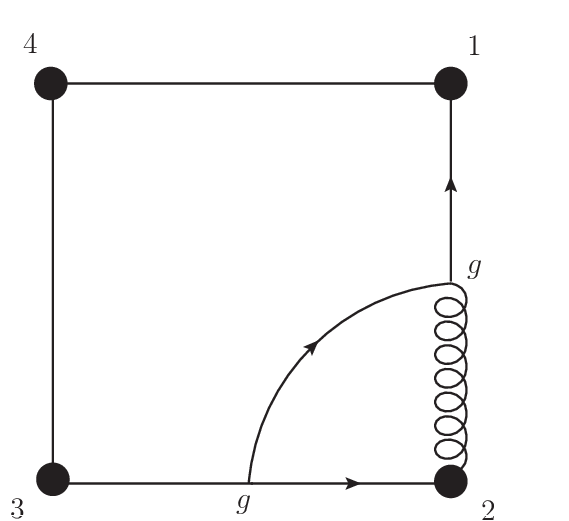} \\ (\Gamma_{4;2}) \end{array} 
\begin{array}{c} \vspace*{-2.5mm} \includegraphics[width = 3 cm]{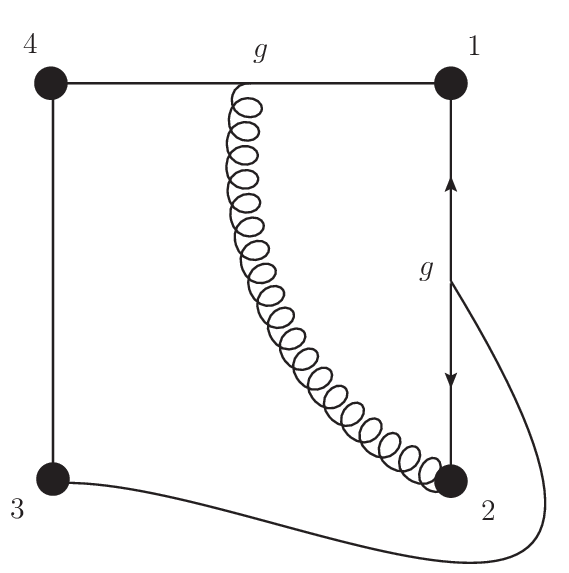} \\ (\Gamma_{4;3}) \end{array} 
\begin{array}{c} \vspace*{-2mm} \includegraphics[width = 3 cm]{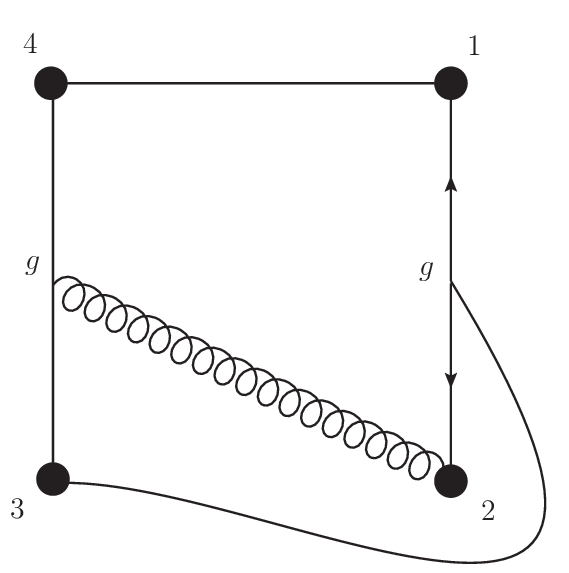} \\ (\Gamma_{4;4}) \end{array}
$$
$$\psfrag{1}[cc][cc]{$\scriptstyle 1$}\psfrag{2}[cc][cc]{$\scriptstyle 2$}\psfrag{3}[cc][cc]{$\scriptstyle 3$}\psfrag{4}[cc][cc]{$\scriptstyle 4$}
\psfrag{g}[cc][cc]{ }
\begin{array}{c}\includegraphics[width = 3 cm]{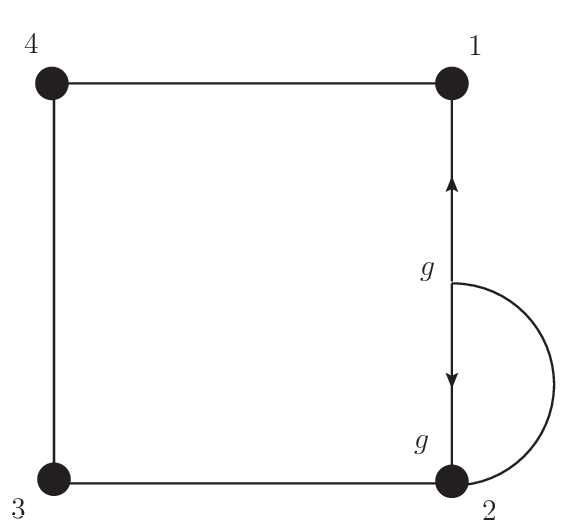} \\ (\Gamma_{4;5}) \end{array} 
\begin{array}{c} \vspace*{-2mm} \includegraphics[width = 3 cm]{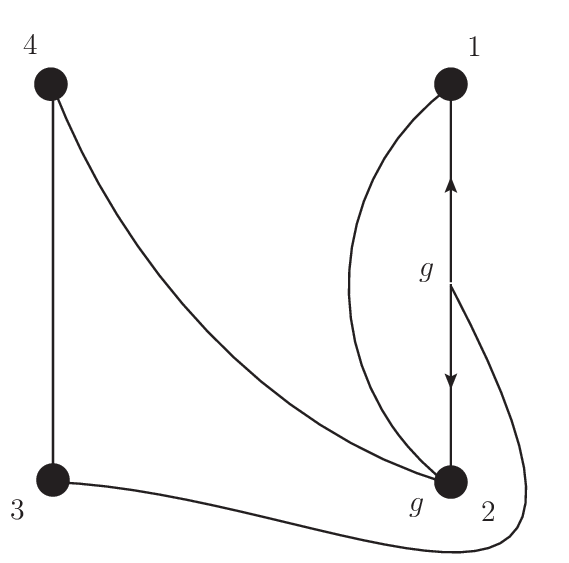} \\ (\Gamma_{4;6}) \end{array} 
\begin{array}{c}\vspace*{-2mm} \includegraphics[width = 3 cm]{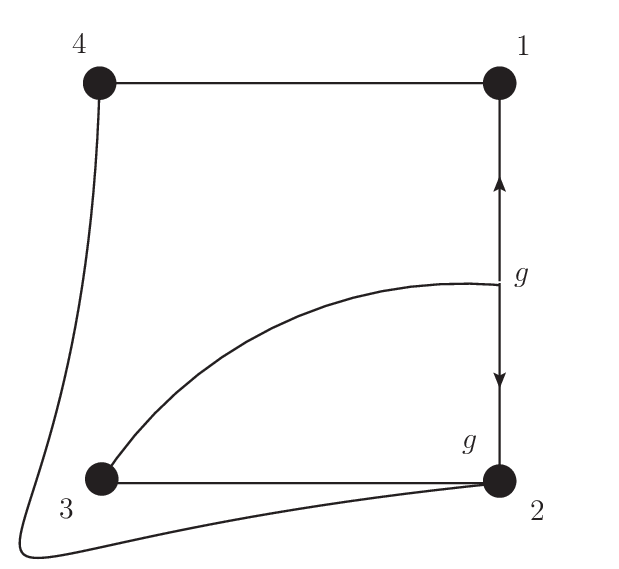} \\ (\Gamma_{4;7}) \end{array}
$$
Here the diagrams in the first and the second lines correspond to the two terms in the expression \re{t4c} for the operator $O^{+,\beta}_b$
at point $2$.
 
The above diagrams involve interaction vertices. We can significantly simplify the calculations of the corresponding
Feynman integrals by defining two simple building blocks which are called bosonic and fermionic  $T-$blocks.
The former represents the interaction of a gluon in the Feynman gauge with a pair of scalars,
\begin{align} \label{T1}\psfrag{1}[cc][cc]{$\scriptstyle 1$}\psfrag{2}[cc][cc]{$\scriptstyle 2$}\psfrag{3}[cc][cc]{$\scriptstyle 3$}\psfrag{4}[cc][cc]{$\scriptstyle 4$}
\psfrag{g}[cc][cc]{ }
\begin{array}{c}\includegraphics[width = 4 cm]{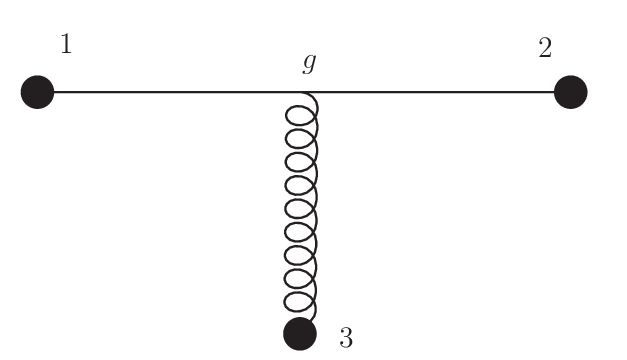}\end{array} = 
\langle \phi^{a,++}(1) \, F_{\alpha\beta}^{b}(3) \,\phi^{c,++}(2) \rangle 
= \frac{2g}{(2\pi)^4} f^{abc} y_{12}^2 \frac{ (x_{31} \widetilde{x}_{32})_{(\alpha\beta)} }{x_{12}^2 x^2_{13} x^2_{23}}\,,
\end{align}
and the latter stands for the Yukawa interaction of a scalar with a pair of chiral fermions,
\begin{align} \label{T2}\psfrag{1}[cc][cc]{$\scriptstyle 1$}\psfrag{2}[cc][cc]{$\scriptstyle 2$}\psfrag{3}[cc][cc]{$\scriptstyle 3$}\psfrag{4}[cc][cc]{$\scriptstyle 4$}
\psfrag{g}[cc][cc]{ }
\begin{array}{c}\includegraphics[width = 3.5 cm]{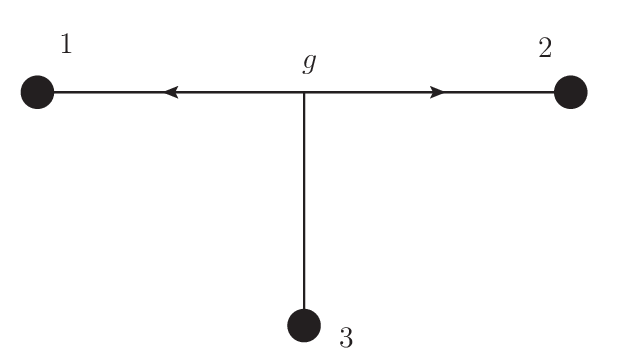}\end{array} = 
\langle \psi^{a,A}_{\alpha}(1) \, \phi^{b,++}(3) \,\psi^{c,B}_{\beta}(2) \rangle 
= -\frac{i \sqrt{2} g}{(2\pi)^4} f^{abc} (\bar{3}^{A}_{-a'} \epsilon^{a'b'} \bar{3}^{B}_{-b'}) \frac{ (x_{31} \widetilde{x}_{32})_{\alpha\beta} }{x_{12}^2 x^2_{13} x^2_{23}}\,.
\end{align}
Here $f^{abc}$ are the $SU(N)$ structure constants and we use the shorthand notation $\bar{3}^{A}_{-a'}\equiv \bar u_{3,-a'}^A$.

We then observe that  diagrams $(\Gamma_{4;3})$ and $(\Gamma_{4;4})$ involve a product of the two $T-$blocks supplemented by scalar propagators
$d_{ij}=y_{ij}^2/x_{ij}^2$, e.g.
\begin{align}
 (\Gamma_{4;4}) \sim  \langle \phi^{++}(3)  F^{\beta\gamma}(2)  \phi^{++}(4) \rangle \langle \psi^{A,\alpha}(1) \, \phi^{++}(3) \,\psi^{B}_{\gamma}(2) \rangle 
 u_{1,A}^{+a} u_{2,B}^{+b} \,d_{14}\,,
\end{align} 
where we suppressed the $SU(N)$ indices.
Going through the calculation of $ (\Gamma_{4;4})$ we find
\begin{align} \label{G44}
(\Gamma_{4;4}) =  -\frac43 c_4\, 
y_{14}^2 y_{34}^2 (y_{13} \widetilde{y}_{32})^{ab}
\frac{(x_{31} \widetilde{x}_{32} x_{24} \widetilde{x}_{23}  - x_{31} \widetilde{x}_{32} x_{23} \widetilde{x}_{24})^{\alpha\beta}}
{x_{12}^2 x_{13}^2 x_{14}^2 x_{23}^4 x_{24}^2 x_{34}^2}\,.
\end{align}
Note that this expression is gauge dependent and, as a consequence, it is not conformally covariant.
Conformal symmetry is restored in the sum of diagrams that is gauge invariant.

Similarly, diagrams $(\Gamma_{4;6})$ and $(\Gamma_{4;7})$ involve only a
single fermionic $T-$block \p{T2}, e.g.
\begin{align} \label{G47}
(\Gamma_{4;7}) = \frac43  c_4 \,
y_{14}^2 y_{34}^2 (y_{13} \widetilde{y}_{32})^{ab}
\frac{(x_{13} \widetilde{x}_{32} )^{\alpha\beta}}
{x_{12}^2 x_{13}^2 x_{14}^2 x_{23}^4 x_{24}^2}\,.
\end{align}
This expression is gauge invariant and, as a consequence, it is conformally covariant. It contains however
the factor of $1/x_{23}^4$ which should disappear in the sum of all Feynman diagrams in order to restore
the expected $1/x_{23}^2$ asymptotic behavior \re{lim} of the correlation function in the short-distance limit $2\to 3$.
 
The remaining diagrams  $(\Gamma_{4;1})$, $(\Gamma_{4;2})$ and $(\Gamma_{4;5})$ cannot be reduced
to   products of $T-$blocks. Moreover, they involve more complicated Feynman integrals that are potentially 
ultraviolet divergent and, in addition, produce a contribution that is not a rational function of $x_{ij}^2$. We recall
however that the correlation function in the Born approximation should be a rational function of $x_{ij}^2$. This suggests that 
the non-rational pieces from the above mentioned diagrams should disappear in the sum of all diagrams. 
Indeed, there exists an efficient way to organise the calculation so that we do not actually need
to compute these complicated integrals.  Instead of considering the `difficult' diagrams one by one, we shall combine 
them into sums that are explicitly rational.

To identify such rational sums, we return to \re{G-mod4} and notice that, in virtue of $\cN=4$ superconformal
symmetry, the  correlation function for $n=4$ only involves the lowest component $G_{n;0}$ given by
\re{Gn0}. This means that $G_{4;1}=0$, so that all coefficient functions in \re{14} vanish for $n=4$. 
In particular, $f^{\a\b}_{ab}(1,2)=0$ for $n=4$. In other words, the sum of all diagrams $\Gamma_{4;k}$ (with $k=1,\dots,7$), symmetrised with respect to the exchange of points $3\leftrightarrow 4$, should vanish.
Since the diagrams $(\Gamma_{4;k})$ have a harmonic structure $y_{13}^2 y_{34}^2 (y_{14} \widetilde{y}_{42})_{ab}$
that is not invariant under the exchange of points $3$ and $4$, this yields the condition 
\begin{align}\label{sum-zero}
\sum_{k=1}^7 (\Gamma_{4;k}) = 0\,.
\end{align}
This relation allows us to express the sum of `difficult' diagrams in terms of  `easy' diagrams $(\Gamma_{4;3})$, $(\Gamma_{4;4})$, $(\Gamma_{4;6})$, $(\Gamma_{4;7})$  that are reduced to  fermionic and bosonic $T-$blocks, Eqs.~(\ref{T1}) and (\ref{T2}). It is convenient to represent \re{sum-zero} in the following diagrammatic form
\begin{equation}\label{BB} \psfrag{1}[cc][cc]{$\scriptstyle 1$}\psfrag{2}[cc][cc]{$\scriptstyle 2$}\psfrag{3}[cc][cc]{$\scriptstyle 3$}\psfrag{4}[cc][cc]{$\scriptstyle 4$}
\begin{array}{c}\includegraphics[width = 2.2 cm]{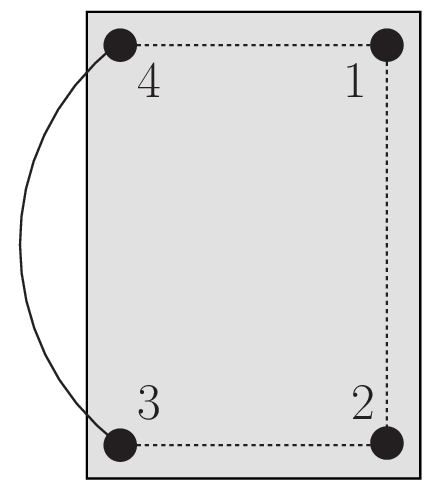}\end{array} = (\Gamma_{4;1}) + (\Gamma_{4;2}) + (\Gamma_{4;3}) + (\Gamma_{4;5}) + (\Gamma_{4;6}) 
= - (\Gamma_{4;4}) - (\Gamma_{4;7}) 
\end{equation}
where the graph on the left-hand side has a shaded block with a free propagator attached to points $3$ and $4$. 
This block stands for the sum of diagrams containing interaction vertices and we shall refer to it as a `black box'. 
It is expressed in terms of the easy diagrams $(\Gamma_{4;4})$ and $(\Gamma_{4;7})$ given by  \p{G44} and \p{G47}
and, therefore, it is a rational function.\footnote{If we were to reproduce \re{BB} without appealing to 
$G_{4;1}=0$, we would need to choose a particular regularisation and to calculate several non-trivial integrals which are not rational. Their sum is rational however.} The main reason for introducing the `black box' is that, as
we show in the next subsection, 
it naturally appears as a non-trivial core of higher-point diagrams.

\subsection{The $O(\rho_1 \rho_2^3)$ component for $5$ points}\label{sect:warm}
 
We are now ready to compute the coefficient function $f^{\a\b}_{ab}(1,2)$  for the $n=5$ correlation function.
We recall that it defines the  $\rho_1 \rho_2^3-$component in the expansion \re{14} of $G_{5;1}$. Unlike the  $n=4$ case examined above, $f^{\a\b}_{ab}(1,2)$ is different from zero for five points. 

Let us first identify the relevant Feynman diagrams. Compared to the $n=4$ case, these diagrams involve the 
additional vertex $5$ with two scalar propagators attached:
\begin{align*}
& \psfrag{1}[cc][cc]{$\scriptstyle 1$}\psfrag{2}[cc][cc]{$\scriptstyle 2$}\psfrag{3}[cc][cc]{$\scriptstyle 3$}\psfrag{4}[cc][cc]{$\scriptstyle 4$}\psfrag{5}[cc][cc]{$\scriptstyle 5$}\psfrag{g}[cc][cc]{ }
\includegraphics[width = 3 cm]{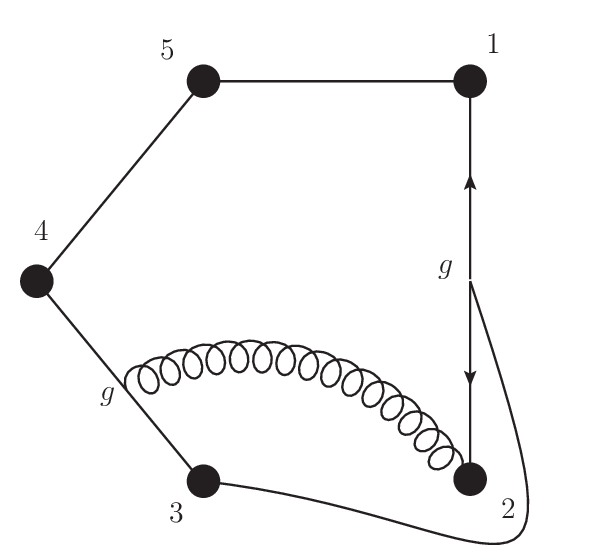} 
&& 
\psfrag{1}[cc][cc]{$\scriptstyle 1$}\psfrag{2}[cc][cc]{$\scriptstyle 2$}\psfrag{3}[cc][cc]{$\scriptstyle 3$}\psfrag{4}[cc][cc]{$\scriptstyle 4$}\psfrag{5}[cc][cc]{$\scriptstyle 5$}\psfrag{g}[cc][cc]{ }
\includegraphics[width = 3 cm]{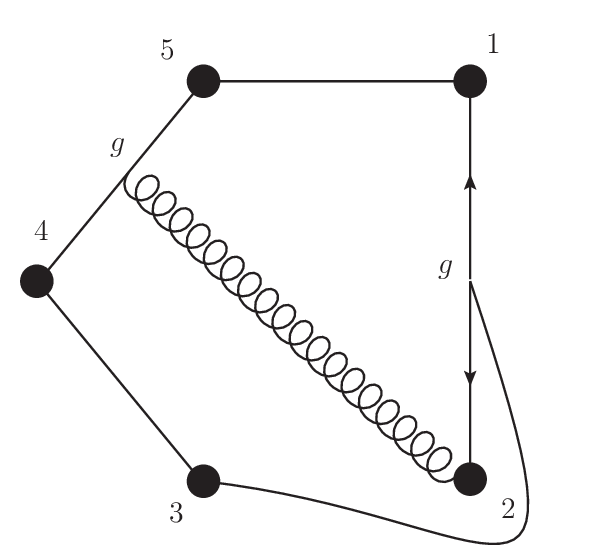} 
&&
\psfrag{1}[cc][cc]{$\scriptstyle 1$}\psfrag{2}[cc][cc]{$\scriptstyle 2$}\psfrag{3}[cc][cc]{$\scriptstyle 3$}\psfrag{4}[cc][cc]{$\scriptstyle 4$}\psfrag{5}[cc][cc]{$\scriptstyle 5$}\psfrag{g}[cc][cc]{ }
\includegraphics[width = 2.75 cm]{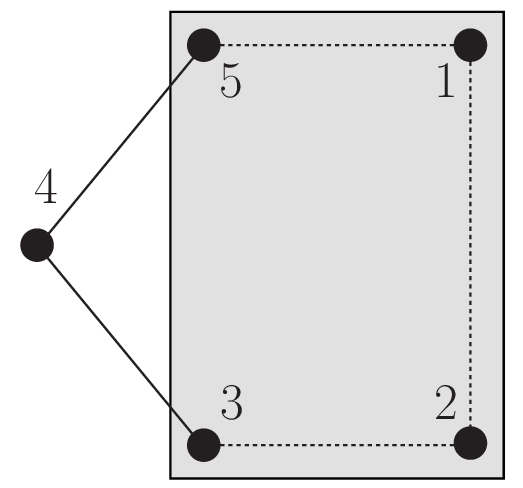} 
&&
\psfrag{1}[cc][cc]{$\scriptstyle 1$}\psfrag{2}[cc][cc]{$\scriptstyle 2$}\psfrag{3}[cc][cc]{$\scriptstyle 3$}\psfrag{4}[cc][cc]{$\scriptstyle 4$}\psfrag{5}[cc][cc]{$\scriptstyle 5$}\psfrag{g}[cc][cc]{ }
\includegraphics[width = 3 cm]{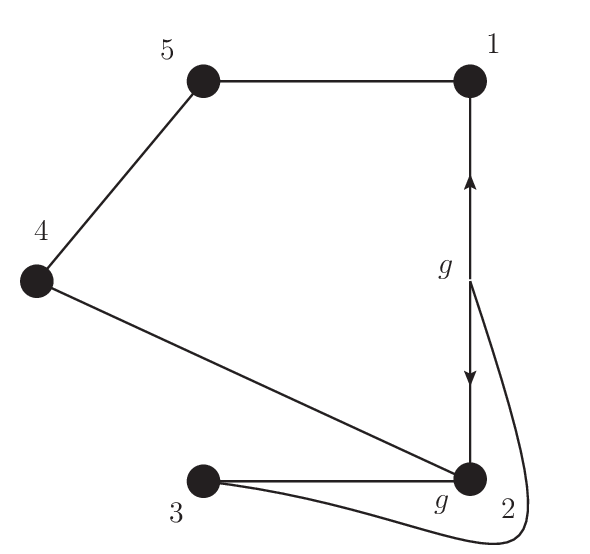} 
&&
\psfrag{1}[cc][cc]{$\scriptstyle 1$}\psfrag{2}[cc][cc]{$\scriptstyle 2$}\psfrag{3}[cc][cc]{$\scriptstyle 3$}\psfrag{4}[cc][cc]{$\scriptstyle 4$}\psfrag{5}[cc][cc]{$\scriptstyle 5$}\psfrag{g}[cc][cc]{ }
\includegraphics[width = 3 cm]{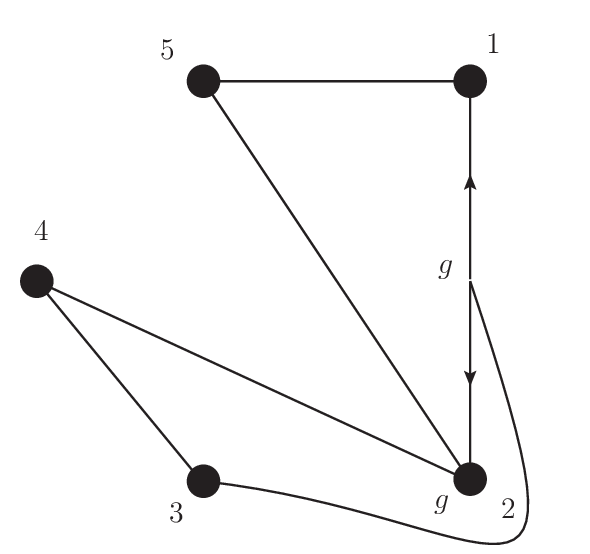}
\\
& \qqquad (\Gamma_{5;1}) &&
\qqquad (\Gamma_{5;2}) &&
\qqquad (\Gamma_{5;3}) &&
\qqquad (\Gamma_{5;4}) &&
\qqquad (\Gamma_{5;5})
\end{align*}
Here the shaded block has the same meaning as in \re{BB}. Namely, it denotes the sum of graphs $ (\Gamma_{4;1}) + (\Gamma_{4;2}) + (\Gamma_{4;3}) + (\Gamma_{4;5}) + (\Gamma_{4;6}) $ with the scalar line between points $3$
and $4$ removed. As a result, the contribution of the diagram $(\Gamma_{5;3})$ can be obtained from \re{BB} by replacing
 the scalar propagator $d_{34}$ with the product of two propagators $d_{34} d_{45}$ in the sum of two `easy' diagrams
 $- [(\Gamma_{4;4}) + (\Gamma_{4;7})]$:
\begin{align}\label{G53}
(\Gamma_{5;3}) = \frac43   c_5\, y_{15}^2 y_{34}^2 y_{45}^2 
(y_{13} \widetilde{y}_{32} )^{ab} 
\frac{(x_{31} \widetilde{x}_{32} x_{25} \widetilde{x}_{23} - x_{31} \widetilde{x}_{32} x_{23} \widetilde{x}_{25})^{\alpha\beta} 
- x_{35}^2 (x_{13} \widetilde{x}_{32})^{\alpha \beta}}
{x_{12}^2 x_{13}^2 x_{15}^2 x_{23}^4 x_{25}^2 x_{34}^2 x_{45}^2}\,.
\end{align}
 
The calculation of $(\Gamma_{5;1})$ and $(\Gamma_{5;2})$ is similar to that of $(\Gamma_{4;4})$. They are given
by  products of fermionic and bosonic $T-$blocks \p{T1} and \p{T2} resulting in
\begin{align}\notag\label{G51}
& (\Gamma_{5;1}) =  -\frac43   c_5\, y_{15}^2 y_{34}^2 y_{45}^2 (y_{13} \widetilde{y}_{32})^{ab} 
\frac{(x_{31} \widetilde{x}_{32} x_{24} \widetilde{x}_{23} - x_{31} \widetilde{x}_{32} x_{23} \widetilde{x}_{24})^{\alpha \beta}}
{x_{12}^2 x_{13}^2 x_{15}^2 x_{23}^4 x_{24}^2 x_{34}^2 x_{45}^2} \,,\notag \\
& (\Gamma_{5;2}) =  -\frac43  c_5\, y_{15}^2 y_{34}^2 y_{45}^2 (y_{13} \widetilde{y}_{32})^{ab}
\frac{(x_{31} \widetilde{x}_{32} x_{25} \widetilde{x}_{24} - x_{31} \widetilde{x}_{32} x_{24} \widetilde{x}_{25})^{\alpha \beta}}
{x_{12}^2 x_{13}^2 x_{15}^2 x_{23}^2 x_{24}^2 x_{25}^2 x_{34}^2 x_{45}^2}\,. 
\end{align}
We  note that $(\Gamma_{5;3})$ contains a double pole $1/(x_{23}^{2})^2$ which should disappear 
in the sum of all Feynman diagrams. In addition,  the expressions in \re{G53} and \re{G51} do not transform 
covariantly under the conformal transformations. In order to recover the conformal symmetry we have to
examine the sum of  all three diagrams. We find after some algebra 
\begin{align}\label{id5}\notag
\sum_{k=1,2,3} (\Gamma_{5;k}) {}& =   -\frac43 c_5\, y_{15}^2 y_{34}^2 y_{45}^2 
(y_{13} \widetilde{y}_{32} )^{ab}  
\\
{}& \times { x_{25}^2 x_{34}^2 (x_{13} \widetilde{x}_{32})^{\alpha \beta}
- x_{23}^2 (x_{13} \widetilde{x}_{35} x_{54} \widetilde{x}_{42} - x_{13} \widetilde{x}_{34} x_{45} \widetilde{x}_{52})^{\alpha \beta} \over x_{12}^2 x_{13}^2 x_{15}^2 x_{23}^4 x_{24}^2 x_{25}^2 x_{34}^2 x_{45}^2}\,.
\end{align}
This example  shows that in a order to obtain a conformal result we have to assemble together 
a gauge invariant set of diagrams with all possible attachments of the gluon propagators.
 
The two remaining diagrams $(\Gamma_{5;4})$ and $(\Gamma_{5;5})$ are conformally covariant.
 The diagram $(\Gamma_{5;4})$ can be obtained from $(\Gamma_{4;7})$ by replacing the scalar propagator 
 $d_{41} \to d_{45} d_{51}$ in \re{G47}. When combined together with \re{id5}, it cancels the first term 
 in the numerator in the second line of \p{id5}.  The resulting expression does not have a double pole $1/(x_{23}^2)^2$
but only a simple pole $1/x_{23}^2$. The diagram $(\Gamma_{5;5})$
is the $5-$point analogue of $(\Gamma_{4;6})$, however its harmonic structure is more complicated due to the higher number of
points,  
\begin{align}
(\Gamma_{5;5}) = \frac43 c_5\,  y_{15}^2 y_{34}^2 
(y_{13} \widetilde{y}_{34} y_{45} \widetilde{y}_{52} - y_{13} \widetilde{y}_{35} y_{54} \widetilde{y}_{42} )^{ab} 
\frac{(x_{13} \widetilde{x}_{32})^{\alpha \beta}}{x_{12}^2 x_{13}^2 x_{15}^2 x_{23}^2 x_{24}^2 x_{25}^2 x_{34}^2}\,.
\end{align}

Finally, to obtain  $f^{\a\b}_{ab}(1,2)$ we add together the contributions of all diagrams $(\Gamma_{5;k})$ (at $k = 1, 2 , \cdots, 5$) and symmetrise over all permutations of the points $3,4,5$
in order to restore the Bose symmetry of the correlation function. The result takes the  remarkably simple form
\begin{align}\label{f1}
f^{\a\b,ab}(1,2) =
\frac{8}{3}c_5 \frac{ x_{14}^2 x_{35}^2 y_{15}^2 y_{34}^2 }{\prod_{1\le i< j\le 5} x^2_{ij}}   \    \biggl[ 
 y_{45}^2  (y_{13} \widetilde{y}_{32})^{ab} 
(x_{13} \widetilde{x}_{35} x_{54} \widetilde{x}_{42})^{\alpha \beta} 
  - 
(x\leftrightarrow y)\biggr] + 
\text{perm}_{345}.
\end{align}
Notice that the product $f^{\a\b,ab}(1,2) \prod_{i<j}  x^2_{ij}$  is symmetric under the exchange  
of spatial and harmonic coordinates  $x_i \rightleftarrows y_i$ (see Appendix~\ref{5ptcomp} for explanation of this property).

Thus, we were able to compute the $O(\rho_1\rho_2^3)$ component of $G_{5;1}$ by using only the  
$T-$blocks \p{T1} and \p{T2} combined with the `black box' relation \re{BB}. We can apply the same approach
to computing the remaining components of the $5-$point correlation function $G_{5;1}$. Their explicit expressions 
can be found in Appendix~\ref{5ptcomp}.

\subsection{Consistency checks}

In this subsection, we compare the obtained result for $G_{5;1}$ with the analogous expression found in \cite{Eden:2011we}.
As was shown in that paper, the $\cN=4$ superconformal symmetry allows us to predict the form of the $5-$point 
correlation function up to an overall normalisation factor
\begin{align}\label{G-I}
G_{5;1} =  c \,\frac{\mathcal{I}_{5;1}(x,\rho,y)}{\prod_{1\le i<j\le 5} x_{ij}^2}\,,
\end{align}
where the dependence on the Grassmann and harmonic variables resides in the function $\mathcal{I}_{5;1}$. It is 
a polynomial in $\rho$ of Grassmann degree $4$, invariant under $Q$ and $\bar{S}$ superconformal transformations.
Its explicit form has been found in \cite{Eden:2011we}
\begin{align} \label{inv5}\notag
\mathcal I_{5;1} {}& = Q^8 \bar S^8 \prod_{i=1}^5 \delta^4(\rho_i)   
\\ \notag
{}&= \int d^4 \epsilon \, d^4 \epsilon'   d^4 \bar \xi \,  d^4 \bar \xi'
\prod_{i=1}^5 \delta^{(4)}\big(\rho_i -(\epsilon +  y_i \epsilon' )-x_i ( \bar \xi+  y_i\bar \xi')  \big)
 \\
{} &=x_{23}^2 x_{24}^2 x_{25}^2x_{34}^2 x_{35}^2 x_{45}^2\times R(2345) \times \bigg( \rho_1 + \sum_{i=2}^5 R_{1i} \, \rho_i \bigg)^4\,,
\end{align}
where $ \delta^4(\rho_i) \equiv \rho_i^4$.  
Here $(R_{1i} \, \rho_i)^{\alpha a}= R_{1i}^{\alpha\beta,a b} \, (\rho_i)_{\beta b}$ involves the matrix $R_{1i}^{\alpha\beta,a b}$ (see Eq.~\re{A-f} below) and the function $R(2345)$ is polynomial in $y_{ij}^2$ and rational in $x_{ij}^2$, 
\begin{align}\notag
R(2345) =       \frac{x_{12}^2 x_{13}^2 x_{14}^2 x_{15}^2}{\prod_{1\le i < j\le 5} x^2_{ij}}   
\biggl[{}& 
(y^2_{23} y_{45}^2 x_{25}^2 x_{34}^2 - x^2_{23} x_{45}^2 y_{25}^2 y_{34}^2)
(y^2_{23} y_{45}^2 x_{24}^2 x_{35}^2 - x^2_{23} x_{45}^2 y_{24}^2 y_{35}^2) 
\\ \notag
+{}& (y^2_{24} y_{35}^2 x_{25}^2 x_{34}^2 - x^2_{24} x_{35}^2 y_{25}^2 y_{34}^2)
(y^2_{24} y_{35}^2 x_{23}^2 x_{45}^2 - x^2_{24} x_{35}^2 y_{23}^2 y_{45}^2)
\\
 +{}&
(y^2_{25} y_{34}^2 x_{23}^2 x_{45}^2 - x^2_{25} x_{34}^2 y_{23}^2 y_{45}^2)
(y^2_{25} y_{34}^2 x_{24}^2 x_{35}^2 - x^2_{25} x_{34}^2 y_{24}^2 y_{35}^2) 
\biggr].
\end{align}
Expanding \re{G-I} in powers of the Grassmann variables and matching the result with \re{14} we can express
 the $f-$coefficient functions in terms of $R(2345)$ and $R_{1i}-$matrices. 
 
 In this way, we examine the $O(\rho_1^4)$ component and obtain
\begin{align}
f(1) = c\,   \frac{R(2,3,4,5)}{x_{12}^2 x_{13}^2 x_{14}^2 x_{15}^2}  \,.
\end{align}
Comparing this relation with \re{f(1)}, we observe perfect agreement and fix the normalisation constant,
$c=2c_5/3$. In a similar manner, for the $O(\rho_2 \rho_1^3)$ component we find
\begin{align}\label{A-f}
f^{\alpha\beta,ab}(2,1) = - 4   R_{12}^{\alpha\beta,a b}  f(1)  \,.
\end{align}
Together with \re{f1} this relation leads to a definite prediction for the matrix $R_{12}$ that we could match
against the integral representation for the same matrix, Eq.~\re{inv5}. Going through the calculation we find agreement. 

The same analysis can be repeated  for the other components of $G_{5;1}$. We verified that for $n=5$
the relation \re{14} with the coefficient functions given in Appendix~\ref{5ptcomp} coincides with \re{G-I}.

\section{Matching the two approaches}

\label{sec:example-component-}

In the preceding section we employed the conventional Feynman diagram technique to compute the 
five-point correlation function $G_{5;1}$. In this section we show that the relation \re{Gn1} obtained 
in the twistor approach correctly reproduces this result. To save space, here we consider the matching of one component only, $(\rho_1^2)^{(ab)} (\rho_3 ^2)^{(cd)}$ in \re{14}, and leave the more detailed discussion
for a future publication.
 
\subsection{Four points}

As a simpler illustration, let us first consider the component $(\rho_1^2)^{ab} (\rho_3 ^2)^{cd}$ in the 
four-point correlation function $G_{4;1}$. As was already mentioned, it should vanish in virtue of $\cN=4$
superconformal symmetry. At the same time, the twistor approach leads to the expression \re{G4-exp}
that involves the product of $3-$point $R-$vertices. In this subsection we  demonstrate  that the 
$(\rho_1^2)^{(ab)} (\rho_3 ^2)^{(cd)}$ contribution to \re{G4-exp} does indeed vanish.

At four points there is only one topology of twistor graphs that contributes to $G_{4;1}$. It is given by:
\begin{align} \label{I1234}\psfrag{1}[cc][cc]{$\scriptstyle 1$} \psfrag{2}[cc][cc]{$\scriptstyle  2$}  
\psfrag{3}[cc][cc]{$\scriptstyle 3$} \psfrag{4}[cc][cc]{$\scriptstyle  4$} 
I_{1234}\  =\   \vcenter{\hbox{\includegraphics[width = 0.15 \textwidth]{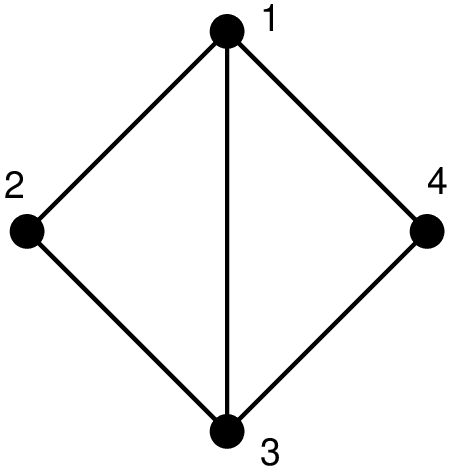}}}  \ =\ 
   d_{12}d_{23}d_{34}d_{41}d_{13}R(1;234)R(3;412)
\end{align}
and is obviously symmetric under the exchange of points $1\leftrightarrow 3$ and $2\leftrightarrow 4$.
The correlation function is given by the sum over the non-trivial permutations of this graph,
\begin{align}\label{G-sum-I}
G_{4;1} \sim I_{1234} + I_{1243}  + I_{2134}+ I_{2143} + I_{1324} + I_{3142}  \,.
\end{align}
To extract the contribution $(\rho_1^2)^{(ab)} (\rho_3 ^2)^{(cd)}$, we have to replace the $R-$invariants
in \re{I1234} by their expansion (see \re{eq:12} in Appendix~\ref{App:R}) and truncate the resulting expression 
to the component we are looking for. In this way, we find after some algebra
\begin{align}  \notag
\label{eq:19}
&  I_{1234}=\Big[ \frac{d_{34}d_{14}(y_{123})_{ab}(y_{123})_{cd}}{x_{12}^{2}x_{23}^{2}x_{13}^{2}y_{13}^{2}}
-\frac{d_{23}d_{14}(y_{123})_{ab}(y_{341})_{cd}}{x_{12}^{2}x_{13}^{2}x_{34}^{2}y_{13}^{2}}
\\\notag
& \hspace*{20mm}+\frac{(y_{12341})_{ac}(y_{34123})_{bd}}{2x_{12}^{2}x_{14}^{2}x_{34}^{2}x_{13}^{2}x_{23}^{2}y_{13}^{2}}
  +(2\leftrightarrow
  4)
  \Big](\rho_{1}^{2})^{ac}(\rho_{3}^{2})^{bd}+ \ldots\,,
\\ \notag
&   
I_{1243}=-\frac{(124)}{(123)}\frac{d_{24}(y_{123})_{ab}(y_{341})_{cd} }{x_{12}^{2}x_{34}^{2}x_{13}^{2}x_{14}^{2}}
(\rho_{1}^{2})^{ad}(\rho_{3}^{2})^{bc} + \ldots\,,
\\
&
I_{2143}=\frac{(124)(324)}{(413)(231)}\frac{d_{24}(y_{321})_{ab}(y_{143})_{cd} }{x_{14}^{2}x_{12}^{2}x_{43}^{2}x_{32}^{2}}
(\rho_{1}^{2})^{bc}(\rho_{3}^{2})^{ad}+ \ldots\,,
\end{align}
where the dots denote the remaining terms and we used the shorthand notations for
\begin{align}
   y_{ijk}&= y_{ij} \tilde y_{jk}\,,\qquad  y_{ijklm}=y_{ij} \tilde y_{jk} y_{kl} \tilde y_{lm}\,,\qquad
(ijk) = \vev{\sigma_{ij} \sigma_{ik}} x_{ij}^2 x_{ik}^2 \ .
\label{eq:171}
\end{align}
The expressions for the remaining terms on the right-hand side of \re{G-sum-I} can be obtained from \re{eq:19}
through permutation of the indices, e.g.  $I_{2134}=I_{1243}[1\leftrightarrow 3,2\leftrightarrow 4]$, $I_{1324}=I_{1243}[{2\leftrightarrow 4}]$ and $I_{3142}=I_{1243}[{1\leftrightarrow 3}]$. 

Note that the contribution to \re{G-sum-I} from $I_{1234}$ is independent of the reference twistor. It is straightforward
to verify that the same is true for the sum of the remaining five terms on the right-hand side of \re{G-sum-I}. Finally, substituting \re{eq:19} into \re{G-sum-I} we find after some algebra
\begin{align}\label{eq:4pntsrhorho}\notag 
& 
G_{4;1}\sim 
\frac{1}{x_{12}^{2}x_{23}^{2}x_{13}^{2}x_{34}^{2}x_{14}^{2}y_{13}^{2}}\Big[y_{34}^{2}y_{14}^{2}(y_{123})_{ab}(y_{321})_{cd}-y_{23}^{2}y_{12}^{2}(y_{143})_{ab}(y_{134})_{cd}-y_{23}^{2}y_{41}^{2}(y_{123})_{ab}(y_{341})_{cd}
\\  
& 
-y_{43}^{2}y_{21}^{2}(y_{143})_{ab}(y_{321})_{cd}-y_{24}^{2}y_{13}^{2}(y_{123})_{ab}(y_{341})_{cd}+(y_{12341})_{ad}(y_{34123})_{bc}\Big](\rho_{1}^{2})^{ad}(\rho_{3}^{2})^{bc} +\ldots
\end{align}
The expression inside the square brackets vanishes via a non-trivial $y-$identity.  
The easiest way to see this is to use the $SU(4)$ covariance of \re{eq:4pntsrhorho} in order
 to fix the $y-$variables at the four points as:
\begin{align}\label{eq:yfixing}
y_{1}\rightarrow \begin{pmatrix} 1&0\\ 0&1\\ \end{pmatrix}\,,\qquad y_{2}\rightarrow \infty\,,\qquad  y_{3}\rightarrow 0\,,\qquad y_{4}\rightarrow \begin{pmatrix}y&0\\ 0&\bar{y}\\ \end{pmatrix}\,.
\end{align} 
Implementing this choice sets (\ref{eq:4pntsrhorho}) to zero. Hence,  the $(\rho_{i}^{2})^{ab}(\rho_{j}^{2})^{cd}$ component of $G_{4;1}$ vanishes
\begin{align}
G_{4;1}\sim 0\times (\rho_{1}^{2})^{ad}(\rho_{3}^{2})^{bc} +\ldots
\end{align}
as it should be.
 
\subsection{Five points}
  
At five points, the correlation function $G_{5;1}$ receives contributions from twistor graphs of three different 
topologies:
\begin{align*}\psfrag{1}[cc][cc]{$\scriptstyle 1$} \psfrag{2}[cc][cc]{$\scriptstyle  2$}  
\psfrag{3}[cc][cc]{$\scriptstyle 3$} \psfrag{4}[cc][cc]{$\scriptstyle  4$}  \psfrag{5}[cc][cc]{$\scriptstyle  5$}  
\psfrag{A}[cc][cc]{$A_{12345}$} \psfrag{B}[cc][cc]{$B_{12345}$} \psfrag{C}[cc][cc]{$C_{12345}$} 
    \includegraphics[width = 0.65 \textwidth]{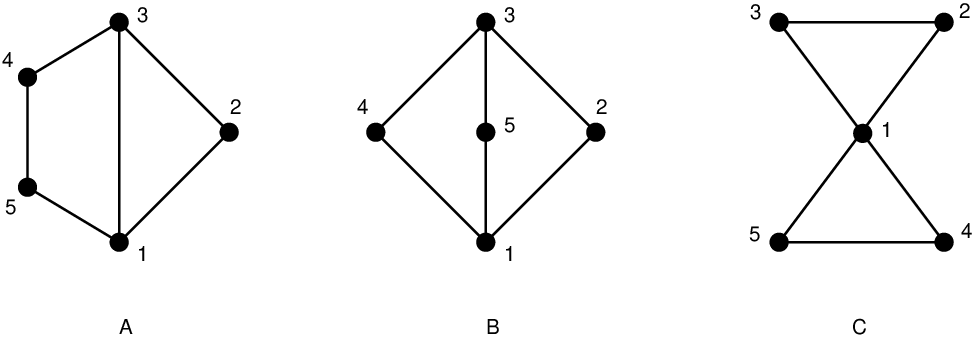}  
\end{align*}
Applying the Feynman rules shown in Fig.~\ref{fig-rules3} we find
\begin{align}\notag\label{ABC}
& A_{12345} = d_{12} d_{23} d_{13} d_{15} d_{45} d_{34} R(1;235) R(3;412) \,,
\\ \notag
& B_{12345} = d_{14} d_{34} d_{15} d_{35} d_{12} d_{23} R(1;452) R(3;254) \,,
\\
& C_{12345} = d_{12} d_{13} d_{14} d_{15} d_{23} d_{45} R(1;345) R(1;234)\,.
\end{align}
$G_{5;1}$ is given by their total sum symmetrised with respect to  the permutations of the five
points. 

Let us examine the contribution of each topology to the  component $(\rho_1^2)^{ab} (\rho_3 ^2)^{cd}$.
Replacing the $R-$invariants in \re{ABC} by their expansion in powers of the Grassmann variables (see Eqs.~\re{eq:18} and
\re{eq:12})
we find that this component does not receive contributions from graphs of type  $C$ for all possible
relabelings of the points. The total set of contributing graphs is  
\begin{align}\label{eq:graphss} 
G_{5;1}\sim  A_{12345}+ \frac12\left(A_{51342}+
 A_{53142}+ A_{41352}    +
 A_{43152}+ B_{53412}\right)+\frac16\,B_{12345}  +\text{perm}_{245}\ .
\end{align}
Here each inequivalent graph appears with coefficient 1, and the numerical factors are introduced to account 
for over-counting in the sum over permutations. We split the computation up in this way, since, as we will see
in  a moment, the linear combination in the parentheses on the right-hand side of \re{eq:graphss} is independent of the 
reference twistor. 

Going through calculations  similar to those performed in the
four-point case, we obtain the following expressions for the  component $(\rho_1^2)^{ab} (\rho_3 ^2)^{cd}$  
\begin{align}&\notag \label{AB-comp}
  A_{12345}=-\frac{y_{45}^{2}(y_{15243})_{ab}(y_{123})_{dc}}{x_{12}^{2}x_{23}^{2}x_{13}^{2}x_{34}^{2}x_{45}^{2}x_{15}^{2}} (\rho_{1}^{2})^{ad}(\rho^{2}_{3})^{bc} +\ldots,
\notag 
 \\[10pt]
&A_{51342}=\frac{(345)}{(341)}\frac{ y_{24}^2y_{25}^2(y_{153})_{ab}(y_{341})_{cd} }{x_{15}^{2}x_{35}^{2}x_{34}^{2}x_{13}^{2}x_{24}^2x_{25}^2}
(\rho_{1}^{2})^{ad}(\rho_{3}^{2})^{bc} + \ldots,
\notag\\[10pt]
&\notag
B_{12345}=\frac{(y_{12541})_{ab}(y_{34523})_{cd} }{x_{12}^{2}x_{23}^{2}x_{34}^{2}x_{41}^{2}x_{15}^{2}x_{35}^{2}}
(\rho_{1}^{2})^{ab}(\rho_{3}^{2})^{cd}+ \ldots,
\\[10pt] &
B_{53412}=\frac{(345)(145)}{(431)(513)}\frac{y_{24}^2y_{25}^2  (y_{341})_{ab}(y_{153})_{cd} }{x_{15}^{2}x_{14}^{2}x_{35}^{2}x_{34}^{2}x_{24}^2x_{25}^2}(\rho_{1}^{2})^{bc}(\rho_{3}^{2})^{ad}+ \ldots
\end{align}
The remaining graphs can be obtained by permuting the indices in these expressions. 

Notice that the expressions for $A_{12345}$ and $B_{12345}$ do not depend on the reference twistor and have the correct
conformal and $SU(4)$ properties. Then, we examine the sum of graphs in the parentheses in \re{eq:graphss}
\begin{align}\label{eq:secondterm}
 \notag A_{51342}{}& +
 A_{53142}+ A_{41352}    +
 A_{43152}+ B_{53412}=
\frac{y_{25}^{2}y_{24}^{2}}{\prod_{1\leq i<j\leq
    5}x_{ij}^{2}}\frac{x_{12}^{2}x_{23}^{2}x_{45}^{2}}{(431)(513)} (y_{341})_{ab}(y_{153})_{cd}(\rho_{1}^{2})^{bc}(\rho_{3}^{2})^{ad}
 \\&\notag
\times \Big[(345)(145)x_{13}^{2}+(451)(351)x_{34}^{2}+(134)(534)x_{15}^{2}+(345)(531)x_{14}^{2}+(451)(143)x_{35}^{2}\Big] +\dots
\notag\\
&= -\frac{y_{25}^{2}y_{24}^{2}}{\prod_{1\leq i<j\leq 5}x_{ij}^{2}}x_{12}^{2}x_{23}^{2}x_{45}^{4}(y_{341})_{ab}(y_{351})_{dc}(\rho_{1}^{2})^{bc}(\rho_{3}^{2})^{ad}\,,
\end{align}
where in the second relation we  made use of the six-term identity \re{eq:id}. We observe that the dependence on
the reference twistor disappears in the sum of graphs.  

Finally, we substitute \re{AB-comp} and \re{eq:secondterm} into \re{eq:graphss} and obtain 
the following expression for the  component $(\rho_{1}^{2})^{ac}(\rho_{3}^{2})^{bd}$ of the correlation function
\begin{align}\notag
G_{5;1}  =\frac{1}{\prod_{1\leq i<j\leq
    5}x_{ij}^{2}}\bigg[ {}& -\frac12 x_{12}^{2}x_{23}^{2}x_{45}^{4}y_{25}^{2}y_{24}^{2}(y_{143})_{ab}(y_{153})_{cd}-x_{14}^{2}x_{24}^{2}x_{25}^{2}x_{35}^{2}y_{45}^{2}(y_{15243})_{ab}(y_{123})_{cd}
\\
&
+\frac16 x_{13}^{2}x_{24}^{2}x_{25}^{2}x_{45}^{2}(y_{12541})_{ac}(y_{34523})_{bd}+\text{perm}_{245}\bigg](\rho_{1}^{2})^{ac}(\rho_{3}^{2})^{bd} + \ldots
\end{align}
We compare this expression with the analogous result \re{su2su25} obtained in the standard Feynman diagram approach 
and find perfect agreement (after appropriate permutations of indices). \footnote{
Note that the harmonic $y-$structure that comes out of the Feynman graph approach for this component is graphically identical to the twistor graph.}

To summarise, we demonstrated by an explicit calculation of a particular component of $G_{5;1}$ that  
the expression \re{Gn1} for the correlation function  in the twistor approach matches that 
obtained in the conventional Feynman  diagram approach.

\section{Conclusions}

We have developed a new approach to computing the correlation function $G_n$ of the chiral part of the stress-tensor supermultiplet in the Born
approximation. It relies on the reformulation of $\cN=4$ SYM in twistor space and gives $G_n$ as a sum of effective diagrams on twistor
space which only involve propagators and no integration vertices. We have used this unusual feature of the twistor diagrams to decompose them into simple building blocks, the $\cN=4$ superconformal invariants $R(i;j_1j_2j_3)$. However, the price to pay for the relative simplicity of the twistor diagrams is the
dependence of these invariants on the reference supertwistor $\mathcal Z_*$  defining the axial gauge condition. This dependence 
cancels in the sum of all twistor diagrams, due to the gauge invariance of $G_n$ but it is present in the contribution of each individual diagram. The situation here is similar to that of the tree-level scattering superamplitudes in planar $\mathcal N=4$ SYM. 

The relation to the scattering amplitudes can be made more precise by examining the asymptotic behaviour of $G_n$ in the light-like
limit. As we have shown, in the simplest case of the NMHV amplitude and the next-to-lowest component $G_{n;1}$, the on-shell NMHV 
invariants are given by the product of two off-shell $R-$invariants evaluated in the light-like limit. The on-shell invariants are known to
possess a larger, dual superconformal symmetry \cite{Drummond:2008vq} which is promoted to a Yangian symmetry \cite{Drummond:2009fd} when combined with the
conventional $\cN=4$ superconformal symmetry. As a consequence, the off-shell invariants also have this extended symmetry, in the
light-like limit at least. Whether this symmetry survives away from the light-like limit is a very interesting question which requires further investigation. 

Knowing $G_n$ in the Born approximation allows us to predict the quantum corrections to the same correlation function using the Lagrangian
insertion method. Namely, integrating the correlation function $G_{n+1}$ over the position of one of the operators, 
$\int d^4 x_{n+1}\, d^4 \theta_{n+1}^+\, G_{n+1}$, produces the order $O(g^2)$ correction to the correlation function $G_{n}$. Continuing this 
procedure, we can interpret  $G_{n+\ell}$ in the Born approximation as the $O(g^{2\ell})$ integrand for the quantum corrections to the correlation function $G_{n}$.
For $n=4$ this procedure, combined with the uniqueness of the top superconformal invariant $\mathcal I_{\ell+4,\ell}$, has been used in \cite{Eden:2011we} to reveal a new permutation symmetry of the four-point correlation function. Starting from $n=5$, the quantum corrections to $G_n$ receive
contributions from several superconformal invariants $\mathcal I_{\ell+n,p}$ (with $p=\ell,\dots,\ell+n-4$) whose explicit form can be found
using the approach presented in this paper. It remains to be seen what these invariants can tell us about the properties of the corresponding integrands.   It would be interesting to establish the relationship with the Grassmannian approach   to the integrand of the amplitude \cite{ArkaniHamed:2010gh} and with the recent `amplituhedron' construction \cite{Arkani-Hamed:2013jha}. 

When computing the correlation function $G_n$, we restricted our analysis to the chiral sector. By putting the antichiral Grassmann variables $\bq$ to zero we explicitly broke half of the supersymmetry. We could ask what happens if we include the dependence of $G_n$ on $\bq$,  thus recovering the full $\cN=4$ superconformal symmetry. In the simplest case  $n=4$ the dependence on $\bq$ can be
restored unambiguously \cite{Belitsky:2014zha}, whereas for $n\ge 5$ the $\cN=4$ superconformal symmetry is not powerful enough to lift the correlation function
from the chiral sector to the full superspace. It would be interesting to extend the twistor space approach to this case.

\section*{Acknowledgements}

B.E. is supported by DFG (``Eigene Stelle" Ed 78/4-2). D.C. is supported by the ``Investissements d'avenir, Labex ENIGMASS'' and partially supported by the RFBR grant 14-01-00341.  R.D. acknowledges support from an STFC studentship,  P.H. from the STFC Consolidated Grant ST/L000407/1. R.D., P.H., B.E. and G.K. also acknowledge support from the Marie Curie network GATIS (gatis.desy.eu) of the European Union's Seventh Framework Programme FP7/2007-2013/ under REA Grant Agreement No 317089. G.K. and E.S. acknowledge partial support by the French National Agency for Research under contract BLANCSIMI-4-2011.

\appendix 

\section{Conventions}\label{app:conv}

We introduce harmonic variables in order to covariantly decompose all quantities carrying indices in the
fundamental representation of $SU(4)$. These variables appear as components of the unitary matrix
\begin{align}
u_A^B \equiv (u_A^{+b}\,, u_A^{-b'})\,,
\end{align}
where the index $A$ transforms under global $SU(4)$ while the other index $B$ splits into two halves $B=(b,b')$ according to the local subgroup $SU(2)\times SU(2)'\times U(1)\in
SU(4)$ with indices $b,b'=1,2$ in the fundamental representation of $SU(2)$ and $SU(2)'$, respectively,
and the signs $+b$ and $-b'$ referring to the $U(1)$ charge. The unitarity conditions for the matrix $u$ and its
conjugate $\bar u$ are
\begin{align}
\bar u_{+a}^A u_A^{+b} = \delta_a^b\,,\qqqquad \bar u_{-a'}^A u_A^{-b'} = \delta_{a'}^{b'}\,,\qqqquad 
\bar u_{-a'}^A u_A^{+b}   =\bar u_{+a}^A u_A^{-b'}  =0\,.
\end{align}
They satisfy the completeness relation
\begin{align}
 u_A^{+a}  \bar u_{+a}^B  + u_A^{-a'} \bar u_{-a'}^B = \delta_A^B\,,
\end{align}
which allows us to decompose $\theta^A$ as
\begin{align}
\theta^A = \theta^{+a} \bar u_{+a}^B + \theta^{-a'} \bar u_{-a'}^A\,,\qqqquad \theta^{+a} = \theta^A u_{A}^{+a}
\,,\qqqquad \theta^{-a'} = \theta^A u_{A}^{-a'}\,.
\end{align}
It is convenient to use a particular parametrisation of the harmonic variables
\begin{align}\label{u-y}
u_B^{+a} = (\delta_b^a,y_{b'}^a)\,,\qquad u_B^{-a'} = (0,\delta_{b'}^{a'}) \,,\qquad 
\bar u_{+a}^B = (\delta_a^b,0)\,,\qquad \bar u_{-a'}^B = (-y_{a'}^b, \delta_{a'}^{b'})\,,
\end{align}
which amounts to choosing a gauge for the local subgroup $SU(2)\times SU(2)'\times U(1)$. 
In this parameterisation, the $SU(4)$ transformations can be reduced to combining a shift of $y$ with
the discrete operation of inversion
\begin{align}\label{y-tran}
y_{a'}^b \to y_{a'}^b + \epsilon_{a'}^b\,,\qqqquad y_{a'}^b \to y^{a'}_b/y^2\,,
\end{align}
with $ y^{a'}_b = y_{b'}^a \epsilon^{b'a'} \epsilon_{ab}$ and $y^2 =   y_{a'}^b y^{a'}_b/2 $, in close analogy with the action
of the conformal group on the space-time coordinates $x_{\alpha\dot\alpha}$
\begin{align}\label{x-tran}
x_{\alpha\dot\alpha} \to x_{\alpha\dot\alpha}+ \epsilon_{\alpha\dot\alpha}\,,\qquad
x_{\alpha\dot\alpha} \to \tilde x^{\dot\alpha\alpha}/x^2\,.
\end{align}
We use the following conventions for rising and lowering Lorentz and $SU(2)$ indices
\begin{align}
\tilde x^{\dot\alpha\alpha} = \epsilon^{\alpha\beta}x_{\beta\dot\beta}\epsilon^{\dot\beta\dot\alpha}= x^{\alpha}_{\dot\beta}\epsilon^{\dot\beta\dot\alpha}\,,
\qqqquad
\tilde y^{a'a} = \epsilon^{ab}y_{bb'}\epsilon^{b'a'} = y_{b'}^a\epsilon^{b'a'} \,,
\end{align}
so that (with $x_{ij}=x_i-x_j$ and $y_{ij}=y_i-y_j$)
\begin{align}
(x_{12} \tilde x_{23})_{\alpha}{}^{\beta} = (x_{12})_{\alpha\dot\beta} (\tilde x_{23})^{\dot\beta\beta}\,,\qquad
(y_{12} \tilde y_{23})_a{}^{b} = (y_{12})_{ab'} (\tilde y_{23})^{b'b}\,.
\end{align}
It is straightforward to verify that these expressions transform covariantly under the $SU(4)$ and conformal
transformations, Eqs.~\re{y-tran} and \re{x-tran}, correspondingly,
\begin{align}\notag
(x_{12} \tilde x_{23})_{\alpha}{}^{\beta} {}& \ \to \  {(x_1)^{\dot\alpha\gamma} (x_{12} \tilde x_{23})_\gamma{}^\delta  (\tilde x_3)_{\delta\dot\beta}\over x_1^2 x_2^2 x_3^2}\,,
\\
(y_{12} \tilde y_{23})_{a}{}^{b}{}& \  \to \ {(y_1)^{a'c} (y_{12} \tilde y_{23})_{c}{}^{d} (\tilde y_3)_{db'}\over y_1^2 y_2^2 y_3^2}\,.
\end{align}

\section{Component form of the $R-$invariants}\label{App:R}

In this appendix we work out the expansion of the three-point $R-$invariants \re{eq:10} in powers of the Grassmann variables.
We start with the definition \re{eq:10}
\begin{align} \label{R-app}
R(i;123) {}&=- {\delta^2\Big(\langle\sigma_{i1}\sigma_{i2}\rangle
A_{i3}+\langle\sigma_{i2}\sigma_{i3}\rangle
A_{i1}+\langle\sigma_{i3}\sigma_{i1}\rangle A_{i2}\Big) \over  \vev{\sigma_{i 1} \sigma_{i 2}}  \,\vev{\sigma_{i 2} \sigma_{i 3}}\,\vev{\sigma_{i 3} \sigma_{i 1}}}\,,
\end{align}
where 
$A_{ij}^{a'}  = \left[\vev{\sigma_{ji}\rho_j^{b}} +\vev{\sigma_{ij}\rho_i^{b}}\right] (y_{ij}^{-1})_{b}^{a'}$
with $\rho_i^a \equiv \theta_i^{+a}$. Compared with \re{A-y}, here we put $\theta_*^A=0$ for simplicity.

Expanding \re{R-app} in powers of $\rho$'s we obtain a sum of five different structures antisymmetrised with
respect to the indices of the external legs
\begin{align}\label{eq:18}
R(i;1 2 3)=&
R_{1}(i;12)+\frac12
R_{2}(i;12)+\frac12R_{3}(i;12)+\frac12R_{4}(i;123)+\frac16R_{5}(i;123)+
\text{antisym}_{123} \ .
\end{align}
Here we have defined
\begin{align}
  \label{eq:12}
{}&  R_1(i;12) =  \frac{\bra{\sigma_{i1}} \rho_i   y_{i12} \rho_2
    \ket{\sigma_{2i}}}{(i12)} {x_{i1}^2\over y_{i1}^2} {x_{i2}^2\over
      y_{i2}^2}\,, \notag\\
{}&  R_2(i;12) =- \frac{\bra{\sigma_{1i}} \rho_1   y_{1i2} \rho_2 \ket{\sigma_{2i}}}{(i12)} {x_{i1}^2\over y_{i1}^2} {x_{i2}^2\over
      y_{i2}^2}\,, \notag\\
 {}& R_3(i;12) = \frac{\bra{\sigma_{i1}} \rho_i^2 \ket{\sigma_{i2}} y^2_{12}}{(i12)}
 {x_{i1}^2\over y_{i1}^2} {x_{i2}^2\over
      y_{i2}^2}\,, \notag\\
{}& R_4(i;123) =- {\bra{\sigma_{1i}} \rho_1^2 \ket{\sigma_{1i}}}\frac{x^2_{i1}(i23)}{(i12)(i31)} 
 {x_{i1}^2\over y_{i1}^2}\,,  \notag\\
{}& R_5(i;123) =  -  (\rho_i^\alpha \,  y_{i123i} \rho_{i,\alpha} ) {1\over
      y_{i1}^2y_{i2}^2y_{i3}^2} \,,
\end{align}
where we used  \re{eq:171} and introduced a shorthand notation for 
$\rho_i^\alpha \, y_{i123i} \rho_{i,\alpha} =  \rho_i^{\alpha a}(y_{i123i})_{a}{}^{b} \rho_{i,\alpha b}$,
$\bra{\sigma_{i1}} \rho_i   y_{i12} \rho_2
    \ket{\sigma_{2i}}= {\sigma_{i1}^\alpha} \rho_{i,\alpha}^a ( y_{i12})_a{}^b \rho_{2,b}^\beta
    {\sigma_{2i,\beta}}$,
etc. 

The functions $R_1$, $R_2$ and $R_3$ depend on two external points  and change sign under their exchange,
$R_k(i,12) = - R_k(i;21)$. The function $R_5(i;123)$ is completely antisymmetric in 1,2,3 and $R_4(i;123)= -R_4(i;132)$. The rational factors are introduced in \re{eq:18} to avoid double counting due to these symmetries. 

We can apply \re{eq:18} to calculate various components in the product of $R-$invariants.  For instance, to
find the component $(\rho_{1}^{2})^{ab}(\rho_{3}^{2})^{cd}$ in \re{I1234} we use
\begin{align} \notag\label{eq:21}
  R(1;234)R(3;412) {}& =-R_{1}(1;23)R_{1}(3;21)+R_{1}(1;23)R_{1}(3;41)
  \\[2mm]
  & -R_{1}(1;43)R_{1}(3;41)
  +R_{1}(1;43)R_{1}(3;21)+R_{5}(1;234)R_{5}(3;412)+\dots
\end{align}
where the dots denote terms that do not produce the above mentioned component. The first term in (\ref{eq:21}) gives:
\begin{align}
R_{1}(1;23)R_{1}(3;21)=\frac{\bra{\sigma_{12}}\rho_1 \tilde y_{123}\rho_3\ket{\sigma_{31}}}{(123)d_{12}d_{13}}\frac{\bra{\sigma_{32}}\rho_3\tilde y_{321}\rho_1\ket{\sigma_{13}}}{(321)d_{23}d_{13}}\, .
\end{align}
We can then decompose the product of two $\rho$'s belonging to the same point into irreducible components
with the help of the identity
\begin{align}
\rho_{\alpha}^{a}\rho_{\beta}^{b}=\frac{1}{2}\epsilon_{\alpha\beta}(\rho^{2})^{ab}+\frac{1}{2}\epsilon^{ab}(\rho^{2})_{\alpha\beta}\,.
\end{align}
To get the  component $(\rho_{1}^{2})^{ab}(\rho_{3}^{2})^{cd}$ we can neglect the second term. In this way, we obtain
 \begin{align}
 R_{1}(1;23)R_{1}(3;21)
 =-\frac{(y_{123})_{ab}(y_{321})_{cd}(\rho^{2}_{1})^{ad}(\rho^{2}_{3})^{bc}}{4x_{12}^{2}x_{13}^{4} x_{32}^{2}d_{13}^{2}d_{12}d_{23}}+\dots\,,
 \end{align}
where we used \re{eq:171} to replace $\vev{\sigma_{12}\sigma_{13}} = (123)/(x_{12}^2 x_{13}^2)$ and
$\vev{\sigma_{32}\sigma_{31}} = (321)/(x_{13}^2 x_{23}^2)$.
Performing similar manipulations we find
\begin{align}\notag
{}& R_{1}(1;23)R_{1}(3;41)=\frac{(y_{123})_{ab}(y_{143})_{dc}(\rho^{2}_{1})^{ad}(\rho^{2}_{3})^{bc}}{4x_{12}^{2}x_{13}^{2}x_{13}^{2}x_{34}^{2}d_{13}^{2}d_{34}d_{12}}+\dots\, ,
\\
{}& R_{5}(1;234)R_{5}(3;412)=\frac{(y_{12341})_{ab}(y_{34123})_{cd}(\rho_{1}^{2})^{ab}(\rho_{3}^{2})^{cd}}{x_{12}^{2}x_{14}^{2}x_{34}^{2}x_{13}^{2}x_{23}^{2}y_{13}^{2}}+\dots\,.
\end{align}
The remaining terms on the right-hand side of \re{eq:21} can be obtained from the last two relations by swapping the indices $2\leftrightarrow 4$. Substituting these expressions into \re{eq:21} we arrive at the first relation in \re{eq:19}.
 
Let us show that the invariants \re{R-app} satisfy relation \re{3pt-new}. We start with the $U(1)$ decoupling relation \re{U(1)}
for the $4-$point vertex
\begin{align}\label{U-4pt}
R(1;abcd) + R(1;acdb) + R(1; adbc) =0
\end{align}
and use \re{eq:11} together with \re{anti} to factor out each term on the left-hand side into a product of $3-$point vertices
\begin{align} \notag
& R(1;abcd) = R(1;abc)R(1;cda)= -R(1;abc)R(1;dca)\,, 
\\ \notag
 & R(1;acdb) =R(1;acb)R(1;cdb) =-R(1;abc)R(1;dbc)  \,,  
\\  
&  R(1; adbc) = R(1;abc) R(1;adb)=-R(1;abc) R(1;dab)\,.
\end{align}
In this way, we obtain from \re{U-4pt}
\begin{align}
 R(1;abc) \big[R(1;dca)+R(1;dbc) + R(1;dab) \big] = 0\,.
\end{align}
It follows from \re{R-app} that $R(1;abc)^2=0$ and, therefore, the general solution to this relation is 
\begin{align}\label{kappa}
 R(1;dca)+R(1;dbc) + R(1;dab) = \kappa R(1;abc)\,.
\end{align} 
We can use \re{eq:44} to verify that the expression on the left-hand side  has zero residue at the poles $(1di)=0$ with $i=a,b,c$,
implying that $\kappa$ does not depend on the choice of point $d$. Putting $d=a$ on both sides and making use of \re{R3=0} we find that $\kappa=1$.  We can obtain the same
result by replacing the $R-$invariants in \re{kappa} by their explicit expressions \re{eq:18} and \re{eq:12}.

\section{The components of the five-point correlator}
\label{5ptcomp}

In this appendix we summarise the expressions for the eight coefficient functions defining the $5-$point correlation
function $G_{5;1}$ in \re{14}. Going through the steps outlined in Sect.~\ref{sect:warm} we can compute them
in terms of bosonic and fermonic $T-$blocks \re{T1} and \re{T2}. One of the coefficient function is given by \re{f1}
and the remaining seven functions are
 \begin{align} \label{f(1)}
 f(1) 
 = \frac{2}{3}  \frac{c_5}{\prod x^2_{ij}}   
\biggl[ {}&
(y^2_{23} y_{45}^2 x_{25}^2 x_{34}^2 - x^2_{23} x_{45}^2 y_{25}^2 y_{34}^2)
(y^2_{23} y_{45}^2 x_{24}^2 x_{35}^2 - x^2_{23} x_{45}^2 y_{24}^2 y_{35}^2) 
\nt
+ {}&  (y^2_{24} y_{35}^2 x_{25}^2 x_{34}^2 - x^2_{24} x_{35}^2 y_{25}^2 y_{34}^2)
(y^2_{24} y_{35}^2 x_{23}^2 x_{45}^2 - x^2_{24} x_{35}^2 y_{23}^2 y_{45}^2)
\nt
+ {}& 
(y^2_{25} y_{34}^2 x_{23}^2 x_{45}^2 - x^2_{25} x_{34}^2 y_{23}^2 y_{45}^2)
(y^2_{25} y_{34}^2 x_{24}^2 x_{35}^2 - x^2_{25} x_{34}^2 y_{24}^2 y_{35}^2) 
\biggr]
\\
f^{(\a\b)(ab)}(1,2)   =
- \frac{c_5}{\prod x^2_{ij}}   \biggl[{}&
y_{34}^2 y_{45}^2 x_{24}^2 x_{35}^2 (y_{23} \widetilde{y}_{31} y_{15} \widetilde{y}_{52} )^{(ab)}
(x_{14} \widetilde{x}_{45} x_{53} \widetilde{x}_{31})^{(\alpha \beta)}    
\notag \\
-{} &  x_{34}^2 x_{45}^2 y_{14}^2 y_{35}^2 (x_{13} \widetilde{x}_{32} x_{25} \widetilde{x}_{51} )^{(\alpha\beta)}
(y_{24} \widetilde{y}_{45} y_{53} \widetilde{y}_{32})^{(ab)} \biggr] + \text{perm}_{345} 
\end{align}
\begin{align} 
 \label{su2su25}
f^{(ab)(cd)}(1,2) =
-2 \frac{c_5}{\prod x^2_{ij}}   \biggl[ {}&
\frac{1}{2} x_{14}^2 x_{24}^2 x_{35}^4 \, y_{34}^2 y_{45}^2 (y_{13} \widetilde{y}_{32})^{{}^{(a}{}_{(c}} (y_{15} \widetilde{y}_{52} )^{{}^{b)}{}_{d)}} 
 \notag \\ 
+ {}&
x_{13}^2 x_{25}^2 x_{34}^2 x_{45}^2 \, y_{35}^2 (y_{15} \widetilde{y}_{54} y_{43} \widetilde{y}_{32} )^{{}^{(a}{}_{(c}} (y_{14} \widetilde{y}_{42})^{{}^{b)}{}_{d)}} 
 \notag\\ +{}&
\frac{1}{6} x_{12}^2 x_{34}^2 x_{35}^2 x_{45}^2 (y_{13} \widetilde{y}_{34} y_{45} \widetilde{y}_{51} )^{(ab)} 
(y_{23} \widetilde{y}_{34} y_{45} \widetilde{y}_{52} )^{(cd)} \biggr]
 + \text{perm}_{345}  
\\
\label{LorLor5}
f^{(\a\b)(\gamma \delta)}(1,2) =
2\frac{c_5}{\prod x^2_{ij}}   \biggl[ {}&
\frac{1}{2} y_{14}^2 y_{24}^2 y_{35}^4 \, x_{34}^2 x_{45}^2 (x_{13} \widetilde{x}_{32})^{{}^{(\alpha}{}_{(\gamma}} (x_{15} \widetilde{x}_{52} )^{{}^{\beta)}{}_{\delta)}} 
 \notag\\
 + {}&
y_{15}^2 y_{23}^2 y_{34}^2 y_{45}^2 \, x_{35}^2 (x_{13} \widetilde{x}_{34} x_{45} \widetilde{x}_{52} )^{{}^{(\alpha}{}_{(\gamma}} (x_{14} \widetilde{x}_{42})^{{}^{\beta)}{}_{\delta)}} 
 \notag\\ 
 + {}&
\frac{1}{6} y_{12}^2 y_{34}^2 y_{35}^2 y_{45}^2 (x_{14} \widetilde{x}_{43} x_{35} \widetilde{x}_{51} )^{(\alpha \beta)} 
(x_{24} \widetilde{x}_{43} x_{35} \widetilde{x}_{52} )^{(\gamma \delta)} \biggr]
 + \text{perm}_{345} 
\\
f^{\a\b\gamma\delta,abcd}(1,2,3,4) = 
8 c_5 {}& \frac{1}{x_{15}^2 x_{25}^2 x_{35}^2 x_{45}^2} \frac{y_{14}^2 y_{23}^2}{x_{12}^2 x_{14}^2 x_{23}^2 x_{34}^2}
(y_{15} \widetilde{y}_{52})^{ab} (x_{15} \widetilde{x}_{52})^{\a\b}
(y_{35} \widetilde{y}_{54})^{cd} (x_{35} \widetilde{x}_{54})^{\gamma\delta}  \notag\\ 
{}& \hspace*{80mm} + \text{graded perm}_{234}
\end{align}%
\begin{align}%
f^{\a\b,ab(cd)}{}& (1,2,3) =
4 \frac{c_5}{\prod x^2_{ij}} \biggl[  
 (x_{14} \widetilde{x}_{42})^{\a\b} 
\bigl( x_{12}^2 x_{35}^2 x_{45}^2 y_{15}^2 y_{25}^2 (y_{14} \widetilde{y}_{43})^{a(c} (y_{24} \widetilde{y}_{43})^{bd)}  
\notag \\
- {}& x_{14}^2 x_{25}^2 x_{35}^2 y_{15}^2 y_{45}^2 (y_{12} \widetilde{y}_{23})^{a(c} (y_{24} \widetilde{y}_{43})^{bd)} 
- x_{15}^2 x_{24}^2 x_{35}^2 y_{25}^2 y_{45}^2 (y_{14} \widetilde{y}_{43})^{a(c} (y_{21} \widetilde{y}_{13})^{bd)} 
\notag\\
+{}& x_{15}^2 x_{25}^2 x_{34}^2 y_{45}^2 (y_{14} \widetilde{y}_{42})^{ab} (y_{31} \widetilde{y}_{12} y_{25} \widetilde{y}_{53})^{(cd)} 
+ x_{13}^2 x_{25}^2 x_{45}^2 y_{15}^2 (y_{14} \widetilde{y}_{42})^{ab} (y_{32} \widetilde{y}_{24} y_{45} \widetilde{y}_{53})^{(cd)} 
 \notag\\
- {}& x_{15}^2 x_{23}^2 x_{45}^2 y_{25}^2 (y_{14} \widetilde{y}_{42})^{ab} (y_{31} \widetilde{y}_{14} y_{45} \widetilde{y}_{53})^{(cd)} \bigr) 
\notag\\
+{}&  \bigl( x_{12}^2 x_{45}^2 y_{15}^2 y_{24}^2 - x_{15}^2 x_{24}^2 y_{12}^2 y_{45}^2 \bigr)
(x_{14} \widetilde{x}_{43} x_{35} \widetilde{x}_{52})^{\a\b} (y_{14} \widetilde{y}_{43})^{a(c} (y_{25} \widetilde{y}_{53})^{bd)}  
\notag\\
+ {}&  x_{45}^2 (x_{15}^2 x_{24}^2 y_{14}^2 y_{25}^2 - x_{14}^2 x_{25}^2 y_{15}^2 y_{24}^2)
(x_{13} \widetilde{x}_{32})^{\a\b} (y_{14} \widetilde{y}_{43})^{a(c} (y_{25} \widetilde{y}_{53})^{bd)}
\biggr]
 + \text{perm}_{45}
\end{align}%
\begin{align}%
f^{\a\b(\gamma\delta),ab} {}& (1,2,3) =
4 \frac{c_5}{\prod x^2_{ij}}   \Bigl[
(y_{14} \widetilde{y}_{42})^{ab} \bigl( y_{12}^2 y_{35}^2 y_{45}^2 x_{15}^2 x_{25}^2 
(x_{14} \widetilde{x}_{43})^{\a(\gamma} (x_{24} \widetilde{x}_{43})^{\b\delta)} 
 \notag \\ 
 - {}&
y_{14}^2 y_{25}^2 y_{35}^2 x_{15}^2 x_{45}^2  (x_{12} \widetilde{x}_{23})^{\a(\gamma} (x_{24} \widetilde{x}_{43})^{\b\delta)} - 
y_{15}^2 y_{24}^2 y_{35}^2 x_{25}^2 x_{45}^2  (x_{14} \widetilde{x}_{43})^{\a(\gamma} (x_{21} \widetilde{x}_{13})^{\b\delta)} 
\notag \\
+{}& y_{15}^2 y_{25}^2 y_{34}^2 x_{45}^2 (x_{14} \widetilde{x}_{42})^{\a\b} (x_{31} \widetilde{x}_{12} x_{25} \widetilde{x}_{53})^{(\gamma\delta)}
+ y_{13}^2 y_{25}^2 y_{45}^2 x_{15}^2 (x_{14} \widetilde{x}_{42})^{\a\b} (x_{32} \widetilde{x}_{24} x_{45} \widetilde{x}_{53})^{(\gamma\delta)} 
\notag\\ 
-{}&   
y_{15}^2 y_{23}^2 y_{45}^2 x_{25}^2 (x_{14} \widetilde{x}_{42})^{\a\b} (x_{31} \widetilde{x}_{14} x_{45} \widetilde{x}_{53})^{(\gamma\delta)} 
\bigr)  
\notag\\ 
+{}& ( y_{12}^2 y_{45}^2 x_{15}^2 x_{24}^2 - y_{15}^2 y_{24}^2 x_{12}^2 x_{45}^2) 
(y_{14} \widetilde{y}_{43} y_{35} \widetilde{y}_{52})^{\a\b} (x_{14} \widetilde{x}_{43})^{\a(\gamma} (x_{25} \widetilde{x}_{53})^{\b\delta)} 
 \notag\\ 
 + {}&
y_{45}^2 (y_{15}^2 y_{24}^2 x_{14}^2 x_{25}^2 - y_{14}^2 y_{25}^2 x_{15}^2 x_{24}^2 ) (y_{13} \widetilde{y}_{32})^{ab} 
(x_{14} \widetilde{x}_{43})^{\a(\gamma} (x_{25} \widetilde{x}_{53})^{\b\delta)} \Bigr] + \text{perm}_{45}
\end{align}
Multiplied by $\prod_{i<j} x_{ij}^2$,
these expressions have a definite parity under the exchange of spatial and harmonic coordinates, $x_i\leftrightarrow y_i$.
Namely, $f$ and $f_{\a\b\gamma\delta,abcd}$ are invariant under this transformation,  
$f_{(\a\b)(ab)}(1,2)$ transforms into $- f_{(\a\b)(ab)}(2,1)$; $f_{(ab)(cd)}$ and $f_{(\a\b)(\gamma\delta)}$ transform
into each other as well as $f_{\a\b(\gamma\delta),ab}$ into $f_{\a\b,ab(cd)}$.
To understand the origin of these properties, we notice that, according to the second relation in \re{inv5},  $\mathcal{I}_{5;1}$ is 
invariant under $x_i \leftrightarrow y_i$.
Consequently the correlation function $G_{5;1}$ (as well as its components) inherit the same symmetry.

\section{Useful identities}
\label{D}
In this appendix we prove some identities that we used in computing the correlation function in the 
twistor approach. They involve the variables $\sigma_{ij}$ defined in \re{sigma}. Using the gauge \re{Z-gauge},
we can express them in terms of the spatial coordinates $x$ as
\begin{align}\label{53}
\sigma_{ij}^\alpha = \epsilon^{\alpha\beta}{\vev{Z_{i,\beta} Z_*Z_{j,1} Z_{j,2}} \over \vev{Z_{i,1} Z_{i,2} Z_{j,1} Z_{j,2}}}= {(x_{ij}^{-1} \tilde x_{j0} |{0}\rangle)^\alpha }\,,
\end{align}
where the auxiliary point $x_0$ and spinor $\ket{0}\equiv \lambda_0$ originate from the expression for the reference twistor
\begin{align}\label{59}
  Z_*^I = (\lambda_{0,\alpha}, i x_0^{\dot\alpha\beta}\lambda_{0,\beta})\,.
\end{align}
Then, we apply \re{53} to obtain the following representation for the brackets $(ijk)$ introduced in \re{66}
\begin{align}\label{any}
(ijk) = \vev{\sigma_{ij} \sigma_{ik}} x_{ij}^2 x_{ik}^2 = \vev{0|x_{0j}\tilde x_{ji} x_{ik}\tilde x_{k0}|0}\,.
\end{align}
It is straightforward to verify that
\begin{align}
(ijk) =  x_{0i}^2 \vev{0|x_{0j}\tilde x_{0k}|0} -x_{0j}^2 \vev{0|x_{0i}\tilde x_{0k}|0} + x_{0k}^2 \vev{0|x_{0i}\tilde x_{0j}|0}
\end{align} 
so that $(ijk)$ is completely antisymmetric in the indices. 
 
Let us show that  the following identities take place
\begin{align} \notag\label{Iden1} 
&   {
  (\sigma_{13}^\alpha \sigma_{21}^\beta)}+{
  (\sigma_{12}^\alpha \sigma_{23}^\beta)}-{
  (\sigma_{13}^\alpha \sigma_{23}^\beta)} = {(x_{13} \tilde x_{32})^{\alpha \beta}\over x_{12}^2 x_{23}^2 x_{31}^2}(123)\,,
  \\[2mm]
&    (i12)(i34) + (i13)(i42) +(i14)(i23) = 0\,.
\end{align} 
To begin with we notice that both relations stay invariant under the conformal transformations acting both on the external
points $1,2,3,4,i$ and on the auxiliary point $0$ defining the reference twistor \re{59}. We can then use the conformal
symmetry to put $x_2=0$ and $x_3\to\infty$ in \re{Iden1}. Under this choice the first relation in \re{Iden1} simplifies as
\begin{align}
  {|{0}\rangle^\alpha (x_{1}^{-1} \tilde x_{10} |{0}\rangle)^\beta }
+({x_{1}^{-1} \tilde x_{0} |{0}\rangle)^\alpha   |{0}\rangle^\beta  }
 -   |{0}\rangle^\alpha   |{0}\rangle^\beta =- \epsilon^{\alpha\beta} \vev{0|x_1^{-1}\tilde x_0 |0}
\end{align} 
and it is obviously satisfied. We can prove the second relation in  \re{Iden1} in a similar manner by choosing $x_i\to\infty$
and $x_2=0$.

Finally, we prove of the non-trivial six-term identity  
\begin{align}\notag\label{eq:id}
 {}& (234)(341)x_{12}^2 - (234)(124)x_{13}^2 + (123)(234)x_{14}^2
 \\[2mm]
 {}& \qquad + (124)(134)x_{23}^2 -(123)(134)x_{24}^2+(123)(124)x_{34}^2=0\,.
\end{align}
It is convenient to introduce an auxiliary dual reference twistor $\tilde{Z}_{*}$ normalised as $\tilde{Z}_{*A}Z_{*}^{A}=1$.
It then allows us to define two sets of dual variables 
\begin{align}\label{dual-var}
\tilde Z_{i A} =   X_{i,AB} Z_*^B\,,\qqqquad 
\hat{Z}_{i}^{A}=X_{i}^{AB}\tilde{Z}_{*B}\,,  
\end{align}
with $X_i^{BC} = Z_{i,1}^{B} Z_{i,2}^{C}-Z_{i,1}^{C} Z_{i,2}^{B}$ and $X_{i,AB} =\frac12\epsilon_{ABCD} X_i^{CD}$ . They satisfy the relations
\begin{align}\label{norm}
\tilde Z_{j A} Z_*^A = \hat{Z}_{i}^{A}\tilde{Z}_{*A}=0\,. 
\end{align}
We also notice that since the $X_{AB}$ takes values in the Clifford algebra of  $\text{SU}(4)$, the following holds true:
\begin{align}\label{Cli}
\hat{Z}^{A}_{i}\tilde{Z}_{jA}+\hat{Z}_{j}^{A}\tilde{Z}_{iA}=-\tilde{Z}_{*A}Z_{*}^{C}(X_{i}^{AB} {X}_{jBC}+X_{j}^{AB} {X}_{iBC})=-(X_{i}\cdot X_{j}) \,.
\end{align}
Using the dual variables \re{dual-var} we can obtain two equivalent representations for $(ijk)$ defined in \re{53} and \re{any}
\begin{align}
(ijk)= \frac{1}{2}\epsilon^{ABCD}\tilde{Z}_{iA}\tilde{Z}_{jB}\tilde{Z}_{kC}\tilde{Z}_{*D}
= \frac{1}{2}\epsilon_{ABCD}\hat{Z}_{i}^{A}\hat{Z}_{j}^{B}\hat{Z}_{k}^{C} {Z}_{*}^{D}
 \equiv    \langle \, i j k * \,\rangle\,.
\end{align}
According to  \re{norm}, the twistors $\tilde Z_{jA}$ with $j=1,\dots,4$ are all orthogonal to $Z_*^A$, therefore, they are linear dependent. The same is true for  $\hat Z_{j}^{A}$ with $j=1,\dots,4$.
This yields two identities
\begin{align}\label{eq:E1} \notag
{}&\tilde{Z}_{1A}\left\langle  {2} {3} {4} {*}\right\rangle+\tilde{Z}_{2A}\left\langle  {3} {4} {*} {1}\right\rangle+\tilde{Z}_{3A}\left\langle  {4} {*} {1} {2}\right\rangle+\tilde{Z}_{4A}\left\langle  {*} {1} {2} {3}\right\rangle=0
\,,
\\
{}&\hat{Z}_{1}^{A}\left\langle  {2} {3} {4} {*}\right\rangle+\hat{Z}_{2}^{A}\left\langle  {3} {4} {*} {1}\right\rangle+\hat{Z}_{3}^{A}\left\langle  {4} {*} {1} {2}\right\rangle+\hat{Z}_{4}^{A}\left\langle  {*} {1} {2} {3}\right\rangle=0
\end{align}
Finally we multiply the expressions on the left-hand side and contract the $SU(4)$ indices to get
\begin{align} \notag
{}&  (234)(341) (X_{1} \cdot X_{2})-(123)(134)(X_{2} \cdot X_{4})-(234)(124)(X_{1}  \cdot X_{3})
\\
{}&
+(123)(234)(X_{1} \cdot X_{4})+(124)(134)(X_{2} \cdot X_{3})+(123)(124)(X_{3}\cdot  X_{4})=0\,.
\end{align}
where we made use of \re{Cli} and took into account that $(X_i \cdot X_i)=0$.
Since the last relation is homogenous in $X$'s we can employ the gauge \re{Z-gauge} and replace
$(X_i\cdot X_j)=x_{ij}^2$ to arrive at \re{eq:id}.


\begin{thebibliography}{99}
 
\bibitem{Eden:2011we}
  B.~Eden, P.~Heslop, G.~P.~Korchemsky and E.~Sokatchev,
  Nucl.\ Phys.\ B {\bf 862} (2012) 193
  [arXiv:1108.3557].


\bibitem{Eden:2012tu}
  B.~Eden, P.~Heslop, G.~P.~Korchemsky and E.~Sokatchev,
  Nucl.\ Phys.\ B {\bf 862} (2012) 450
  [arXiv:1201.5329].

\bibitem{Freedman:1998tz}
  D.~Z.~Freedman, S.~D.~Mathur, A.~Matusis and L.~Rastelli,
  Nucl.\ Phys.\ B {\bf 546} (1999) 96
  [hep-th/9804058].
\\
  D.~Z.~Freedman, S.~D.~Mathur, A.~Matusis and L.~Rastelli,
  Phys.\ Lett.\ B {\bf 452} (1999) 61
  [hep-th/9808006].
  
\bibitem{Arutyunov:2000py}
  G.~Arutyunov and S.~Frolov,
  Phys.\ Rev.\ D {\bf 62} (2000) 064016
  [hep-th/0002170].

\bibitem{GonzalezRey:1998tk}
  F.~Gonzalez-Rey, I.~Y.~Park and K.~Schalm,
  Phys.\ Lett.\ B {\bf 448} (1999) 37
  [hep-th/9811155].
 
\bibitem{Eden:2000mv}
  B.~Eden, C.~Schubert and E.~Sokatchev,
  Phys.\ Lett.\ B {\bf 482} (2000) 309
  [hep-th/0003096].

\bibitem{Bianchi:2000hn}
  M.~Bianchi, S.~Kovacs, G.~Rossi and Y.~S.~Stanev,
  Nucl.\ Phys.\ B {\bf 584} (2000) 216
  [hep-th/0003203].

\bibitem{Drummond:2013nda}
  J.~Drummond, C.~Duhr, B.~Eden, P.~Heslop, J.~Pennington and V.~A.~Smirnov,
  JHEP {\bf 1308} (2013) 133
  [arXiv:1303.6909].
 
\bibitem{Dolan:2000ut}
  F.~A.~Dolan and H.~Osborn,
  Nucl.\ Phys.\ B {\bf 599} (2001) 459
  [hep-th/0011040].
  
\bibitem{Alday:2010zy}
  L.~F.~Alday, B.~Eden, G.~P.~Korchemsky, J.~Maldacena and E.~Sokatchev,
  JHEP {\bf 1109} (2011) 123
  [arXiv:1007.3243].
 
\bibitem{Eden:2010zz}
  B.~Eden, G.~P.~Korchemsky and E.~Sokatchev,
  JHEP {\bf 1112} (2011) 002
  [arXiv:1007.3246].


\bibitem{Mason:2010yk}
  L.~J.~Mason and D.~Skinner,
  JHEP {\bf 1012} (2010) 018
  [arXiv:1009.2225].


\bibitem{Eden:2011yp}
  B.~Eden, P.~Heslop, G.~P.~Korchemsky and E.~Sokatchev,
  Nucl.\ Phys.\ B {\bf 869} (2013) 329
  [arXiv:1103.3714].

\bibitem{Eden:2011ku}
  B.~Eden, P.~Heslop, G.~P.~Korchemsky and E.~Sokatchev,
  Nucl.\ Phys.\ B {\bf 869} (2013) 378
  [arXiv:1103.4353].


\bibitem{Adamo:2011dq}
  T.~Adamo, M.~Bullimore, L.~Mason and D.~Skinner,
  JHEP {\bf 1108} (2011) 076
  [arXiv:1103.4119].


\bibitem{Drummond:2008vq}
  J.~M.~Drummond, J.~Henn, G.~P.~Korchemsky and E.~Sokatchev,
  Nucl.\ Phys.\ B {\bf 828} (2010) 317
  [arXiv:0807.1095].
  
\bibitem{Mason:2009qx}
  L.~J.~Mason and D.~Skinner,
  JHEP {\bf 0911} (2009) 045
  [arXiv:0909.0250].



\bibitem{Boels:2006ir}
  R.~Boels, L.~J.~Mason and D.~Skinner,
  JHEP {\bf 0702} (2007) 014
  [hep-th/0604040].

\bibitem{HSS} A.~Galperin, E.~Ivanov, S.~Kalitsyn, V.~Ogievetsky and E.~Sokatchev,
  Class.\ Quant.\ Grav.\  {\bf 1} (1984) 469;
\\
  A.~S.~Galperin, E.~A.~Ivanov, V.~I.~Ogievetsky and E.~S.~Sokatchev,
  ``Harmonic Superspace,''
{\it  Cambridge, UK: Univ. Pr. (2001) 306 p}


\bibitem{HH}
  P.~S.~Howe and G.~G.~Hartwell,
  Class.\ Quant.\ Grav.\  {\bf 12} (1995) 1823.

\bibitem{HH2}
   P.~J.~Heslop  and  P.~S.~Howe,
  Phys.\ Lett.\  {\bf B516 } (2001)  367-375.
  [hep-th/0106238];
\\
  P.~J.~Heslop  and P.~S.~Howe,
  Nucl.\ Phys.\  {\bf B626 } (2002)  265-286.
  [hep-th/0107212];
  \\
   P.~J.~Heslop and P.~S.~Howe,
  JHEP {\bf 0401} (2004) 058
  [arXiv:hep-th/0307210].

\bibitem{Cachazo:2004kj}
  F.~Cachazo, P.~Svrcek and E.~Witten,
  JHEP {\bf 0409} (2004) 006
  [hep-th/0403047].
  
\bibitem{Bullimore:2010pj}
  M.~Bullimore, L.~J.~Mason and D.~Skinner,
  JHEP {\bf 1012} (2010) 032
  [arXiv:1009.1854].

\bibitem{Adamo:2011pv}
  T.~Adamo, M.~Bullimore, L.~Mason and D.~Skinner,
  J.\ Phys.\ A {\bf 44} (2011) 454008
  [arXiv:1104.2890].

\bibitem{Adamo:2011cd}
  T.~Adamo,
  JHEP {\bf 1112} (2011) 006
  [arXiv:1110.3925 [hep-th]].
  
\bibitem{Koster:2014fva}
  L.~Koster, V.~Mitev and M.~Staudacher,
  arXiv:1410.6310.


\bibitem{Brandhuber:2014pta}
  A.~Brandhuber, B.~Penante, G.~Travaglini and D.~Young,
  Phys.\ Rev.\ Lett.\  {\bf 114} (2015) 071602
  [arXiv:1412.1019 [hep-th]].

  
\bibitem{Dixon:1996wi}
  L.~J.~Dixon,
  In *Boulder 1995, QCD and beyond* 539-582
  [hep-ph/9601359].
  

\bibitem{Gatheral:1983cz}
  J.~G.~M.~Gatheral,
  Phys.\ Lett.\ B {\bf 133} (1983) 90.


\bibitem{Frenkel:1984pz}
  J.~Frenkel and J.~C.~Taylor,
  Nucl.\ Phys.\ B {\bf 246} (1984) 231.
  
\bibitem{Drummond:2009fd}
  J.~M.~Drummond, J.~M.~Henn and J.~Plefka,
  JHEP {\bf 0905} (2009) 046
  [arXiv:0902.2987].

\bibitem{ArkaniHamed:2010gh}
  N.~Arkani-Hamed, J.~L.~Bourjaily, F.~Cachazo and J.~Trnka,
  JHEP {\bf 1206} (2012) 125
  [arXiv:1012.6032].
  
\bibitem{Arkani-Hamed:2013jha}
  N.~Arkani-Hamed and J.~Trnka,
  JHEP {\bf 1410} (2014) 30
  [arXiv:1312.2007].
 
 
\bibitem{Belitsky:2014zha}
  A.~V.~Belitsky, S.~Hohenegger, G.~P.~Korchemsky and E.~Sokatchev,
  arXiv:1409.2502.
 
\end{thebibliography}
\end{document}